\documentclass[aps,twocolumn,amsmath,amssymb,showpacs,prb,floatfix]{revtex4-1}

\usepackage{epsfig}
\usepackage{amsmath}
\usepackage{graphics}
\usepackage{slashed}
\usepackage{graphicx}
\usepackage{color, soul}
\usepackage{cleveref}
\usepackage{csquotes}
\usepackage{ragged2e}
\usepackage{csquotes}
\usepackage{natbib}

\usepackage{indentfirst}
\usepackage{amssymb}
\usepackage{cancel}
\usepackage{leftidx}
\usepackage{multirow}
\usepackage{epstopdf}
\usepackage{float}

\newcommand{\bq}{\bf{q}}
\newcommand{\vect}[1]{\boldsymbol{\mathbf{#1}}}
\newcommand{\B}{\Bigg}
\begin{document}

\title{Emergent Ising degrees of freedom above double-stripe magnetism }
\author{Guanghua Zhang, Rebecca Flint}
\affiliation{Department of Physics and Astronomy, Iowa State University,
  12 Physics Hall, Ames, Iowa 50011, USA}
\affiliation{Division of Materials Science and Engineering, Ames
  Laboratory, U.S.  DOE, Ames, Iowa 50011, USA}

\begin{abstract}
Double-stripe magnetism $[\mathbf{Q}=(\pi/2, \pi/2)]$ has been proposed as the magnetic ground state for both the iron-telluride and BaTi$_2$Sb$_2$O families of superconductors. Double-stripe order is captured within a $J_1-J_2-J_3$ Heisenberg model in the regime $J_3 \gg J_2 \gg J_1$. Intriguingly, besides breaking spin-rotational symmetry, the ground state manifold has three additional Ising degrees of freedom associated with bond-ordering. Via their coupling to the lattice, they give rise to an orthorhombic distortion and to two non-uniform lattice distortions with wave-vector $(\pi, \pi)$. Because the ground state is four-fold degenerate, modulo rotations in spin space, only two of these Ising bond order parameters are independent. Here we introduce an effective field theory to treat all Ising order parameters, as well as magnetic order, and solve it within a large-$N$ limit. All three transitions, corresponding to the condensations of two Ising bond order parameters and one magnetic order parameter are simultaneous and first order in three dimensions, but lower dimensionality, or equivalently weaker interlayer coupling, and weaker magnetoelastic coupling can split the three transitions, and in some cases allows for two separate Ising phase transitions above the magnetic one.
\end{abstract}

\maketitle
\section{Introduction}

Long range order that breaks both discrete and continuous symmetries can, in the presence of strong fluctuations, be melted in stages, whereby the discrete symmetries may remain broken well above the continuous symmetry breaking\cite{Kivelson98}. 
The most famous example is the spin-driven nematicity that occurs in the iron-based superconductors. The single-stripe(SS) magnetic ground state\cite{FeAs2008, Ma2008} breaks both continuous spin rotation symmetry and discrete $C_4$ lattice rotation symmetry, allowing a nematic phase breaking only the rotation symmetry to develop above the magnetic transition where the spin-rotation symmetry is broken\cite{ccl}. 
Essentially, this nematic order can be understood as an Ising bond-order, where ferromagnetic or antiferromagnetic correlations develop along one direction, but not the other. As this bond order breaks rotational symmetry, it couples to the development of an orthorhombic lattice distortion that occurs coincidently with the nematic phase transition\cite{Fang2008, Xu2008}. There is now a clear consensus that the orthorhombic phase in the iron-pnictides is just such a spin-driven nematic phase, where the primary order parameter is this Ising bond order\cite{Fernandes2014}. This order has been found in both local\cite{Si2008, Si2011, Fang2008, Xu2008} and itinerant\cite{Qi2009, Fernandes2010} models, and appears to be quite generic.
Indeed, this phenomena is relevant beyond the iron-pnictides, and has recently been explored above the charge density wave phase proposed in the cuprates\cite{Chubukov2014, Kivelson2014}, and in tetragonal Kondo insulators\cite{Galitski2015}. The nematic degrees of freedom themselves may be important for driving higher temperature superconducting transitions\cite{Yoshizawa2012, Gallais2013, Fisher_arxiv, Lederer2015}. In this paper, we explore the nematicity that can occur above the double-stripe (DS) magnetic state, which breaks not one, but two distinct discrete symmetries in addition to the spin-rotation symmetry.  Here, fluctuations can melt the magnetic order via up to three distinct phase transitions: one magnetic and two Ising bond order transitions associated with the two discrete symmetries\cite{Zhang2017}.

\begin{figure}
\begin{centering}
\includegraphics[width=0.9\columnwidth]{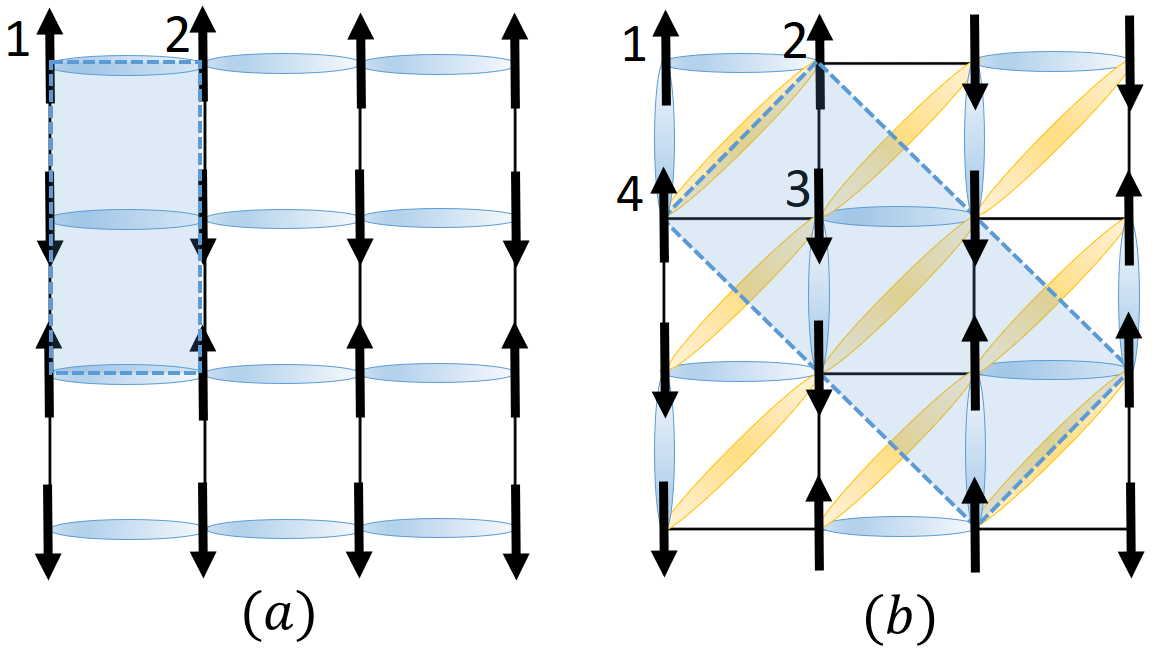} 
\par\end{centering}
\caption{(Color online) Comparison of (a) SS magnetic order in $\mathrm{FeAs}$, with an orthorhombic lattice distortion and (b)  DS magnetic order in $\mathrm{FeTe}$, with a monoclinic lattice distortion, rotated 45$^\circ$ from the SS distortion. The sublattices are as labeled. The NN and NNN ferromagnetic bonds are indicated by blue and yellow ovals respectively. The shaded area included by blue dashed line indicates the unit cell.}
\label{fig:FeAs_FeTe}
\end{figure}

The DS magnetic ground state has been proposed in $\mathrm{BaTi_2Sb_2 O}$\cite{Yajima2012, Doan2012, Frandsen2014, Singh2012} and found in the 11 system $\mathrm{Fe}_{1+y}\mathrm{Se}_x \mathrm{Te}_{1-x}$\cite{Fruchart1975, Martinelli2010}, which exhibits magnetic order with the commensurate ordering vector $\vect{Q}=(\pi/2, \pi/2)$\cite{Xu2009_arXiv, Li2009, Bao2009}. DS order can be understood as the N\'{e}el ordering of a four spin plaquette with three up- and one down-spins, which results in double width ferromagnetic(FM) stripes along one diagonal direction and double width antiferromagnetic(AFM) stripes along the other, see Fig \ref{fig:FeAs_FeTe}(b).  These stripes are rotated by $45^\circ$ from the SS magnetism, in addition to being double the width, and they break the tetragonal symmetry down to monoclinic rather than orthorhombic symmetry via coupling to the lattice.

\begin{figure*}
\begin{centering}
\includegraphics[width=1.8\columnwidth]{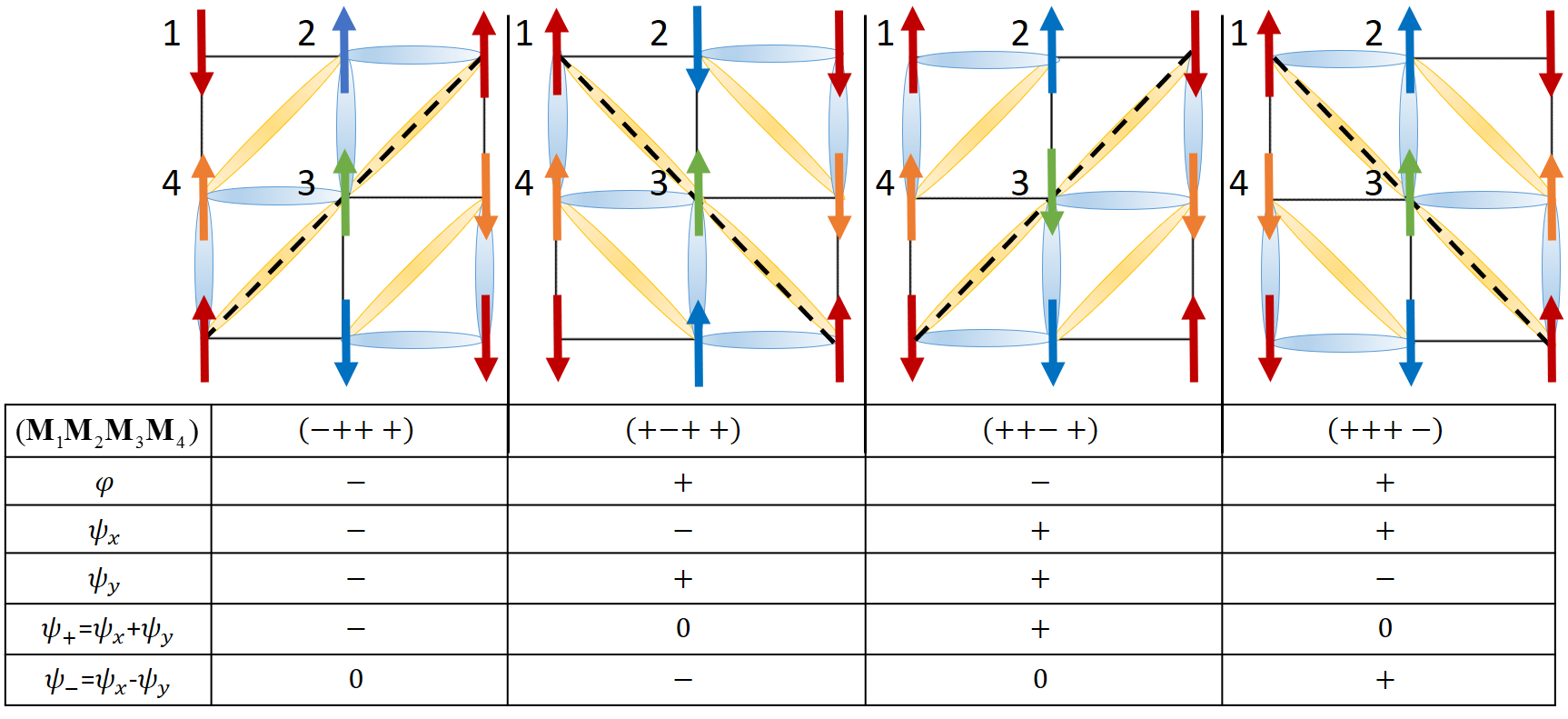} 
\par\end{centering}

\caption{(Color online) Representation of the four-fold degenerate ground states and the corresponding order parameters $\bf{M}$, $\varphi, \psi_x$ and $\psi_y$. The FM bonds are indicated with blue and yellow ovals for NN and NNN, respectively. The black dashed line indicates the diagonal mirror symmetry broken in each state. This figure has been reproduced from Zhang et al, Phys. Rev. B 95, 174402 (2017)\cite{Zhang2017}.}
\label{fig:GSs}
\end{figure*}

For the purpose of contrasting the DS ordered state with the SS one, we first briefly review SS magnetism and the associated nematicity. SS magnetism can be captured within a $J_1 -J_2$ Heisenberg model on the square lattice, with an additional biquadratic coupling \cite{ccl,Ma2008},
\begin{equation}
  H=J_{1}\sum_{\langle ij\rangle}\mathbf{S}_{i}\cdot\mathbf{S}_{j}+J_{2}
  \sum_{\langle\langle ij\rangle\rangle}\mathbf{S}_{i}\cdot\mathbf{S}_{j}
  -K_{1}\sum_{\langle ij\rangle}\left(  \mathbf{S}_{i}\cdot\mathbf{S}
    _{j}\right)  ^{2},
\end{equation}
where $J_{1}$ and $J_{2}>0$ are nearest(NN) and next-nearest neighbor(NNN) exchange couplings, and $K_{1}>0$ is the NN biquadratic coupling, which can be generated by order from disorder\cite{ccl}, but is more likely to arise from itinerant magnetism. For $J_2 \gg J_1$, two N\'{e}el sublattices are given by the antiferromagnetic $J_2$ coupling.  For $K_1 = 0$, the two N\'{e}el order parameters $\mathbf{M}_1$ and $\mathbf{M}_2$ are fully decoupled in the classical, zero temperature limit. $K_1$ then couples them together, favoring collinear spin states and leading to FM stripes along either the $\hat x$ or $\hat y$ directions[Fig \ref{fig:FeAs_FeTe}(a)]. Depending on the orientation of the FM stripes, the ground state is doubly degenerate with wave-vector $(\pi, 0)$ or $(0, \pi)$. This SS magnetism breaks both continuous spin rotational symmetry and discrete $C_4$ rotational symmetry. While the continuous spin-rotational symmetry cannot be broken at any finite temperature in two dimensions, the $C_4$ rotational symmetry breaking can.  It can be described by an Ising-nematic order parameter:
\begin{align}
  \varphi & = \frac{1}{N_{s}} \sum_{i} \langle \mathbf{S}_{i} \cdot
            \mathbf{S}_{i+\hat{x}} - \mathbf{S}_{i} \cdot \mathbf{S}_{i+\hat{y}} \rangle \\
          & = \langle \mathbf{M}_{1} \cdot \mathbf{M}_{2} \rangle,
\end{align}
where $N_{s}$ is the number of sites. Essentially, $\varphi$ is positive (negative) for NN FM correlations along $\hat x$ ($\hat y$), making it a NN bond order. The coupling of $\varphi$ to the lattice gives rise to a orthorhombic structural distortion. 
We shall see that DS magnetism contains both NN and NNN bond orders(see Fig. \ref{fig:FeAs_FeTe} for comparison).

In order to model the DS magnetism, we take the $J_1-J_2-J_3$ Heisenberg model in the regime $J_3 \gg J_2 \gg J_1$. Really, this model is a low energy effective model that can describe either local or itinerant moments. The third neighbor exchange coupling, $J_3$ partitions the spins into four interpenetrating N\'{e}el sublattices $\vect{M}_i(i=1,2,3,4)$.  Since the exchange fields due to both $J_1$ and $J_2$ cancel out at each site, the four sublattices are decoupled in the classical ground state.  $J_1$ drives the classical ground state into a spiral state and away from DS magnetism\cite{Ducatman2012}, so we neglect $J_1$ in this paper, which is valid for sufficiently large four-spin interactions.  As in the SS case, four spin interactions are required to couple together the four sublattices.  Indeed, we can consider the $J_2-J_3-K_2$ model as two copies of $45^\circ$ rotated $J_1-J_2-K_1$ SS magnetism, where $K_2$ will couple together two pairs of sublattices, $( \vect{M}_1, \vect{M}_3 )$ and $( \vect{M}_2 , \vect{M}_4 )$.   As in SS, $K_2$ can be derived from order by disorder \cite{Villain1977, Villain1980, Shender1982, Henley1989,ccl} or itinerant terms \cite{Fazekas2003}.  We can define Ising bond-order parameters for both pairs of sublattices capturing the direction of the ferromagnetic bonds, however, only the two  particular linear combinations of these order parameters break well-defined symmetries.  The first, which we call $\varphi$ in analogy with SS nematicity is defined as:
\begin{equation}
\varphi \propto \langle \mathbf{M}_{1}\cdot\mathbf{M}_{3} - \mathbf{M}_{2}\cdot\mathbf{M}_{4}\rangle.
\end{equation}
Like in the SS case, $\varphi$ breaks the $C_4$ rotational symmetry of the lattice, and couples to the orthorhombic component of the uniform strain $\varepsilon_{xy}$, which would lead to a uniform orthorhombic distortion with short and long NNN $\mathrm{Fe}$-$\mathrm{Fe}$ bonds \cite{Paul2011,Ducatman2014}.  $\varphi$ will be nonzero in the DS ground state.  The second order parameter, 
\begin{equation}
\zeta \propto \langle \mathbf{M}_{1}\cdot\mathbf{M}_{3} + \mathbf{M}_{2}\cdot\mathbf{M}_{4}\rangle,
\end{equation}
preserves the $C_4$ rotation symmetry, but breaks translation symmetry. $\zeta$ is zero in the DS state, but nonzero in the related plaquette ordered state, which consists of antiferromagnetically arranged plaquettes of four ferromagnetic spins and breaks translation symmetry.  Indeed, the NNN biquadratic exchange, $K_2$ favors collinear alignment of the four sublattices, but will not distinguish between DS ($\varphi$) and plaquette ($\zeta)$ orders. However, NNN ring-exchange terms ($R_2$) may be added to the Hamiltonian to select $\varphi$, and thus the DS ground state\cite{Si2016_arXiv, Zhang2017}. In what follows, we will therefore neglect $\zeta$.

While $\varphi$ fixes the relative orientations of the NNN FM bonds, at this point, the two pairs of sublattices are still able to rotate freely with respect to one another. A NN biquadratic exchange $K_1$ will couple these two pairs together.  Again, $( \vect{M}_1, \vect{M}_3 )$ and $( \vect{M}_2 , \vect{M}_4 )$ may be parallel or antiparallel along either $\hat x$ or $\hat y$. We introduce two more Ising bond order parameters to describe this alignment:
\begin{align} 
& \psi_{x} \propto \langle \mathbf{M}_{1}\cdot\mathbf{M}_{2} - \mathbf{M}_{3}\cdot\mathbf{M}_{4}\rangle  \\
& \psi_{y} \propto \langle \mathbf{M}_{1}\cdot\mathbf{M}_{4} - \mathbf{M}_{2}\cdot\mathbf{M}_{3}\rangle.
\end{align}
$\psi_x$ and $\psi$ break both diagonal mirror symmetry and translation symmetry, and couple to the nonuniform, $(\pi, \pi)$ lattice distortions $u_{x/y}$, which distort the lattice with alternating short and long NN $\mathrm{Fe}$-$\mathrm{Fe}$ bonds\cite{Paul2011,Ducatman2014}.

Moreover, $\psi_x$ and $\psi_y$ will generally break the $C_4$ rotational symmetry, and therefore must couple to $\varphi$. Indeed, the signs of the three Ising-bond order parameters are not independent, as shown in Fig. \ref{fig:GSs}, but must satisfy $\varphi \psi_x \psi_y <0$, implying that  $\psi_x \psi_y$ acts like a field for $\varphi$.  Therefore, $\varphi$ will always turns on above or simultaneous to $\psi_x$ and $\psi_y$. As $\psi_{x/y}$ are both associated with $K_1$, they will turn on simultaneously, and we must consider $\psi_\pm = \psi_x \pm \psi_y$ as the true order parameters associated with well-defined broken symmetries.  Assuming that $\varphi$ is already non-zero, $\psi_\pm$ will both double the unit cell [as $(\pi, \pi)$] and break the diagonal mirror symmetry shown in Fig \ref{fig:GSs}.

The full magnetic order will break the $C_4$ and mirror symmetries above, but will also quadruple the unit cell (or double, compared to the $\psi_{\pm}$ unit cell), and break the spin-rotational symmetry.  It can be described in momentum space as a superposition of wave-vectors $\mathbf{Q}=(\pm \frac{\pi}{2}, \pm \frac{\pi}{2})$. When DS magnetism melts via thermal fluctuations, it can therefore do so via three distinct stages: first, melting the magnetism to a state with nonzero $\varphi$ and $\psi_{\pm}$; second, melting $\psi_\pm$ to regain the translation and mirror symmetries, but not the rotation symmetry, in a nematic state; and finally, by melting the nematic state, $\varphi$ to regain the rotation symmetry. In momentum space, below $T_{\varphi}$ the fluctuations at one pair of $\mathbf{Q}$ grow stronger, thus breaking the rotation symmetry; while below $T_{\psi_{\pm}}$, the fluctuations at different $\mathbf{Q}$'s become phase correlated. These stages need not be distinct - for example, in the three-dimensional limit, all three transitions will be simultaneous and first order. However, this is not the case for quasi-two-dimensional systems, leading to rich phase diagrams. In this paper, we develop an effective field theory description based on the $J_1-J_2-J_3-K_1 -K_2$ Heisenberg model, and use it to explore possible phase diagrams with varying degrees of localization, relative ratios of the NN/NNN biquadratic couplings, and dimensionality.

We organize this paper as follows. In section \ref{sec2}, we develop the effective field theory by deriving an effective action via Hubbard-Stratonovich transformations of the quartic spin terms.  We then obtain a set of saddle-point equations by minimizing this effective action with respect to all order parameters, and discuss the conditions for the emergence of magnetic order. In section \ref{sec3}, we solve these equations for the Ising-bond and magnetic order parameters as we vary the dimensionality and other parameters, and we conclude in section \ref{sec4} by discussing the relevance to real materials.

\section{Effective field-theory model}\label{sec2}

\subsection{Model}

In this section, we develop the appropriate effective field theory describing the DS magnetic state, and any related Ising bond-orders.  We begin with the $J_{1}-J_{2}-J_{3}$ Heisenberg model, in the regime $J_{3}\gg J_{2}\gg J_{1}$ where the system can be divided into four interpenetrating N\'{e}el sublattices, with order parameters $\mathbf{M}_{i}$, $i=1,...,4$ (see Fig. \ref{fig:GSs}). In the classical ground state of this model, these sublattices remain decoupled, but they are coupled together by higher order four-spin couplings.  These couplings may originate from order by disorder, magnetoelastic coupling, or simply from the itinerant nature of the relevant spins.  In the spirit of Landau-Ginsburg theory, we will expand the action to fourth order in the N\'{e}el order parameters, with the most general form:
\begin{align}
S\left[\mathbf{M}_{i}\right] \! = \! &\sum_{i,j=1}^{4} \! \int_{\mathbf{q}}\mathbf{M}_{i,\mathbf{q}}\chi_{ij}^{-1}\left(\mathbf{q}\right)\mathbf{M}_{j,-\mathbf{q}}
\! + \! \frac{u}{2} \B(  \sum_{i=1}^4 \mathbf{M}_{i}^2 \B)^2 \cr
& -  \sum_{\left\{ i\neq j,k\neq l\right\} } \int_{\mathbf{r}}\lambda_{ij,kl}\left(\mathbf{M}_{i}\cdot\mathbf{M}_{j}\right)\left(\mathbf{M}_{k}\cdot\mathbf{M}_{l}\right),  \label{S} 
\end{align}
where $\mathbf{M}_1= \langle \sum_{n \in \text{ sublattice } 1} (-1)^{(n_x +n_y)/2} \vect S_n  \rangle$ is the N\'{e}el order parameter on sublattice one, and $\mathbf{M}_i(i=2,3,4)$ are similarly defined. $\int_{\bq} = \int \frac{d^d \bq}{(2 \pi)^d}$, where we keep the dimension, $d$ arbitrary for now.

While at first sight, there are many biquadratic terms, we will neglect those with either $i = j$ or $k = l$.  We will, however, keep the $i = j = k = l$ terms, as these govern the overall softness of the spins, with $u \rightarrow \infty$ describing hard, Heisenberg spins.  For our purposes, we consider the terms that satisfy either $(i, j)\neq (k, l)$ or $(i, j)= (k, l)$, which reduce by symmetry to
\begin{eqnarray} 
& & S\left[\mathbf{M}_{i}\right]  =  \sum_{i,j=1}^{4}\int_{\mathbf{q}}\mathbf{M}_{i,\mathbf{q}}\chi_{ij}^{-1}\left(\mathbf{q}\right)\mathbf{M}_{j,-\mathbf{q}} + \frac{u}{2} \B(  \sum_{i=1}^4 \mathbf{M}_{i}^2 \B)^2 \cr
& & \!-\!\lambda_{1} \! \left[\! \left(\mathbf{M}_{1} \! \cdot\!\mathbf{M}_{2}\right)^{2} \! + \! \left(\mathbf{M}_{1}\!\cdot\!\mathbf{M}_{4}\right)^{2} \! + \! \left(\mathbf{M}_{2}\!\cdot\!\mathbf{M}_{3}\right)^{2} \! + \!\left(\mathbf{M}_{3}\!\cdot\!\mathbf{M}_{4}\right)^{2} \!\right] \cr
& & -\lambda_{2}\left[\left(\mathbf{M}_{1}\cdot\mathbf{M}_{3}\right)^{2}+\left(\mathbf{M}_{2}\cdot\mathbf{M}_{4}\right)^{2}\right] \cr
& & -\lambda_{3}\left[\left(\mathbf{M}_{1}\cdot\mathbf{M}_{2}\right)\left(\mathbf{M}_{3}\cdot\mathbf{M}_{4}\right)+\left(\mathbf{M}_{1}\cdot\mathbf{M}_{4}\right)\left(\mathbf{M}_{2}\cdot\mathbf{M}_{3}\right)\right] \cr
& &  -\lambda_{4}\left(\mathbf{M}_{1}\cdot\mathbf{M}_{3}\right)\left(\mathbf{M}_{2}\cdot\mathbf{M}_{4}\right).
\label{S2}
\end{eqnarray}
We define the coefficients for NN biquadratic exchange, $\lambda_{1}\equiv\lambda_{12,12}=\lambda_{14,14}=\lambda_{23,23}=\lambda_{34,34}$; NNN biquadratic exchange, $\lambda_{2}\equiv\lambda_{13,13}=\lambda_{24,24}$; NN ring exchange\cite{Glasbrenner2015}  $\lambda_{3}=\lambda_{12,34}=\lambda_{14,32}$; and $\lambda_{4}=\lambda_{13,24}$ involving
a ``diagonal'' ring exchange. Motivated by the Ising bond-order parameters discussed in the previous section, we may rewrite these quartic terms as squares,
\begin{align}
&S\left[\mathbf{M}_{i}\right]  =  \sum_{i,j=1}^{4}\int_{\mathbf{q}}\mathbf{M}_{i,\mathbf{q}}\chi_{ij}^{-1}\left(\mathbf{q}\right)\mathbf{M}_{j,-\mathbf{q}} + \frac{u}{2} \B(  \sum_{i=1}^4 \mathbf{M}_{i}^2 \B)^2 \cr
&\qquad \! -\!\frac{g_{1}}{2}\left(\mathbf{M}_{1}\!\cdot\!\mathbf{M}_{3}\!-\!\mathbf{M}_{2}\!\cdot\!\mathbf{M}_{4}\right)^{2} 
\! -\!\frac{g_{2}}{2}\left(\mathbf{M}_{1}\!\cdot\!\mathbf{M}_{3}\!+\!\mathbf{M}_{2}\!\cdot\!\mathbf{M}_{4}\right)^{2}\nonumber \\
 &\qquad \!-\!\frac{g_{3}}{2}\left[\left(\mathbf{M}_{1}\!\cdot\!\mathbf{M}_{2}\!-\!\mathbf{M}_{3}\!\cdot\!\mathbf{M}_{4}\right)^{2}\!+\!\left(\mathbf{M}_{1}\!\cdot\!\mathbf{M}_{4}\!-\!\mathbf{M}_{2}\!\cdot\!\mathbf{M}_{3}\right)^{2}\right]\nonumber \\
 &\qquad \!-\!\frac{g_{4}}{2}\left[\left(\mathbf{M}_{1}\!\cdot\!\mathbf{M}_{2}\!+\!\mathbf{M}_{3}\!\cdot\!\mathbf{M}_{4}\right)^{2}\!+\!\left(\mathbf{M}_{1}\!\cdot\!\mathbf{M}_{4}\!+\!\mathbf{M}_{2}\!\cdot\!\mathbf{M}_{3}\right)^{2}\right],   \label{S_final}
\end{align}
where we have:
\begin{align}
g_{1}  = & \lambda_{2}-\frac{\lambda_{4}}{2}; \qquad
g_{2}  =  \lambda_{2}+\frac{\lambda_{4}}{2}. \cr
g_{3}  = & \lambda_{1}-\frac{\lambda_{3}}{2}; \qquad
g_{4}  =  \lambda_{1}+\frac{\lambda_{3}}{2}. \label{g_def}
\end{align}

The quartic exchange terms will lead to collinear alignments of the four sublattices, assuming positive $g$'s.  We can treat the possible ground states by fixing $\mathbf{M}_1$ and examining the relative orientations of the three other sublattices, which we label with $+/-$. In total there are eight possible configurations, which can be split into those with an odd number of $+$'s and those with an even number: $\{(+---),(+-++),(++-+),(+++-)\}$ and  $\{(++++),(++--),(+--+),(+-+-)\}$.  The first four correspond to the four degenerate ground states of double-stripe order (see Fig. \ref{fig:GSs}), and the last four to the four degenerate ground states of plaquette order.  The energies of these two orders are
\begin{align}
F_{\left\{+---\right\}} & =-2g_{1}-4g_{3}+8 u\cr
F_{\left\{++++\right\}} & =-2g_{2}-4g_{4} + 8 u
\end{align}

Therefore, if $g_{1} + 2 g_{3} > g_2 + 2 g_4$, the DS configuration will be the ground state. We can therefore
ignore the quartic terms $g_{2}$ and $g_{4}$, which correspond to plaquette order and we finally arrive at:
\begin{eqnarray}  \label{S[Mi]}
S\left[\mathbf{M}_{i}\right] & = &  \sum_{i,j=1}^{4}\int_{\mathbf{q}}\mathbf{M}_{i,\mathbf{q}}\chi_{ij}^{-1}\left(\mathbf{q}\right)\mathbf{M}_{j,-\mathbf{q}} + \frac{u}{2} \B(  \sum_{i=1}^4 \mathbf{M}_{i}^2 \B)^2  \cr
&  & -\frac{g_{1}}{2}\left(\mathbf{M}_{1}\cdot\mathbf{M}_{3}-\mathbf{M}_{2}\cdot\mathbf{M}_{4}\right)^{2}  \cr
 &  & -\frac{g_{3}}{2}[ (\mathbf{M}_{1}\cdot\mathbf{M}_{2}-\mathbf{M}_{3}\cdot\mathbf{M}_{4})^{2} \cr
 & & \qquad +(\mathbf{M}_{1}\cdot\mathbf{M}_{4}-\mathbf{M}_{2}\cdot\mathbf{M}_{3})^{2}].
\end{eqnarray}

In order to examine the possible Ising bond-orders, we will decouple all four quartic terms via Hubbard-Stratonovich transformations, introducing the following scalar fields:
\begin{align}\label{def_order_parameters}
 \varphi = & g_{1}\left(\left\langle \mathbf{M}_{1}\cdot\mathbf{M}_{3}\right\rangle -\left\langle \mathbf{M}_{2}\cdot\mathbf{M}_{4}\right\rangle \right)  \cr
 \psi_{x}   = & g_{3}\left(\left\langle \mathbf{M}_{1}\cdot\mathbf{M}_{2}\right\rangle -\left\langle \mathbf{M}_{3}\cdot\mathbf{M}_{4}\right\rangle \right) \cr
\psi_{y}  = & g_{3}\left(\left\langle \mathbf{M}_{1}\cdot\mathbf{M}_{4}\right\rangle -\left\langle \mathbf{M}_{2}\cdot\mathbf{M}_{3}\right\rangle \right)  \cr
\eta = & u \sum_{i=1}^4 \langle \mathbf{M}_{i}^2\rangle .
\end{align}
The resulting effective action then becomes:
\begin{eqnarray}
& & S_{\mathrm{eff}}\left[\mathbf{M}_{i},\psi_{x},\psi_{y},\varphi,\eta\right]  =  \sum_{i,j=1}^{4}\int_{\mathbf{q}}\mathbf{M}_{i,\mathbf{q}}\chi_{ij}^{-1}\left(\mathbf{q}\right)\mathbf{M}_{j,-\mathbf{q}} \cr
& &\qquad \qquad -\varphi\left(\mathbf{M}_{1}\cdot\mathbf{M}_{3}-\mathbf{M}_{2}\cdot\mathbf{M}_{4}\right)\nonumber \\
 &  &\qquad\qquad -\psi_{x}\left(\mathbf{M}_{1}\cdot\mathbf{M}_{2}-\mathbf{M}_{3}\cdot\mathbf{M}_{4}\right)\cr
&  &\qquad\qquad -\psi_{y}\left(\mathbf{M}_{1}\cdot\mathbf{M}_{4}-\mathbf{M}_{2}\cdot\mathbf{M}_{3}\right)\nonumber \\
 &  &\qquad\qquad +\eta \sum_{i=1}^4 \mathbf{M}_{i}^2 +\frac{\varphi^{2}}{2g_{1}}+\frac{\psi_{x}^{2}}{2g_{3}}+\frac{\psi_{y}^{2}}{2g_{3}}-\frac{ \eta^2}{2 u}. \label{eq4}
\end{eqnarray}
We can now interpret these fields: $\eta$ is the uniform magnetic fluctuations; $\varphi$ is the NNN Ising bond-order that breaks the $C_4$ rotational symmetry, and couples to the uniform
orthorhombic distortion $\partial_{x}u_{y}+\partial_{y}u_{x}$; $\psi_{x/y}$ are the NN Ising bond-orders along the $x$- and $y$- directions that give rise to staggered FM/AFM bonds, and couple to the non-uniform distortions, $u_{x/y}\mathrm{e}^{i\left(\pi,\pi\right)\cdot\mathbf{R}_{j}}$\cite{}.
Thus, we have three Ising bond-order parameters: $\varphi, \psi_x$ and $\psi_y$. Because the ground state is four-fold degenerate, they cannot be independent. Indeed, by inspection
of the possible ground states and the values of corresponding order parameters (shown in Fig. \ref{fig:GSs}), one can see that if $\varphi>0$, then $\psi_{x}\psi_{y}<0$, whereas if $\varphi<0$, $\psi_{x}\psi_{y}>0$. That is, the three bond-order parameters must satisfy $\varphi \psi_x \psi_y <0$.

In order to proceed, we will need the correct quadratic terms for DS magnetism. While we will ultimately work with the real space definition of the four sublattices used above, the quadratic term is best derived using the momentum space definition of the four sublattices, $\Delta_\alpha$\cite{Paul2011}, where $\Delta_\alpha$ is the magnetic order parameter at the four $\mathbf{Q}_{\alpha}$'s: $\mathbf{Q}_{1}=(\pi/2,\pi/2)$,
$\mathbf{Q}_{2}=(\pi/2,-\pi/2)$, $\mathbf{Q}_{3}=(-\pi/2,\pi/2)$
and $\mathbf{Q}_{4}=-(\pi/2,\pi/2)$.  The inverse susceptibility, $\chi_{\alpha}^{-1}(\mathbf{q})= r_0 + f_{\alpha}(\vect q)$, which is diagonal in $\alpha$, consists of a $\bq$-independent mean-field component, $r_0 = b (T -T_0)$ ($b>0$), and a $\bq-$dependent part coming from spatial fluctuations of the four sublattice order parameters, $f_{\alpha}(\mathbf{q})=J_{\alpha}(\mathbf{q})$.  We shall expand $J_\alpha(\bq)$ in $\delta \bq$, for $\bq = \mathbf{Q}_\alpha+ \delta\bq$. For conciseness, in the next expression, we write $\mathbf{Q}_{\alpha}=(\eta_{1}\pi/2,\eta_{2}\pi/2)$
($\eta_{1,2}=\pm1$), and we find
\begin{eqnarray}
J_{\alpha}(\mathbf{q}) & = & 2 J_{1}(\cos q_{x}a+\cos q_{y}a)+ 4 J_{2}\cos q_{x}a\cos q_{y}a \cr
& & + 2 J_{3}(\cos2q_{x}a+\cos2q_{y}a)\cr
 & = & - 2 J_{1}(\eta_{1}\delta q_{x}+\eta_{2}\delta q_{y})+4\eta_{1}\eta_{2}J_{2}\delta q_{x}\delta q_{y} \cr
 & & +4J_{3}(\delta q_{x}^{2}+\delta q_{y}^{2})-4J_{3} +O(\delta q^{3})
\end{eqnarray}
where $a$ is the lattice constant, which we set to unity in what follows. 

We can see that fluctuations about the $\mathbf{Q}_\alpha$ cost energy
via $J_{2}$ and $J_{3}$, as expected, while $J_{1}$ drives the system
away from these states (towards a spiral state, as it turns out)\cite{Ducatman2012}.
In the following, we set $J_{1}=0$. So now we have the quadratic
susceptibility term as $\Delta_{\alpha}^{*}\chi_{\alpha}^{-1}(\mathbf{q})\Delta_{\alpha}$. We can convert this term to
$\mathbf{M}_{i}$'s using the matrix: 
\begin{eqnarray}
\left(\! \begin{array}{c}
\mathbf{M}_{1}\\
\mathbf{M}_{2}\\
\mathbf{M}_{3}\\
\mathbf{M}_{4}
\end{array} \!\right)\!=\!\mathcal{O}^{-1} \!\left(\! \begin{array}{c}
\Delta_{1}\\
\Delta_{2}\\
\Delta_{3}\\
\Delta_{4}
\end{array}\!\right)\!,\ \mathcal{O}^{-1}\!=\!\left(\!\begin{array}{cccc}
1 & 1 & 1 & 1\\
i & i & -i & -i\\
-1 & 1 & 1 & -1\\
i & -i & i & -i
\end{array}\!\right)\!.\quad 
\end{eqnarray}
The constraint that the $\mathbf{M}_{i}$'s
must be real imposes that $\Delta_1 =\Delta_4^\ast$ and $\Delta_2 =\Delta_3^\ast$.

Using the transformation $\chi_{ij}^{-1}(\mathbf{q})=\mathcal{O}_{i\alpha}^{\dagger}\chi_{\alpha}^{-1}(\mathbf{q})\mathcal{O}_{\alpha j}$, the susceptibility becomes, 
\begin{align}
\!\chi_{ij}^{-1}(\mathbf{q})\! = \!\! \left[ \!\!\begin{array}{cccc}
J_{3}\delta q^{2} \!+\! r_0 & 0 & -J_{2}\delta q_{x}\delta q_{y} & 0\\
0 & J_{3}\delta q^{2}\!+\!r_0 & 0 & J_{2}\delta q_{x}\delta q_{y}\\
 -J_{2}\delta q_{x}\delta q_{y} & 0 & J_{3}\delta q^{2}\!+\!r_0 & 0\\
0 & J_{2}\delta q_{x}\delta q_{y} & 0 & J_{3}\delta q^{2}\!+\!r_0
 \end{array}\!\!\right] \!.
 \end{align}
For simplicity, we have rescaled $r_0/2 \rightarrow r_0 $, absorbed the $-J_3$ into $r_0$, and defined $\delta q^2 = \delta q_x^2 + \delta q_y^2$.

It is illuminating to examine our bond-order parameters in terms of the momentum space sublattice order parameters, where all the bond-order parameters defined in eq. \eqref{def_order_parameters} become
\begin{align}
\varphi \propto &\; \Delta_2 \Delta_3 -\Delta_1 \Delta_4  \cr
\psi_x \propto &\;  i (\Delta^2_1+\Delta^2_2-\Delta^2_3-\Delta^2_4)  \cr
\psi_y \propto &\;  i (\Delta^2_1-\Delta^2_2+\Delta^2_3-\Delta^2_4)   \cr
\eta \propto & \;\Delta_2 \Delta_3 + \Delta_1 \Delta_4.
\end{align}
An analysis of the $\vect{Q}_\alpha$ associated with each $\Delta_\alpha$ reveals that $\varphi$ and $\eta$ carry zero total momentum, while $\psi_x$ and $\psi_y$ carry a $(\pi,\pi)$ momentum transfer, in agreement with \citet{Paul2011}, and consistent with the translation symmetries identified above.

As a final note in this section, even though we ignore the $g_2$ and $g_4$ terms in the effective action $S_{\mathrm{eff}}[\mathbf{M}_i]$, in order to focus on only the DS order, this model could equally well treat the complementary order parameters, with $\varphi$, $\psi_x$ and $\psi_y$ replaced with the plaquette bond-order parameter, $\left\langle \zeta\right\rangle =  g_{2}\left(\left\langle \mathbf{M}_{1}\cdot\mathbf{M}_{3}\right\rangle + \left\langle \mathbf{M}_{2}\cdot\mathbf{M}_{4}\right\rangle\right)$. As the plaquette order breaks only translation symmetry, $\zeta$ is the only relevant Ising bond-order parameter.

We shall now proceed to minimize the effective action to obtain the behavior of $\varphi, \psi_x, \psi_y$ and $\mathbf{M}$ as functions of temperature and $g_1$, $g_3$ and $u$.  We must consider two separate cases: first, when magnetic order is absent we can integrate out the $\mathbf{M}_i$'s and obtain saddle point equations by minimizing the action with respect to $\varphi, \psi_x, \psi_y$ and $\eta$; second, when magnetic order is present, we will need to carefully integrate out the magnetic fluctuations only, again yielding a set of saddle point equations.  We treat these two cases separately in the following sections.

\subsection{Saddle-point equations in the absence of magnetic order}

We first examine how the Ising bond-orders develop above magnetic order, where $\langle M_i\rangle=0$. This regime will be valid at all temperatures for two dimensions, where the magnetic order is suppressed due to the Mermin-Wagner theorem, and possibly for a finite range of temperatures in higher dimensions. In the next section, we will reincorporate $M$ into the effective action to find the magnetic transition. 

We consider the large-$N$ limit\cite{Fernandes2012} where the number of components of $\vect M_i$ is extended from $N=3$ to $N=\infty$. In this limit, the saddle point approximation becomes exact, and we will use it
to find self-consistent equations for these parameters and solve them.  After integrating out the $\mathbf{M}_{i}$'s, we obtain the effective action
\begin{eqnarray}
S_{\mathrm{eff}}\left[\psi_{x},\psi_{y},\varphi,\eta\right] & = & \frac{T}{2}\sum_{\mathbf{q}}\log\left[\det\mathcal{G}^{-1}\right] \cr
&  & +\frac{\varphi^{2}}{2g_{1}} +\frac{\psi_{x}^{2}}{2g_{3}}+\frac{\psi_{y}^{2}}{2g_{3}}- \frac{ \eta^2}{2 u},  \label{final2_Seff}
\end{eqnarray}
with $\mathcal{G}_{ij}^{-1}(\mathbf{q})$, the inverse Green's function for the $\mathbf{M}_{i}$'s, given by: 
\begin{align} 
\!\!\!\!\!\!\left[\!\!\begin{array}{cc}
(r+J_{3}\delta q^{2}) \mathbb{I} -\frac{\psi_{x}}{2} \sigma_1 &-\frac{i\psi_{y}}{2} \sigma_2 \!-\!(J_{2}\delta q_{x}\delta q_{y}\!+\!\frac{\varphi}{2}) \sigma_3\!\!  \\
  \frac{i\psi_{y}}{2} \sigma_2\!-\!(J_{2}\delta q_{x}\delta q_{y}\!+\!\frac{\varphi}{2}) \sigma_3\!\!\! & (r+J_{3}\delta q^{2}) \mathbb{I} +\frac{\psi_{x}}{2} \sigma_1
\end{array}\right]\!,
\end{align}
where $r\equiv r_0 + \eta $. For compactness, we have used Pauli matrices to write this 4$\times$4 matrix as a 2$\times$2 matrix.  As before, the matrix acts on the space of $(\mathbf{M}_1, \mathbf{M}_2,\mathbf{M}_3,\mathbf{M}_4)$.

The determinant of the inverse Green's function is: 
\begin{eqnarray}
\det\mathcal{G}^{-1} & = & \frac{1}{16}\left(2\tilde{J}_{2}+2\tilde{J}_{3}+2 r +\varphi-\psi_{x}-\psi_{y}\right) \cr
& & \times \left(2\tilde{J}_{2}-2\tilde{J}_{3}-2 r +\varphi+\psi_{x}-\psi_{y}\right)\cr
&  & \times\left(2\tilde{J}_{2}-2\tilde{J}_{3}-2 r +\varphi-\psi_{x}+\psi_{y}\right)\cr
& & \times \left(2\tilde{J}_{2}+2\tilde{J}_{3}+2 r +\varphi+\psi_{x}+\psi_{y}\right) \cr
& =& \left(\tilde{J}_{2}-\tilde{J}_{3}-r \right)^{2}\left(\tilde{J}_{2}+\tilde{J}_{3}+r \right)^{2}\cr
&  & +2\tilde{J}_{2}\left(\tilde{J}_{2}-\tilde{J}_{3}-r\right)\left(\tilde{J}_{2}+\tilde{J}_{3}+r \right)\varphi \cr
 &  & +\frac{1}{2}\left(3\tilde{J}_{2}^{2}-(\tilde{J}_{3}+r)^{2}\right)\varphi^{2}+\frac{\tilde{J}_{2}}{2}\varphi^{3}+\frac{\varphi^{4}}{16}\cr
 &  & -\frac{1}{2}\left(\tilde{J}_{2}^{2}+(\tilde{J}_{3}+r)^{2}\right)\left(\psi_{x}^{2}+\psi_{y}^{2}\right) \cr
 &  & +2\tilde{J}_{2}(\tilde{J}_{3}+r)\psi_{x}\psi_{y}+\frac{1}{16}\left(\psi_{x}^{2}-\psi_{y}^{2}\right)^{2}\cr
 &  & +(\tilde{J}_{3}+r)\varphi\psi_{x}\psi_{y} +\frac{\tilde{J}_{2}}{2}\varphi\left(\psi_{x}^{2}+\psi_{y}^{2}\right) \cr
 &  & -\frac{1}{8}\varphi^{2}\left(\psi_{x}^{2}+\psi_{y}^{2}\right),
\end{eqnarray}
where we have introduced $\tilde{J}_{3}=J_{3}\delta q^{2}$ and $\tilde{J}_{2}=J_2 \delta q_{x}\delta q_{y}$,
for conciseness.  If we do a Landau expansion, we expand $\log\det\mathcal{G}^{-1}$ by assuming that everything involving $\varphi$, $\psi_{x}$ and $\psi_{y}$ is small in comparison to the first term. By doing so, we get a new Landau theory in terms of $\varphi$ and $\psi_{x/y}$. The $\sum_{\mathbf{q}}\tilde{J}_{2}^{2n+1}$ type terms will vanish once the integral over $\bq$ is done. So the linear and cubic $\varphi$ terms vanish, as do the $\varphi(\psi_{x}^{2}+\psi_{y}^{2})$ and $\psi_{x}\psi_{y}$ term. However, the $\varphi\psi_{x}\psi_{y}$ term is really there, as expected. As $\psi_{x}\psi_{y}$ acts like an external field for $\phi$, either $\varphi$ turns on first, or $\psi_x, \psi_y$ and $\varphi$ must all turn on at the same time.

It is convenient to rewrite the action as:
\begin{align}
&  S_{\mathrm{eff}}\left[\psi_{x},\psi_{y},\varphi,\eta\right]  =  \frac{\varphi^{2}}{2g_{1}}+\frac{\psi_{x}^{2}}{2g_{3}} +\frac{\psi_{y}^{2}}{2g_{3}}-\frac{ \eta^2}{2 u} \cr
&  \quad+\frac{T}{2}\sum_{\mathbf{q}}\log\left(J_{3}q^{2}+J_{2}q_{x}q_{y}+r+\varphi-\psi_{x}-\psi_{y}\right)\cr
 &  \quad +\frac{T}{2}\sum_{\mathbf{q}}\log\left(J_{3}q^{2}-J_{2}q_{x}q_{y}+r-\varphi-\psi_{x}+\psi_{y}\right)\cr
 &	\quad +\frac{T}{2}\sum_{\mathbf{q}}\log\left(J_{3}q^{2}-J_{2}q_{x}q_{y}+r-\varphi+\psi_{x}-\psi_{y}\right)\cr
 &  \quad +\frac{T}{2}\sum_{\mathbf{q}}\log\left(J_{3}q^{2}+J_{2}q_{x}q_{y}+r+\varphi+\psi_{x}+\psi_{y}\right), \qquad
\end{align}
where we have renormalized $\left(\varphi,\psi_{x},\psi_{y}\right)\rightarrow2\left(\varphi,\psi_{x},\psi_{y}\right)$
and $g_{i}\rightarrow4g_{i}$ for convenience.

The next step is to minimize the effective action by taking the derivative of $S_{\mathrm{eff}}[\psi_{x}, \psi_y, \varphi, \eta]$ with respect to $\psi_{x}$, $\psi_{y}$, $\varphi$ and $\eta$, setting these to zero. The saddle point equations
$\frac{\partial S_{\mathrm{eff}}[x_{i}]}{\partial x_{i}}=0$($x_i = \eta, \varphi, \psi_x$ and $\psi_y$) become:
\begin{eqnarray}
\eta & = & \frac{T u }{2}\sum_{\mathbf{q}}\left[I_{1}\left(\mathbf{q}\right)+I_{2}\left(\mathbf{q}\right)+I_{3}\left(\mathbf{q}\right)+I_{4}\left(\mathbf{q}\right)\right]\nonumber \\
\varphi & = & \frac{Tg_{1}}{2}\sum_{\mathbf{q}}\left[-I_{1}\left(\mathbf{q}\right)+I_{2}\left(\mathbf{q}\right)+I_{3}\left(\mathbf{q}\right)-I_{4}\left(\mathbf{q}\right)\right]\nonumber \\
\psi_{x} & = & \frac{Tg_{3}}{2}\sum_{\mathbf{q}}\left[I_{1}\left(\mathbf{q}\right)+I_{2}\left(\mathbf{q}\right)-I_{3}\left(\mathbf{q}\right)-I_{4}\left(\mathbf{q}\right)\right]\nonumber \\
\psi_{y} & = & \frac{Tg_{3}}{2}\sum_{\mathbf{q}}\left[I_{1}\left(\mathbf{q}\right)-I_{2}\left(\mathbf{q}\right)+I_{3}\left(\mathbf{q}\right)-I_{4}\left(\mathbf{q}\right)\right],\label{aux_sp}
\end{eqnarray}
where we introduce four convenient integrands $I_{l}(\vect q)(l=1,2,3,4)$. We rotate the coordinate system in the $\vect q$ space by $45^\circ$ to define the effective coupling constant $J\equiv\sqrt{J_{3}^{2}-\frac{J_{2}^{2}}{4}}$, which allows us to rewrite $I_{l}(\vect q)$ in the convenient form:
\begin{eqnarray}
I_{1}\left(\mathbf{q}\right) & = & \frac{1}{Jq^{2}+ r +\varphi-\psi_{x}-\psi_{y}}\nonumber \\
I_{2}\left(\mathbf{q}\right) & = & \frac{1}{Jq^{2}+ r -\varphi-\psi_{x}+\psi_{y}}\nonumber \\
I_{3}\left(\mathbf{q}\right) & = & \frac{1}{Jq^{2}+ r -\varphi+\psi_{x}-\psi_{y}}\nonumber \\
I_{4}\left(\mathbf{q}\right) & = & \frac{1}{Jq^{2}+ r +\varphi+\psi_{x}+\psi_{y}}.\label{aux_aux_sp_2}
\end{eqnarray}

To proceed further, we will need to fix the dimension. While the real materials are quasi-two-dimensional, with an interlayer coupling, $J_z$, for ease of calculation, we will mimic this varying $J_z$ by working in an effective fractional dimension $2\leqslant d\leqslant 3$. The integrals of $I_{l}(\vect q)$ diverge for  $2 < d \leqslant 3$, which we may treat by subtracting and adding the counter-term $\frac{1}{Jq^{2}} $ from each $I_{l}(\vect q)$.  This term absorbs all the ultra-violet divergences and is infra-red convergent for $d>2$. The two dimensional case will be treated separately. The integrands will then be replaced by,
\begin{align}
\frac{1}{J} \tilde{I}_{l}\left(\mathbf{q}\right) \equiv I_{l}\left(\mathbf{q}\right)-\frac{1}{Jq^{2}},
\end{align}
where we have introduced the dimensionless integrands $\tilde{I}_{l}(\mathbf{q})=-\frac{a_l/J}{q^2(q^2 +a_l/J)}$, with the divergent term kept track of separately. $a_l(l=1,2,3,4)$ are the $\bq$-independent parts of the denominators:
\begin{align}
a_1=r+\varphi-\psi_x - \psi_y; \quad a_2=r-\varphi-\psi_x + \psi_y; \cr
a_3=r-\varphi+\psi_x - \psi_y; \quad a_4=r+\varphi+\psi_x + \psi_y. 
\end{align}

The divergent term will cancel out of the last three equations in (\ref{aux_sp}), allowing us to simply replace $I_{l}(\vect q)\rightarrow \frac{1}{J}\tilde{I}_{l}\left(\mathbf{q}\right)$. However, the first equation becomes
\begin{align}
\eta=\frac{T u }{2 J }\sum_{\mathbf{q},l}\tilde{I}_{l}\left(\mathbf{q}\right)+\frac{2 T u }{J}\sum_{\mathbf{q}}\frac{1}{q^{2}}.
\end{align}

We can absorb the second, UV divergent term into the effective mass,
\begin{align}
r= r_0 + \eta = \bar{r}_0 + \frac{T u }{2 J}\sum_{\mathbf{q},l}\tilde{I}_{l}\left(\mathbf{q}\right),
\end{align}
where $\bar{r}_0 = r_0 + \frac{2 T u }{J}\sum_{\mathbf{q}}\frac{1}{q^{2}}$. $\bar{r}_0$ absorbs the ultra-violet divergence. In real materials, this divergence will be cutoff by some higher energy scale, however the microscopic details are irrelevant here, and we will work with $\bar{r}_0$ as the rescaled temperature.

In the spirit of Landau theory, we now approximate $T$ with $T_0$ everywhere, except in $r_0 \propto T-T_0$. We may make all quantities dimensionless by rescaling $\frac{T_0}{2J^{2}} (u, g_1, g_3)  \rightarrow (u, g_1, g_3)$ and 
$\frac{1}{J}\left(r, \bar{r}_0, \varphi,\psi_{x},\psi_{y}, \eta\right)  \rightarrow \left(r, \bar{r}_0,\varphi,\psi_{x},\psi_{y}, \eta\right)$. With this rescaling, $\tilde{I}_l(\bq)$ becomes
\begin{align}
\tilde{I}_{l}(\mathbf{q})=-\frac{a_l}{q^2(q^2 +a_l)}.
\end{align}

Finally, we obtain the saddle-point equations:
\begin{eqnarray}
r & = & \bar r_{0}+ u \sum_{\mathbf{q}}\left[\tilde{I}_{1}\left(\mathbf{q}\right)+\tilde{I}_{2}\left(\mathbf{q}\right)+\tilde{I}_{3}\left(\mathbf{q}\right)+\tilde{I}_{4}\left(\mathbf{q}\right)\right]\cr
\varphi & = & g_{1}\sum_{\mathbf{q}}\left[-\tilde{I}_{1}\left(\mathbf{q}\right)+\tilde{I}_{2}\left(\mathbf{q}\right)+\tilde{I}_{3}\left(\mathbf{q}\right)-\tilde{I}_{4}\left(\mathbf{q}\right)\right]\cr
\psi_{x} & = & g_{3}\sum_{\mathbf{q}}\left[\tilde{I}_{1}\left(\mathbf{q}\right)+\tilde{I}_{2}\left(\mathbf{q}\right)-\tilde{I}_{3}\left(\mathbf{q}\right)-\tilde{I}_{4}\left(\mathbf{q}\right)\right]\cr
\psi_{y} & = & g_{3}\sum_{\mathbf{q}}\left[\tilde{I}_{1}\left(\mathbf{q}\right)-\tilde{I}_{2}\left(\mathbf{q}\right)+\tilde{I}_{3}\left(\mathbf{q}\right)-\tilde{I}_{4}\left(\mathbf{q}\right)\right].
\end{eqnarray}

It is now straightforward to evaluate the momentum integrals for fractional dimensions,
\begin{eqnarray}
\sum_{\mathbf{q}}\tilde{I}_{l}\left(\mathbf{q}\right) & = & -\int\frac{d^{d}q}{\left(2\pi\right)^{d}}\,\frac{a_{l}}{q^{2}\left(q^{2}+a_{l}\right)} \cr
 & = & -\left[\frac{S_{d}}{\left(2\pi\right)^{d}}\int_{0}^{\infty}dx\,\frac{x^{d-3}}{x^{2}+1}\right]a_{l}^{\frac{d-2}{2}},
\end{eqnarray}
where $a_l$ represents the $\vect q$ independent part of the denominator, and $S_d = \int d\Omega_q = \frac{2 \pi^{d/2}}{\Gamma(d/2)}$ is the surface area of a $d-$dimensional sphere with unit radius.

Since the prefactor converges for $2<d<4$, we absorb it too, into the $g$'s and $u$,
 in order to obtain a set of simple algebraic equations:
\begin{align}
&\frac{\bar r_{0} -r} {u}  =  \left(r+\varphi-\psi_{x}-\psi_{y}\right)^{\frac{d-2}{2}}+\left(r-\varphi-\psi_{x}+\psi_{y}\right)^{\frac{d-2}{2}} \cr
& \qquad +\left(r-\varphi+\psi_{x}-\psi_{y}\right)^{\frac{d-2}{2}}+\left(r+\varphi+\psi_{x}+\psi_{y}\right)^{\frac{d-2}{2}}\nonumber \\
& \frac{\varphi}{g_{1}}  =  \left(r+\varphi-\psi_{x}-\psi_{y}\right)^{\frac{d-2}{2}}-\left(r-\varphi-\psi_{x}+\psi_{y}\right)^{\frac{d-2}{2}} \cr
& \qquad -\left(r-\varphi+\psi_{x}-\psi_{y}\right)^{\frac{d-2}{2}}+\left(r+\varphi+\psi_{x}+\psi_{y}\right)^{\frac{d-2}{2}}\nonumber \\
& \frac{\psi_{x}}{g_{3}}  =  -\left(r+\varphi-\psi_{x}-\psi_{y}\right)^{\frac{d-2}{2}}-\left(r-\varphi-\psi_{x}+\psi_{y}\right)^{\frac{d-2}{2}} \cr
& \qquad +\left(r-\varphi+\psi_{x}-\psi_{y}\right)^{\frac{d-2}{2}}+\left(r+\varphi+\psi_{x}+\psi_{y}\right)^{\frac{d-2}{2}}\nonumber \\
& \frac{\psi_{y}}{g_{3}}  =  -\left(r+\varphi-\psi_{x}-\psi_{y}\right)^{\frac{d-2}{2}}+\left(r-\varphi-\psi_{x}+\psi_{y}\right)^{\frac{d-2}{2}} \cr
&\qquad -\left(r-\varphi+\psi_{x}-\psi_{y}\right)^{\frac{d-2}{2}}+\left(r+\varphi+\psi_{x}+\psi_{y}\right)^{\frac{d-2}{2}}.  \label{final_sp_eq}
\end{align}

These equations define how the parameters $\eta$ (now hidden within  $r$), $\varphi$,
$\psi_{x}$, and $\psi_{y}$ depend on the control parameter $r_{0}\propto T-T_{0}$. We can then solve these as a function of $r_0$ to find the transition temperatures for the various bond-orders.  The magnetic transition takes place when the mass of the renormalized magnetic action vanishes, i.e. when:
\begin{equation} \label{mag_transition}
r=-\varphi\pm\left(\psi_{x}+\psi_{y}\right)\quad\mathrm{or}\quad r =\varphi\pm\left(\psi_{x}-\psi_{y}\right).
\end{equation}
We can use this criterion to resolve the location of the magnetic transition, but resolving the order of the transition will require the more involved calculations of the next section.

As discussed previously, $\psi_x$ and $\psi_y$ enter in the same fashion, governed by the same $g_3$, and we expect them to develop the same magnitude $|\psi_x|=|\psi_y|$ at the same temperature. In fact, the correct  pair of order parameters $\psi_{\pm} = \psi_x \pm \psi_y $ are the only legitimate order parameters breaking well-defined symmetries.  $|\psi_x|=|\psi_y|$ implies that only one of $\psi_{\pm}$ can be nonzero.  In terms of $\psi_+$ and $\psi_-$, the constraint $\varphi \psi_x \psi_y <0$ becomes $\varphi(\psi_+^2-\psi_-^2)<0$.  So the nonzero order parameter is selected by the sign of $\varphi$. That is, for $\varphi<0$, $\psi_+$ can be nonzero with the converse true for $\varphi>0$.

Replacing $\psi_x$ and $\psi_y$ with $\psi_{\pm}$, we decouple the last two saddle-point equations,
\begin{align}
& \frac{\bar r_{0} -r} {u}  =  \left(r+\varphi-\psi_{+}\right)^{\frac{d-2}{2}}+\left(r-\varphi-\psi_{-}\right)^{\frac{d-2}{2}} \cr
& \qquad\qquad +\left(r-\varphi+\psi_{-}\right)^{\frac{d-2}{2}}+\left(r+\varphi+\psi_{+}\right)^{\frac{d-2}{2}} \cr
&\frac{\varphi}{g_{1}}  =  \left(r+\varphi-\psi_{+}\right)^{\frac{d-2}{2}}-\left(r-\varphi-\psi_{-}\right)^{\frac{d-2}{2}} \cr
&  \quad\quad -\left(r-\varphi+\psi_{-}\right)^{\frac{d-2}{2}}+\left(r+\varphi+\psi_{+}+\right)^{\frac{d-2}{2}} \cr
& \frac{\psi_{+}}{2 g_{3}}   =    -\left(r+\varphi-\psi_{+}\right)^{\frac{d-2}{2}}+\left(r+\varphi+\psi_{+}\right)^{\frac{d-2}{2}}  \cr
& \frac{\psi_{-}}{2 g_{3}}  =   -\left(r-\varphi-\psi_{-}\right)^{\frac{d-2}{2}}+\left(r-\varphi+\psi_{-}\right)^{\frac{d-2}{2}}.
\end{align}
Up to the sign of $\varphi$, the two cases $\psi_+ \neq 0$, $\psi_-= 0$ (DS order in $y=x$ direction) or $\psi_+ = 0$, $\psi_-\neq 0$ (DS order in $y=-x$ direction) give equivalent sets of saddle-point equations. We will adopt the former($\psi_+ \neq 0$) and further define $\psi_+ \equiv \psi$ in order to simplify the notation. The remaining three saddle-point equations become
\begin{eqnarray}\label{eq8}
\frac{\bar r_{0} - r}{u} & = & \left(r +\varphi-\psi\right)^{\frac{d-2}{2}}+\left(r +\varphi+\psi\right)^{\frac{d-2}{2}}+ 2 \left(r -\varphi\right)^{\frac{d-2}{2}}\nonumber \\
\frac{\varphi}{g_{1}} & = & \left(r +\varphi-\psi\right)^{\frac{d-2}{2}}+\left(r +\varphi+\psi\right)^{\frac{d-2}{2}}-2 \left(r -\varphi\right)^{\frac{d-2}{2}}\nonumber \\
\frac{\psi}{2 g_{3}} & = & -\left(r+\varphi-\psi\right)^{\frac{d-2}{2}}+\left(r+\varphi+\psi\right)^{\frac{d-2}{2}}.
\end{eqnarray}
 
In this case, $\varphi<0$ while $\psi$ can be either sign. However, eqs. \eqref{eq8} are invariant under $\psi \rightarrow -\psi$, and so all the physics will be independent of the sign of $\psi$. From Fig. \ref{fig:GSs}, it can be seen that the DS order for $\psi>0$ is just the mirror of that with $\psi<0$ along the $y=x$ direction. Or equivalently, one can shift the DS ground state with $\psi>0$ by one lattice constant along either the $x$ or $y$ direction to obtain the DS ground state with $\psi<0$. 
So it is sufficient to take $\psi>0$, which corresponds to the ground state $(++-+)$ once $M$ condenses.

\subsection{Saddle-point equations in the presence of magnetic order}
\label{subsec2_3}

In dimensions greater than two, magnetic order will always develop at sufficiently low temperatures, and in this case, we must use the saddle-point equations with the magnetic order included to determine the order of the magnetic transition. 
We begin with the effective action in eq.\eqref{eq4}, and replace $\mathbf M_{i, \vect q} $ with $\mathbf M_{i, \vect q} =  \langle \mathbf M_i\rangle \delta(\vect q) + \delta \mathbf M_{i, \vect q} $.  Here, the magnetic order parameters, $\langle \mathbf M_i\rangle$ are collinear, and all have the same magnitude, $M$.  We keep $\langle \mathbf M_i\rangle$, but integrate out the fluctuations about magnetic order, $\delta\mathbf{M}_{i}$. The resulting effective action is:
\begin{align} \label{eq23}
& S_{\mathrm{eff}}\left[\langle \mathbf M_i\rangle ,\psi_{x},\psi_{y},\varphi,\eta\right]  =  S_{\mathrm{eff}}\left[\psi_{x},\psi_{y},\varphi,\eta\right] \cr
& \qquad \qquad\qquad\qquad  +(r - |\varphi| -|\psi_x| -|\psi_y| ) M^2
\end{align}
where we have rescaled   $(\varphi, \psi_x, \psi_y) \rightarrow 2 (\varphi, \psi_x, \psi_y)$, $g_i \rightarrow 4 g_i$ and $M \rightarrow M/(2\sqrt{2})$.

The differentiation of the effective action over $\eta, \phi$, $\psi_x$, $\psi_y$ and $M$ gives the five coupled equations. 
\begin{align} \label{eq20}
& \eta  \!=\! \frac{T u }{2}\sum_{\mathbf{q}}\left[I_{1}\left(\mathbf{q}\right)\!+\!I_{2}\left(\mathbf{q}\right)\!+\! I_{3}\left(\mathbf{q}\right)\!+\! I_{4}\left(\mathbf{q}\right)\right] \!+\! u M^2 \cr
& \varphi  \!=\! \frac{Tg_{1}}{2}\sum_{\mathbf{q}}\left[-I_{1}\left(\mathbf{q}\right)\!+\! I_{2}\left(\mathbf{q}\right)\!+\! I_{3}\left(\mathbf{q}\right)\!-\! I_{4}\left(\mathbf{q}\right)\right] \!-\! g_1 M^2 \cr
& \psi_{x}\!  = \! \frac{Tg_{3}}{2}\sum_{\mathbf{q}}\left[I_{1}\left(\mathbf{q}\right)\!+\!I_{2}\left(\mathbf{q}\right)\!-\!I_{3}\left(\mathbf{q}\right)\!-\!I_{4}\left(\mathbf{q}\right)\right]\!+\! g_3 M^2 \cr
& \psi_{y} \! = \! \frac{Tg_{3}}{2}\sum_{\mathbf{q}}\left[I_{1}\left(\mathbf{q}\right)\!-\!I_{2}\left(\mathbf{q}\right)\!+\!I_{3}\left(\mathbf{q}\right)\!-\!I_{4}\left(\mathbf{q}\right)\right] \!+\! g_3 M^2  \cr
& (r - |\varphi| -|\psi_x| -|\psi_y| ) M 	=  0.
\end{align}

For $d>2$, we again subtract $\frac{1}{Jq^{2}} $ from each $I_{l}(\vect q)$. For the choice of $\varphi \leqslant 0$, $\psi \geqslant 0$ (corresponding to the ground state $(++-+)$), these equations become:
\begin{align} \label{eq10}
&\frac{\bar r_{0} - r}{u}  =  \left(r +\varphi-\psi\right)^{\frac{d-2}{2}}+\left(r +\varphi+\psi\right)^{\frac{d-2}{2}} \cr
& \qquad\qquad   + 2 \left(r -\varphi\right)^{\frac{d-2}{2}} - M^2 \cr
&\frac{\varphi}{g_{1}}  =  \left(r +\varphi-\psi\right)^{\frac{d-2}{2}}+\left(r +\varphi+\psi \right)^{\frac{d-2}{2}} \cr
& \ \qquad -2 \left(r -\varphi\right)^{\frac{d-2}{2}} - M^2 \cr
& \frac{\psi}{2 g_{3}}  =  -\left(r+\varphi-\psi\right)^{\frac{d-2}{2}}+\left(r+\varphi+\psi\right)^{\frac{d-2}{2}} + M^2 \cr
& (r - |\varphi| -|\psi_x| -|\psi_y| ) M 	=  0.
\end{align}
where we have further rescaled $\frac{T_0}{2J^{2}} (u, g_1, g_3)  \rightarrow (u, g_1, g_3)$, 
$\frac{1}{J}\left(r, \bar{r}_0, \varphi,\psi, \eta\right)  \rightarrow \left(r, \bar{r}_0,\varphi,\psi, \eta\right)$ as before, and also $M \rightarrow \sqrt{\frac{T_0}{2 J}} M$. This rescaled $M$ is dimensionless.

The last equation in \eqref{eq20} is particularly simple: with $M$ nonzero, the only solution is $r = |\varphi| +|\psi| = -\varphi +\psi$, which is the condition for the onset of magnetic order obtained in the previous section.

Without Ising-bond order, the ``bare" magnetic transition occurs at $r=0$. If $\varphi$ turns on first (without $\psi$), the magnetic transition will occur at a larger $r=|\varphi|>0$. If both $\varphi$ and $\psi$ turn on above magnetic order, the transition will be still higher, $r=|\varphi|+|\psi|>0$. Remember that $r$ increases linearly with the temperature. Thus, both Ising-bond orders increase the temperature at which the magnetic order appears. The coexistence of Ising-bond and magnetic order enhances the magnetic ordering temperature; this stabilization of the magnetic order via Ising bond-order has been seen, for example, in Fe$_{1+y}$Te\cite{Tranquada2014}, and will be enhanced if the bond order is further stabilized via coupling to the lattice\cite{Paul2011,Moreo2016}.

\section{Results} \label{sec3}

In this section, we solve the saddle point equations and present the resulting phase diagrams. In general, as temperature is lowered, NNN bond order ($\varphi$) appears first, breaking the $C_{4}$ rotational symmetry, followed by NN bond-order ($\psi$), breaking the translation and mirror symmetries of the lattice, followed by magnetic order that breaks spin-rotational symmetry. The ordering of these transitions is fixed by their respective symmetries, however, the nature and spacing of these transitions can vary widely, from three distinct second order transitions to one simultaneous first order transition.  Our action, eq. \eqref{S[Mi]} contains three tuning parameters: $u$, which governs the overall scale of the magnetic fluctuations; $g_1$, which favors $\varphi$; and $g_3$, which favors $\psi$. We combine these three dimension-full parameters into two dimensionless parameters, $\alpha \equiv u/g_1$ and $\beta \equiv g_3/g_1$, where roughly speaking decreasing $\alpha$ favors $\varphi$ bond-order and increasing $\beta$ favors $\psi$ bond-order. Note that for our model to make sense, $u > g_1$ and so $\alpha > 1$.  As $\varphi$ turns on automatically once $\psi$ turns on, we generally restrict our analysis to the more interesting region of $g_3 < g_1$, or $\beta < 1$.

We can tune the inter-layer coupling strength by changing the fractional dimensionality, $d$.  If $\beta = 0$, our model becomes two copies of single-stripe magnetism, and we recover all the results of \citet{Fernandes2012}; we reproduce some of these results here in order to illustrate our solution techniques.  For nonzero $\beta$, the resulting phase diagrams become much richer. We will first present our results for the two limiting cases: 2D and 3D, and then examine the intermediate dimensionalities $2<d<3$. For each case, we examine the transitions into each phase as a function of $r_0$, which acts as temperature, and show how the behavior evolves in the $(\alpha, \beta)$ plane.

\subsection{Two dimensions} \label{subsec3_1}

Two dimensions is special, as the magnetic order is completely suppressed at any finite temperature due to strong thermal fluctuations. In addition, the ultra-violet divergence in $I_{l}(\vect q)$ cannot be removed by $\frac{1}{J q^2}$ subtraction in 2D, so we evaluate the momentum integrals in eq.\eqref{aux_sp} directly:
\begin{align} \label{eq21}
\sum_{\vect q} I_l(\vect q) = & \int \frac{d^2 q}{(2\pi)^2} \frac{1}{J (q^2 + a_l/J)} \cr
& = \frac{1}{4 \pi J} [\ln(\Lambda^2+a_l/J) - \ln (a_l/J)] \cr
& \approx \frac{1}{4 \pi J} [2 \ln\Lambda - \ln (a_l/J)],
\end{align}
where we have introduced an explicit momentum cutoff, $\Lambda$. The approximation in the third line is valid when $a_l$ is small compared to $\Lambda$.  We can then substitute these results into eq.\eqref{aux_sp}, rescale
$
 \frac{T}{2  J^2} (u, g_1, g_3) \rightarrow  (u, g_1, g_3),
\frac{1}{J}\left(r, \varphi,\psi_{x},\psi_{y}, \eta\right)  \rightarrow \left(r,\varphi,\psi_{x},\psi_{y}, \eta \right)
$
as before, and absorb the pre-factor of the integration $1/(4\pi)$ in the temperature $T_0$, in order to obtain a new set of saddle-point equations:
\begin{eqnarray} \label{eq9}
\frac{\bar r_0 - r}{u} & = & \ln(r +\varphi-\psi)+\ln(r +\varphi+\psi)+ 2 \ln(r -\varphi)\nonumber \\
\frac{\varphi}{ g_1} & = & \ln(r +\varphi-\psi)+\ln(r +\varphi+\psi)- 2 \ln(r -\varphi)\nonumber \\
\frac{\psi}{2 g_3} & = & -\ln(r +\varphi-\psi) +\ln(r +\varphi+\psi),
\end{eqnarray}
where we introduce $\bar r_0 = r_0 + 8 u \ln \Lambda$, and $r = r_0 + \eta$, as before.   Note that we can already see the absence of magnetic order here, as in the absence of bond-orders, magnetic order emerges when $r = 0$. In this limit, the first equation becomes $r =  r_0-4u\ln r$, where the right hand side diverges as $r \rightarrow 0$, implying that $r$ can never reach zero, and thus the system cannot order.

In solving these equations, we first consider the simpler limit $g_3 = 0$, in which $\psi = 0$, and the equations reduce to those in \citet{Fernandes2012}. For completeness, we reproduce those results here. The saddle point equations in \eqref{eq9} simplify into two equations:
\begin{align} \label{eq27}
 r = & \bar r_0 - 2 u \ln(r^2 - \varphi^2) \cr
 r = & \varphi \coth \Big( \frac{\varphi}{4 g_1} \Big).
\end{align}
We can introduce $\varphi^\ast \equiv \varphi/(4 g_1)$ to eliminate $r$ and simplify to a single equation,
\begin{align}
 \label{eq27p}
\varphi^{\ast} \coth \varphi^{\ast} + \alpha \ln \Big( \frac{\varphi^{\ast}}{\sinh \varphi^{\ast}}  \Big) = \bar{\bar{r}}_0
\end{align}
where we introduce $\bar{\bar{r}}_0 \equiv \bar r_0 /(4 g_1) - \alpha \ln(4 g_1)$ and $\alpha \equiv u/g_1$. 

\begin{figure}[ht]\centering
   \begin{minipage}{\columnwidth}
     \includegraphics[width=\linewidth]{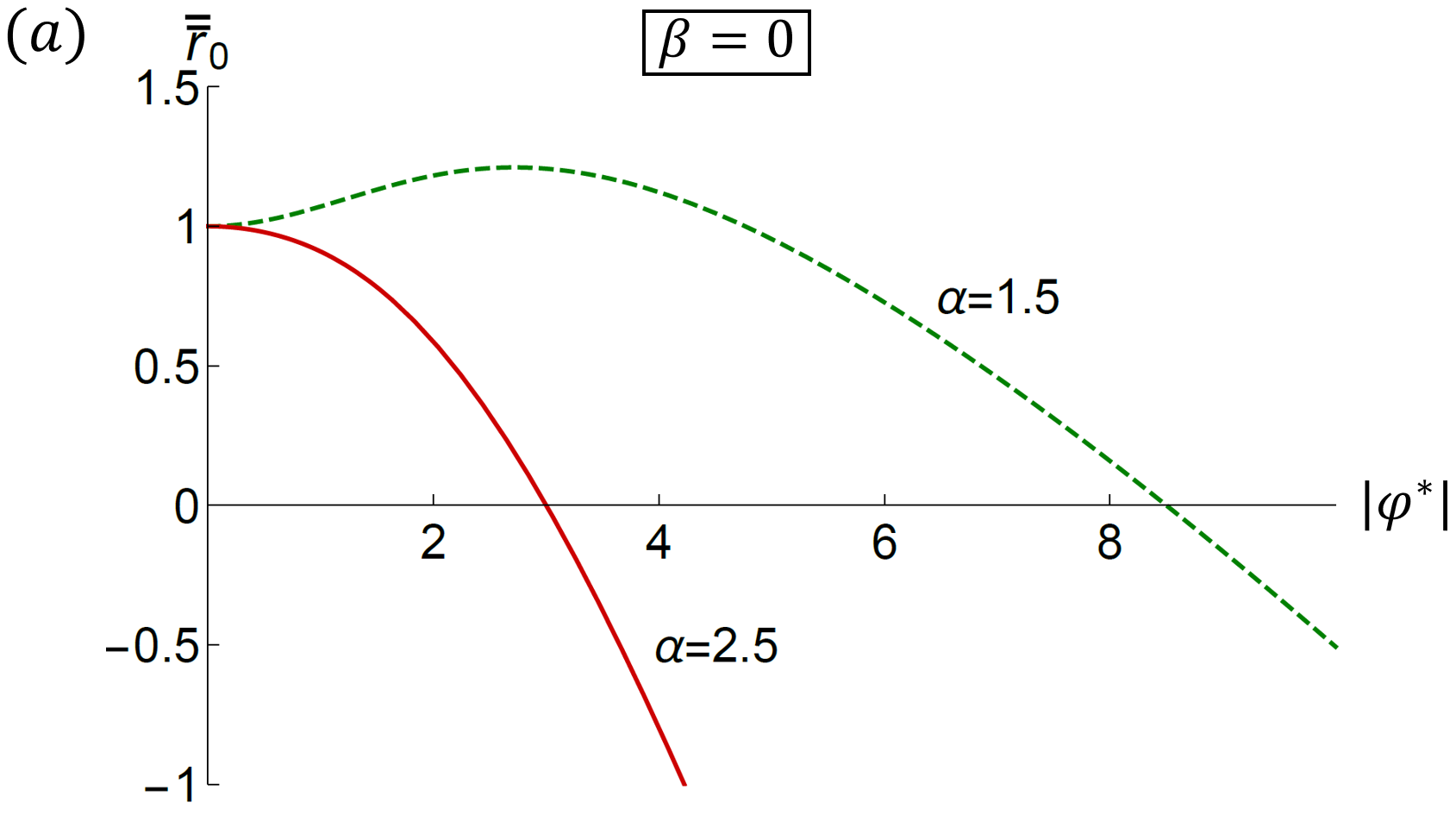}
   \end{minipage}
   \begin {minipage}{\columnwidth}
     \includegraphics[width=\linewidth]{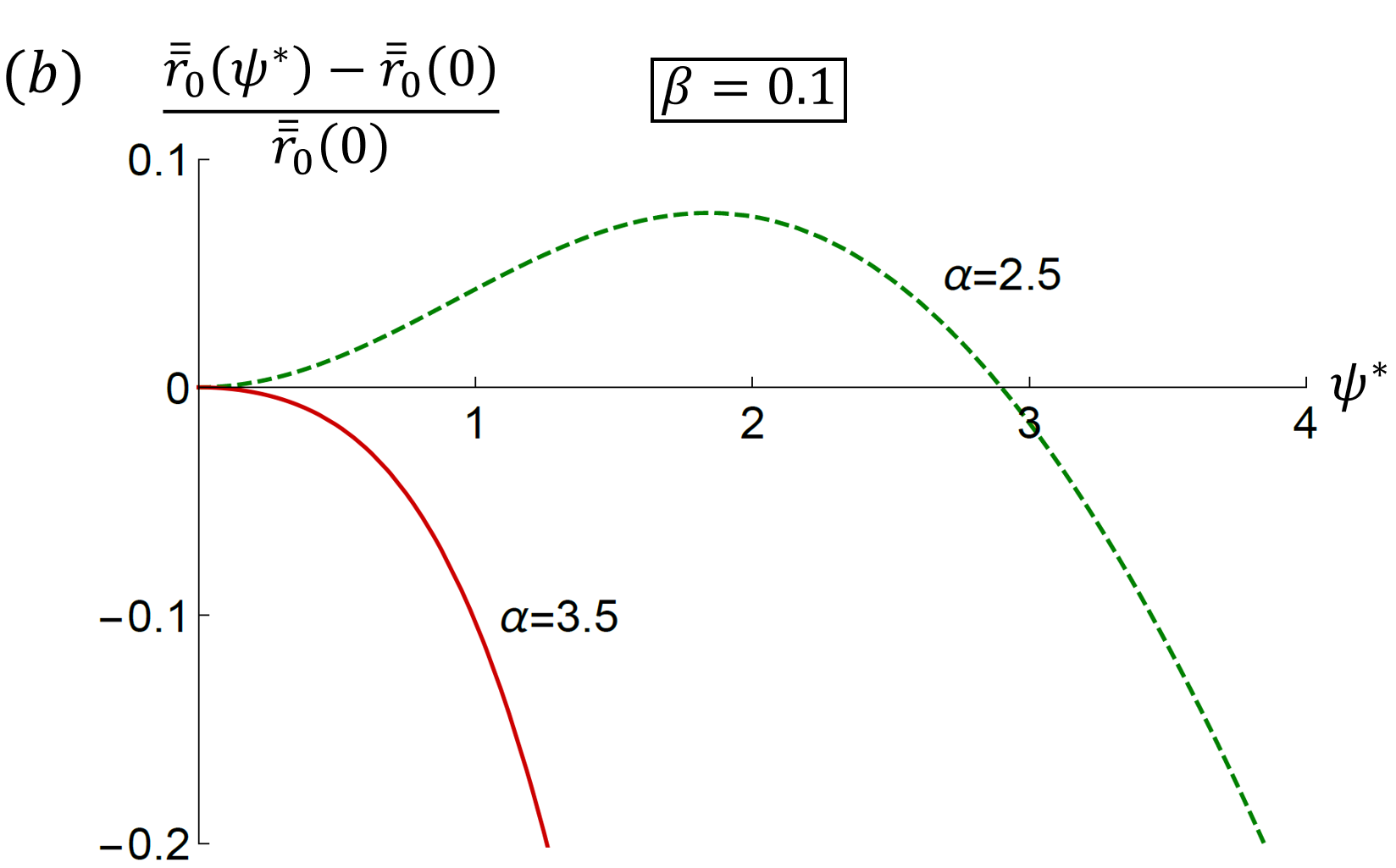}
   \end{minipage}
\caption{(Color online) Here we show the first and second order transitions for $\varphi$ and $\psi$. In (a), $g_3 = 0$, and $\psi = 0$.  We plot $\bar{\bar{r}}_{0 }$ as a function of the Ising-bond order $\varphi^\ast = \varphi /(4 g_1)$ in 2D for two representative values of $\alpha \equiv u/g_1$ in the region $1<\alpha<\alpha_{\varphi}$(green dashed) and $\alpha >\alpha_{\varphi} $(red solid) where $\alpha_{\varphi}=2$. For $1<\alpha<\alpha_\varphi$, the $\varphi$ transition is first order, as $\bar{\bar{r}}_{0 }$ is maximized at a finite $\varphi^\ast$. For $\alpha >\alpha_\varphi $, the $\varphi$ transition is second order as $\bar{\bar{r}}_{0 }$ is maximized at $\varphi^\ast =0 $. In
(b), we show the $g_3 > 0$ results for the $\psi$ transition.  We plot the rescaled $\bar{\bar{r}}_{0}$ as a function of $\psi^\ast=\psi/(4 g_3)$ in 2D for $\beta\equiv g_3/ g_1 =0.1$ and for two representative values of $\alpha$ in the region $1<\alpha<\alpha_\psi$(green dashed) and $\alpha >\alpha_\psi $(red solid) where $\alpha_\psi=3.3$. These describe first and second order transitions of $\psi$.} \label{fig:r0bar_vs_varphi_Sigma_2D}
\end{figure}

Recall that $\bar{\bar{r}}_0$ decreases with decreasing temperature, just as $r_0$ does. The leading instability of the system with decreasing temperature can be found from the maximum of the left hand side of \eqref{eq27p}, where the value of $\varphi^\ast$ at the transition will be the location of the maximum.  When the maximum occurs at $\varphi^\ast = 0$, as it does for sufficiently large $\alpha$, the transition is second order. For smaller $\alpha$, the maximum occurs at a finite $\varphi^\ast$ and the transition is first order. By investigating the slope of the $\bar{\bar{r}}_{0 }$ vs $\varphi^\ast$ plot at $\varphi^\ast =0$, we find that there is a critical value of $\alpha$, i.e. $\alpha_{\varphi}=2$, beyond which the $\varphi$ transition changes from first to second order, as shown in Fig. \ref{fig:r0bar_vs_varphi_Sigma_2D}(a).

According to the discussion in Sec. \ref{subsec2_3}, magnetic order will occur if $r = |\varphi|$.  However, the second equation in eqs.\eqref{eq27} implies that $r$ can only reach $-\varphi$ as $-\varphi \rightarrow \infty$, and therefore magnetic order will not occur even in the presence of a preemptive nematic transition.

For finite $g_3$, we now consider the $\psi$ transition.  As $\psi$ acts as a field for $\varphi$, $\varphi$ will either already be nonzero, governed by the equations above, or will turn on with $\psi$.  In either case, it is necessary and sufficient to explore the transitions of $\psi$. By eliminating $r$, eqs.\eqref{eq9} now yields two equations instead of one.
\begin{align} \label{eq25}
\bar{\bar{r}}_0 = &\alpha \ln ( \beta \psi^\ast \ \mathrm{csch} \psi^{\ast} ) + \beta \psi^\ast \coth \psi^{\ast} -  (\alpha + 1) \varphi^\ast  \cr
\beta = & \frac{ 2 \varphi^\ast}{\psi^\ast [ \coth \psi^{\ast} - \ \mathrm{csch} \psi^{\ast} \mathrm{exp} ( - 2 \varphi^\ast)   ] }
\end{align}
where we have defined $\beta\equiv g_3/g_1$,  rescaled $\psi^{\ast} \equiv \frac{\psi}{4 g_3}$ and $\varphi^\ast$ and $\bar{\bar{r}}_0$ are defined as above.

To examine the nature of the $\psi$ transition, we need to find $\bar{\bar{r}}_{0}$ as a function of $\psi^\ast$. To do so, we first solve $\varphi^\ast$ from the second equation in \eqref{eq25} for $\psi^\ast$. Then we substitute it into the first equation in \eqref{eq25}.  For simplicity, $\bar{\bar{r}}_0$ is rescaled to $\bar{\bar{r}}_{0 res}\equiv \bar{\bar{r}}_0 (\psi^\ast)/\bar{\bar{r}}_0 (0)-1$ and plotted as a function of $\psi^\ast$ in Fig. \ref{fig:r0bar_vs_varphi_Sigma_2D}(b) for two representative $\alpha$'s. Again, the transition will occur at the $\psi^\ast$ that maximizes $\bar{\bar{r}}_{0 res}$, and will be second order if that $\psi^\ast$ is zero, and first order otherwise.

For any given $\beta$, the maximum of $\bar{\bar{r}}_{0 res}$ approaches infinity as $\alpha \rightarrow 1$, meaning that $\alpha=1$ is unphysical. As $\alpha$ increases, the maximum of $\bar{\bar{r}}_{0 res}$ moves towards smaller $\psi^\ast$. There is a critical value $\alpha_\psi(\beta)$ separating the first and second order transition of $\psi$. For $1<\alpha<\alpha_{\psi}$, the maximum of $\bar{\bar{r}}_{0 res}$ is at a finite $\psi^\ast$, which means $\psi^\ast$ turns on discontinuously. For $\alpha >\alpha_{\psi}$, the maximum of $\bar{\bar{r}}_{0 res}$ is at $\psi^\ast = 0$, which implies a second order transition. 

As before, the absence of the magnetic order can be verified by checking that $r$ can never reach $ -\varphi +\psi$. From the last equation in \eqref{eq9}, we find $r+\varphi = \psi \coth \big(\frac{\psi}{4g_3}  \big) > \psi$, which means $r>-\varphi +\psi $. So again there is no magnetic order. 

\begin{figure}[!ht]
\begin{centering}
\includegraphics[width=0.95\columnwidth]{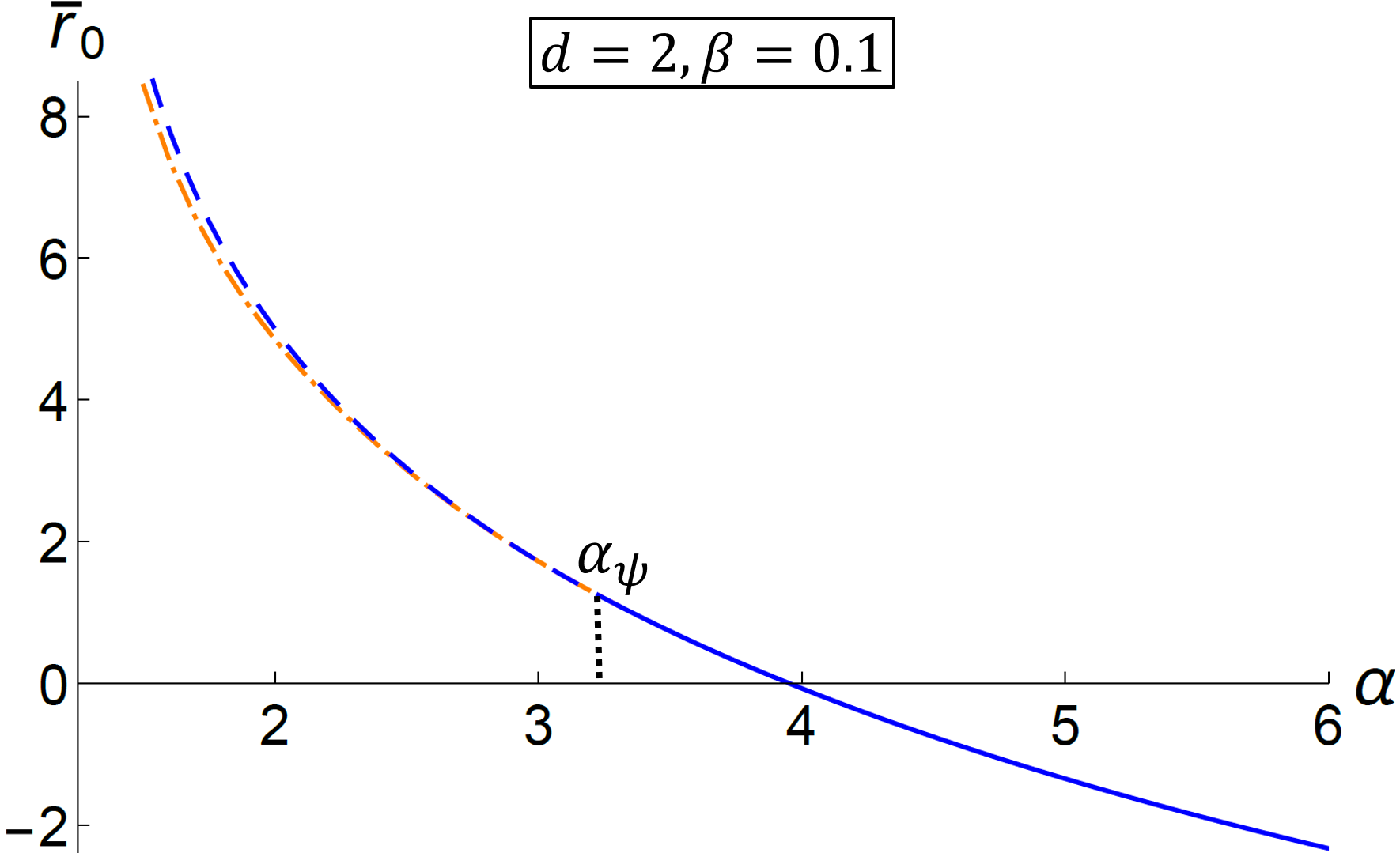} 
\par\end{centering}
\caption{The phase diagram of $\psi$ in the $(\alpha, \bar r_0)$ plane for $d=2$ and $\beta =0.1$. The upper spinodal (blue line) shows $\bar{r}_{0}^{max}$ with the lower one (dot-dashed orange line) showing $\bar{r}_{0}^{cr}$, which is the actual first order transition line where the global minimum of the effective action shifts from $\psi=0$ to a finite $\psi$. } \label{fig:spinodal}
\end{figure}

Regarding the first order transition of $\psi$, the actual $\bar{\bar{r}}_{0}^{cr}$ at which the first order $\psi^\ast$ occurs is actually slighter lower than $\bar{\bar{r}}_{0}^{max}$. The reason is that the effective action $S_{\mathrm{eff}}$ develops a local minimum  at $\psi^\ast =0$. We have found where the local minimum develops at $\bar{\bar{r}}_{0}= \bar{\bar{r}}_{0}^{max}$, $\psi^\ast=\psi^\ast_{cr}$. However, for this local minimum to be the global minimum, the condition $S_{\mathrm{eff}}(\psi^\ast_{cr})\leqslant S_{\mathrm{eff}}(\psi^\ast=0)$ must be satisfied. So we must evaluate the effective action at both local minima $\psi^\ast =0$ and $\psi^\ast = \psi^\ast_{cr}$, and find the actual $\bar{\bar{r}}_{0}^{cr}$ at which $S_{\mathrm{eff}}(\psi^\ast_{cr}) = S_{\mathrm{eff}}(\psi^\ast=0)$. In Fig. \ref{fig:spinodal}, we present the phase diagram of $\psi$ in the $(\alpha, \bar r_0)$ plane with both the actual $\bar{r}_{0}^{cr}$ and $\bar{r}_{0}^{max}$ plotted. Clearly, the difference between $\bar{r}_{0}^{cr}$ and $\bar{r}_{0}^{max}$ is negligible. In the rest of paper, we neglect this difference and approximate $\bar{r}_{0}^{cr}$ with $\bar{r}_{0}^{max}$. The same argument applies to the first order transition of $\varphi$ and the actual $\bar{r}_{0}^{cr}$ as a function of $\alpha$ is presented in Fig. 5 by \citet{Fernandes2012}, and is also negligible. Again, we neglect this difference in the rest of the paper.

Now we can combine the $\varphi$ and $\psi$ results to present the phase diagram in $\bar r_0$ and $\alpha$ for two representative $\beta$'s, shown in Fig. \ref{figtrans_2D}.  There are several characteristic regions of behavior classified by the nature and splitting of the two transitions, $T_\varphi$ and $T_\psi$.
\textbf{\begin{figure}[!h]\centering
   \begin{minipage}{\columnwidth}
     \includegraphics[width=\linewidth]{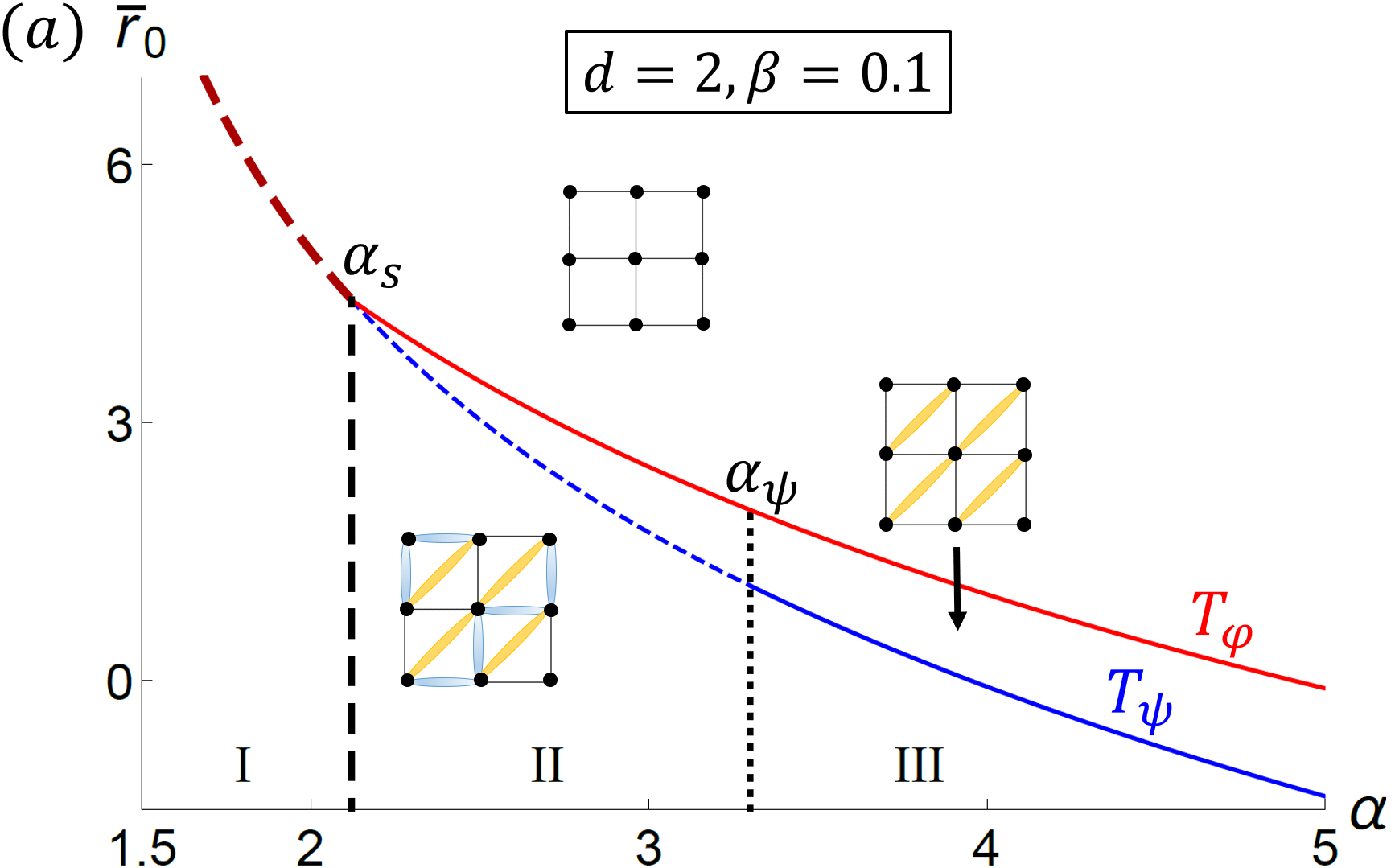}
   \end{minipage}
   \begin {minipage}{\columnwidth}
     \includegraphics[width=\linewidth]{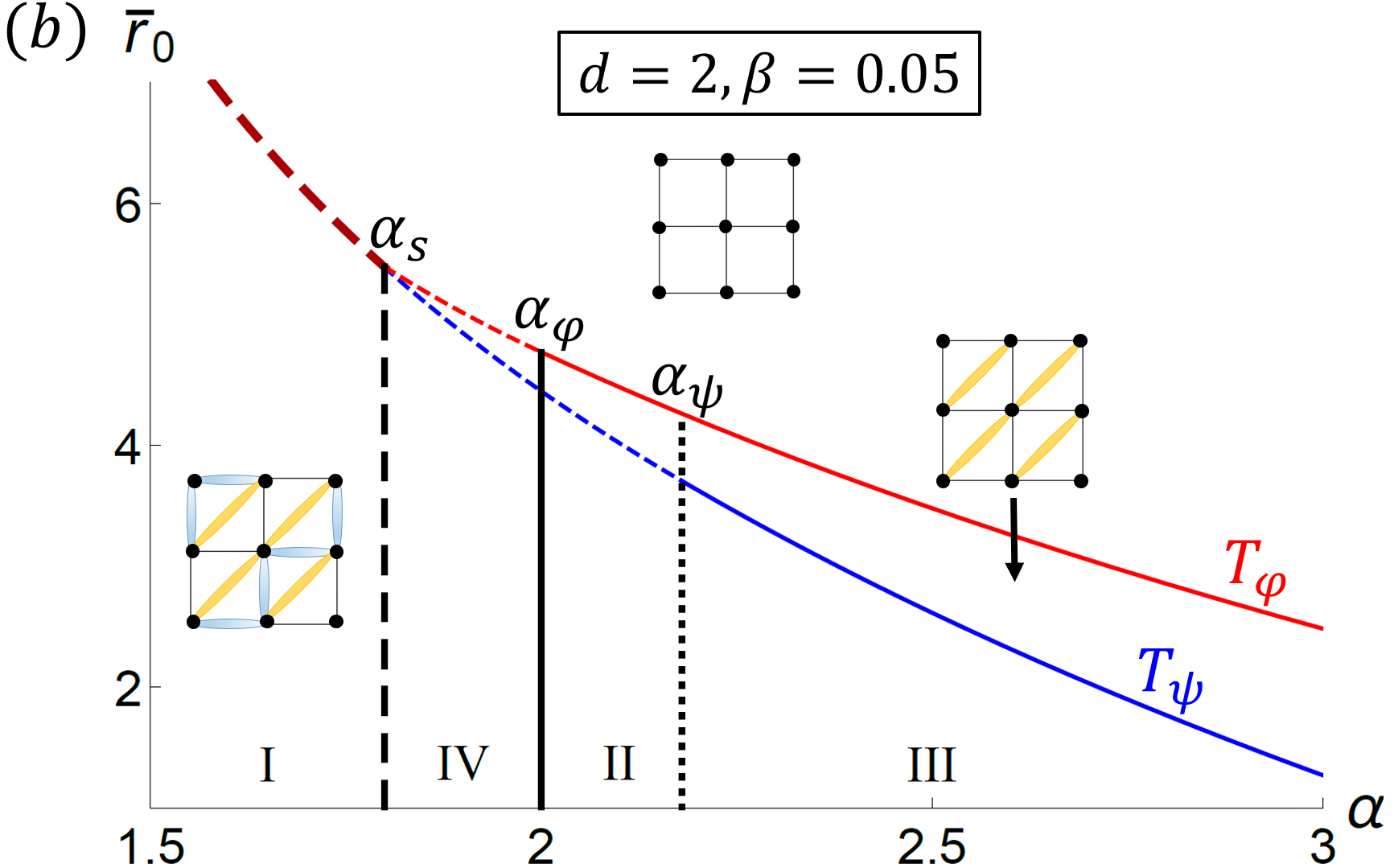}
   \end{minipage}
   \caption{(Color online) Two example phase diagrams of the onset of $\varphi$(red) and $\psi$(blue) with $\bar{r}_0$ plotted versus $\alpha$ for two values of $\beta$. Since $\bar{r}_0$ is linear in $T$, it can be thought of as a proxy for the transition temperature. $\alpha = u/g_1$ tunes the relative strength of uniform fluctuation and NNN biquadratic coupling, while $\beta = g_3/g_1$ tunes the relative strength of the NN and NNN biquadratic couplings. $T_\varphi$(red) indicates rotational symmetry breaking ($\varphi$), while $T_\psi$(blue) indicates dimerization ($\psi$), which breaks the diagonal mirror mirror symmetry. Solid lines indicate second order transitions; dashed lines indicate first order transitions; and the double-dashed line indicates simultaneous first order transitions. The three critical values of $\alpha$ are indicated with vertical black lines: $\alpha_\varphi$ with a solid line, $\alpha_\psi$ with a dotted line and $\alpha_s$ with a dashed line. In part (a) $\alpha_s$=2.12 and $\alpha_\psi=3.3$; in part (b) $\alpha_s = 1.8 $, $\alpha_\varphi =2$ and $\alpha_\psi = 2.18$. Different regions of behavior are labeled with Roman numerals, and their extent in $\alpha$ and $\beta$ is indicated in Fig \ref{fig:alpha1_alpha2_2D}.}\label{figtrans_2D}
\end{figure}}
 
We find that for any given $\beta$, the two transition lines will intersect at $\alpha=\alpha_s$: for $\alpha<\alpha_s$, $\varphi$ and $\psi$ turn on simultaneously, while for $\alpha>\alpha_s$, the two transitions split. In total, there are three critical values of $\alpha$ that separate four possible regions of transitions: $\alpha_s$, and $\alpha_\varphi$ and $\alpha_\psi$ which mark the change from first to second order transitions of $\varphi$ and $\psi$, respectively. Depending on the relative magnitude of $\alpha_s$ and $\alpha_\varphi$, there are two possible phase diagram topologies. For $\alpha_s<\alpha_\varphi$, typically there are four phase regions as shown in Fig. \ref{figtrans_2D}(a). While for $\alpha_s>\alpha_\varphi$, there are three possible phase regions as shown in Fig. \ref{figtrans_2D}(b).  

\begin{figure}[t]
\begin{centering}
\includegraphics[width=\columnwidth]{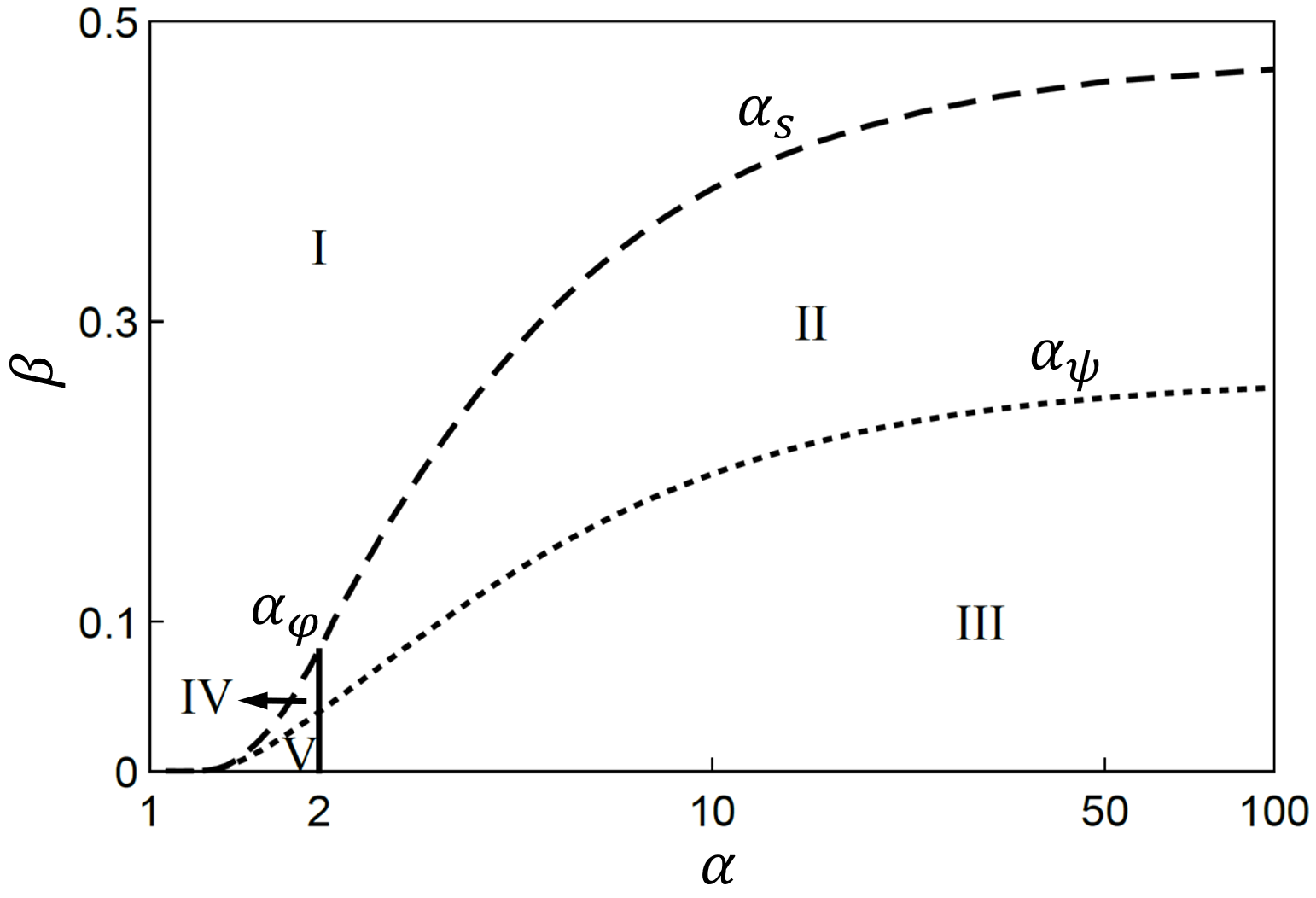} 
\par\end{centering}
\caption{The five regions of behavior in the $(\alpha, \beta)$ plane.  {\bf I}: $T_{\varphi 1} =T_{\psi 1}$; {\bf II}: $T_{\varphi 2} >T_{\psi 1}$; {\bf III}: $T_{\varphi 2} > T_{\psi 2}$; {\bf IV}: $T_{\varphi 1} > T_{\psi 1}$; {\bf V}:$T_{\varphi 1} > T_{\psi 2}$. $T_{O i}$ stands for the $i$-th($i=1,2$) order transition temperature of the order parameter $O(= \varphi, \psi)$. The asymptotic value of $\beta$ as $\alpha_\psi$(dotted) and $\alpha_s$(dashed) approaching infinity is $\beta_\psi=0.26$ and $\beta_s=0.48$ respectively. The vertical solid line strands for $\alpha_{\varphi} = 2$. It intercepts with $\alpha_\psi$ and $\alpha_s$ at $\beta_{\varphi \psi}=0.04$ and $ \beta_{\varphi s}=0.08$ respectively. Note that $\alpha_s$ and $\alpha_\psi$ stop at $\alpha=1$ since the effective action $S_{\mathrm{eff}}$ is unbounded below for $\alpha<1$.  } 
\label{fig:alpha1_alpha2_2D}
\end{figure}

\begin{figure}[htbp!]\centering
   \begin{minipage}{0.45\columnwidth}
     \includegraphics[width=\linewidth]{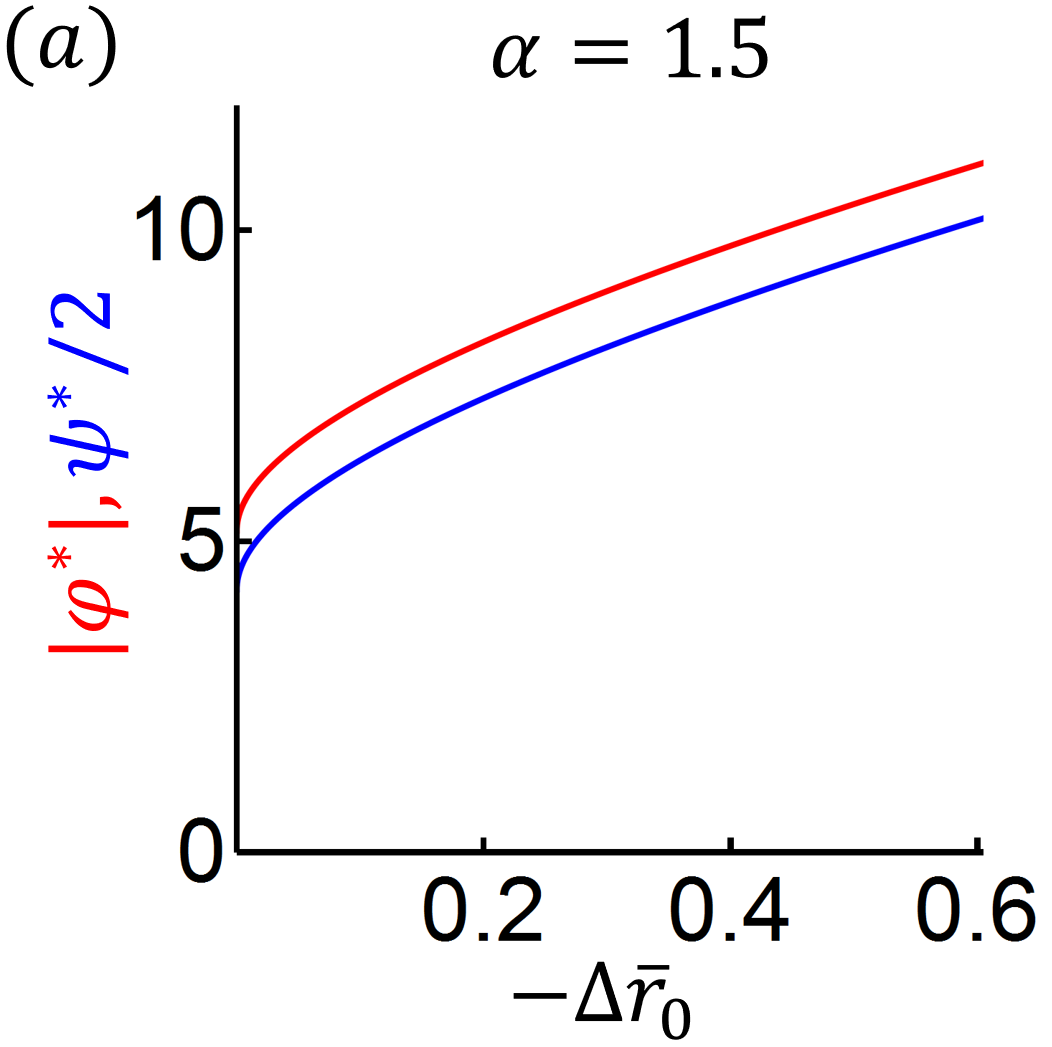}
   \end{minipage}
   \begin {minipage}{0.45\columnwidth}
     \includegraphics[width=\linewidth]{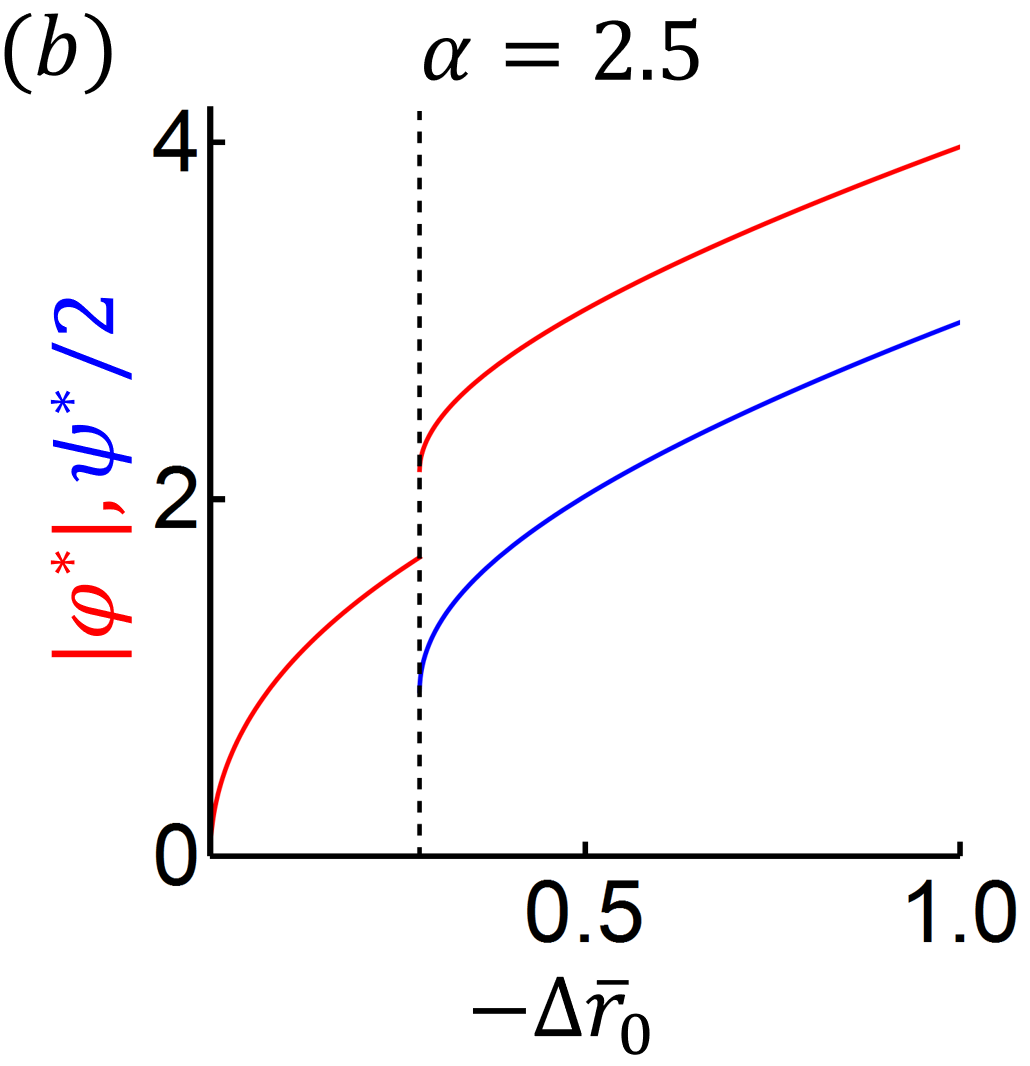}
   \end{minipage}
   \begin {minipage}{0.45\columnwidth}
     \includegraphics[width=\linewidth]{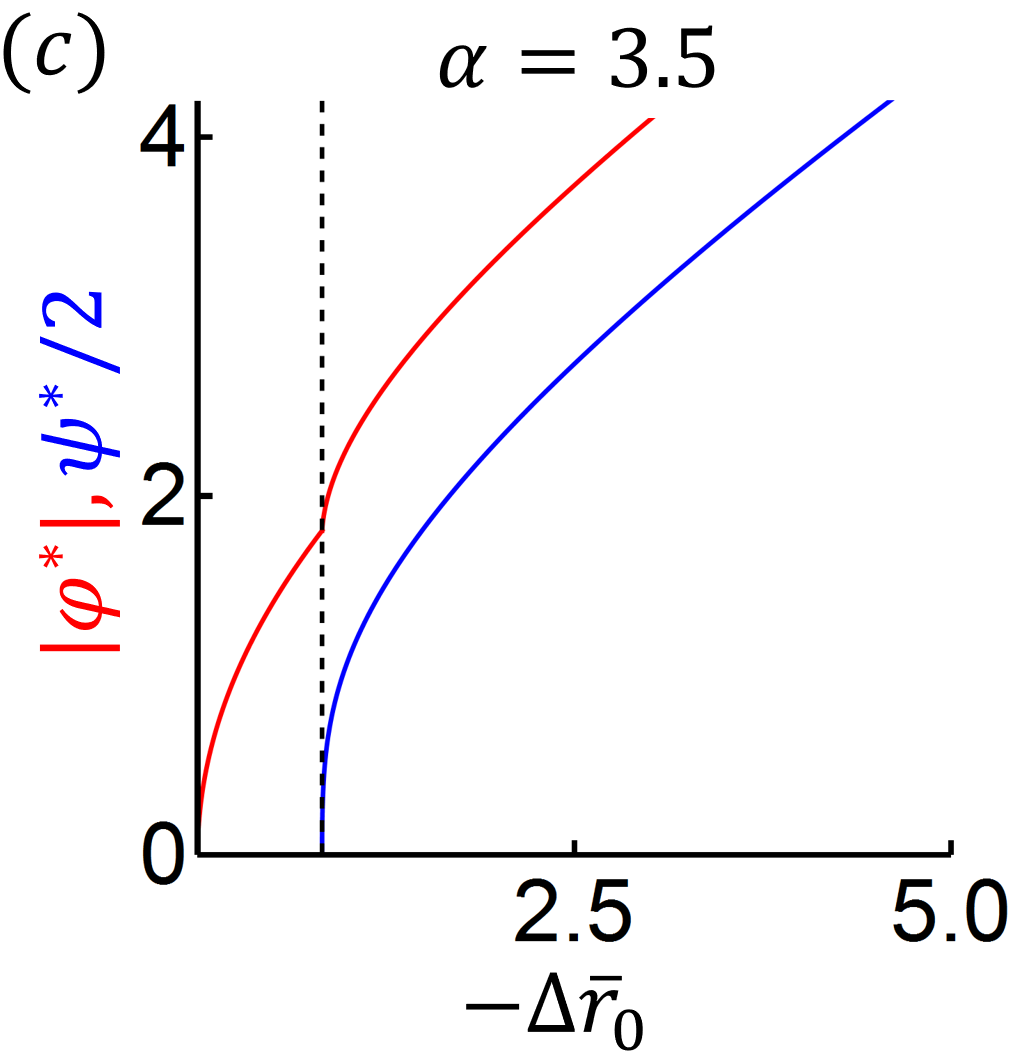}
   \end{minipage}
   \begin {minipage}{0.45\columnwidth}
     \includegraphics[width=\linewidth]{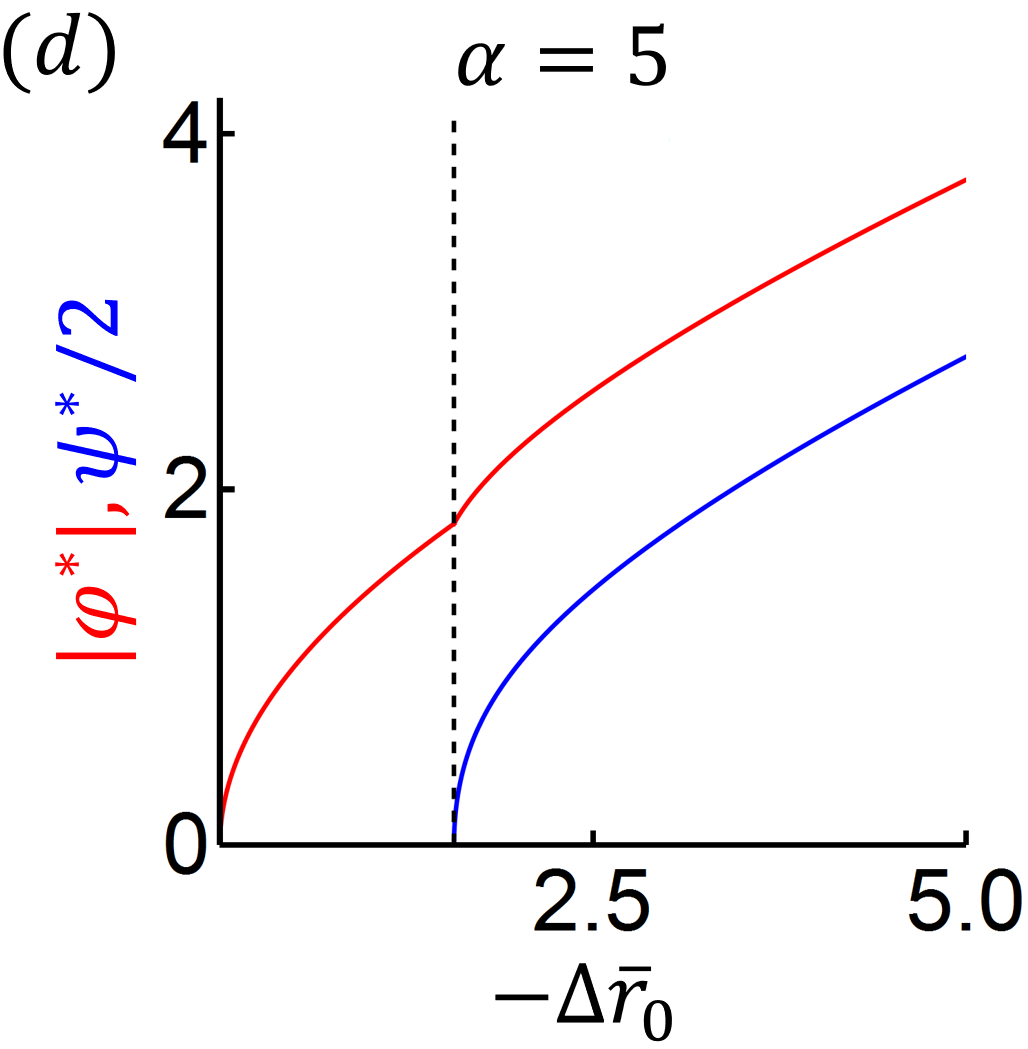}
   \end{minipage}
   \caption{(Color online) The onset of $\varphi$ (red) and $\psi$ (blue) as functions of $-\Delta \bar r_0$, for $\alpha$ in the three different regions for $\beta = 0.1$, $d = 2$ as shown in Fig. \ref{figtrans_2D}(a). Black dashed lines indicate where $\psi$ first turns on. In figure (a), $\alpha=1.5 < \alpha_s$(region \textbf{I}) (b) $\alpha_s<\alpha=2.5<\alpha_\psi $(region \textbf{II} of the phase diagram); (c)$\alpha=3.5>\alpha_\psi$(region \textbf{III}), which shows $\psi$ is almost first order at $\alpha$ slightly larger than $\alpha_\psi=3.3$; (d) $\alpha = 5>\alpha_\psi$(region \textbf{III}).} \label{fig:OP_vs_T_2D_a}
\end{figure}

$\alpha_\varphi$ is independent of $\beta$, but both $\alpha_\psi$ and $\alpha_s$ vary with $\beta$. We present all three values in a ``phase diagram'' in the $(\alpha, \beta)$ plane in Fig. \ref{fig:alpha1_alpha2_2D}. Both $\alpha_{\psi}$ and $\alpha_{s}$ increase monotonically with $\beta$, and both approach $1$ as $\beta\rightarrow 0$, and $\infty$ as $\beta\rightarrow\beta_\psi=0.26$ and $\beta_s=0.48$ respectively.  There are five regions of behavior. Utilizing the short-hand notation $T_{O i}$ to stand for the $i$-th($i=1,2$) order transition temperature of the order parameter $O(= \varphi, \psi)$, the five regions are, {\bf I}: $T_{\varphi 1} =T_{\psi 1}$, meaning simultaneous first order transitions for $\varphi$ and $\psi$; {\bf II}: $T_{\varphi 2} >T_{\psi 1}$, meaning a second order transition for $\varphi$ followed by a first order transition for $\psi$; {\bf III}: $T_{\varphi 2} > T_{\psi 2}$, meaning distinct second order phase transitions for $\varphi$ and $\psi$; {\bf IV}: $T_{\varphi 1} > T_{\psi 1}$, meaning distinct first order transitions for $\varphi$ and $\psi$; {\bf V}: $T_{\varphi 1} > T_{\psi 2}$, meaning a first order transition for $\varphi$ followed by a second order transition for $\psi$.

In Fig. \ref{fig:OP_vs_T_2D_a}, we plot the onset of $\varphi^\ast$ and $\psi^\ast$ for $\beta = 0.1$ and several values of $\alpha$ as functions of $\bar r_0$ to illustrate the generic behavior of these order parameters at the transitions. We plot $-\Delta \bar r_0 = \bar r_{0, cr} - \bar r_0$ along the $x$-axis, where we have shifted $\bar r_0$ by the $\bar r_{0, cr}$ where $\varphi$ onsets, and changed the sign so that increasing $x$ corresponds to decreasing temperature. One point of interest is the large jump in $\varphi^\ast$ as $\psi^\ast$ undergoes a first order transition, as shown in Fig. \ref{fig:OP_vs_T_2D_a}(b). This jump originates from the linear $\varphi \psi^2$ coupling that causes $\psi^2$ to act as a field for $\varphi$.

\subsection{Three dimensions} \label{subsec3_2}

Next we treat the three-dimensional limit, where we find no pre-emptive nematic transitions, just a single, simultaneous first order transition. For $d=3$, the saddle-point equations in eqs.\eqref{eq8} become:
\begin{eqnarray} \label{eq28}
\frac{\bar r_{0} - r}{u} & = & \sqrt{r +\varphi-\psi}+\sqrt{r +\varphi+\psi}+ 2 \sqrt{ r -\varphi}\nonumber \\
\frac{\varphi}{g_{1}} & = & \sqrt{r +\varphi-\psi}+\sqrt{r +\varphi+\psi}- 2 \sqrt{ r -\varphi}\nonumber \\
\frac{\psi}{2 g_{3}} & = & -\sqrt{r+\varphi-\psi}+\sqrt{r+\varphi+\psi}.
\end{eqnarray}
We follow the same steps as for 2D, solving the above saddle-point equations for both $g_3=0$ and $g_3\neq 0$, and obtaining the overall phase diagram.

For $g_3 = 0$, $\psi = 0$, and we only need to solve the saddle point equations in eqs.\eqref{eq28} for $r$ and $\varphi$.
\begin{eqnarray}
\frac{\bar r_{0} - r}{2 u} & = & \sqrt{r +\varphi}+\sqrt{r -\varphi} \nonumber \\
\frac{\varphi}{ 2 g_{1}} & = & \sqrt{r +\varphi}-\sqrt{r -\varphi}.
\end{eqnarray}
We can define $z \equiv \varphi/r$ in order to eliminate $r$ from the above equations,
\begin{align} 
\bar r_0 = 8 g^2_1 \B( \alpha +\frac{1}{1+\sqrt{1-z^2}}   \B).
\end{align}

As before the transition will occur for the $z$ where $\bar r_0$ is maximized.  In 3D, this is clearly always at $|z| = 1$, where $r = -\varphi$.  As this maximum is at a nonzero $\varphi$, the transition is first order, and the condition for magnetic order is satisfied at the transition, and so the two transitions will be simultaneous.  In order to examine the nature of the magnetic transition, we return to the saddle-point equations including $M$, \eqref{eq20}, which simplify for $d=3$ and $g_3 = 0$:
\begin{eqnarray} 
&	& \frac{\bar r_{0} - r}{u}  =  2 \sqrt{r +\varphi}+ 2 \sqrt{r -\varphi} - M^2 \nonumber \\
&	& \frac{\varphi}{g_{1}}  =  2 \sqrt{r +\varphi}-2 \sqrt{r -\varphi} - M^2 \nonumber \\
&	& (r + \varphi ) M =0.
\end{eqnarray}
From the final equation, we find that either $r=-\varphi$ or $M=0$. Setting $r=-\varphi$ and substituting it into the first two equations, we obtain:
\begin{eqnarray} 
\frac{\bar r_{0} + \varphi}{u}  &=&   2 \sqrt{ -2 \varphi} - M^2 \nonumber \\
 \frac{\varphi}{g_{1}} &=& -2 \sqrt{ -2 \varphi} - M^2.
\end{eqnarray}
from which we get the relationship between $\bar r_0$ and $M$,
\begin{align}
\bar r_0 \!=\! g^2_1  \B[ (1-\alpha)\frac{M^2}{g_1} \! +\! 4 (1+\alpha) \B( 1 + \sqrt{ 1 + \frac{M^2}{2 g_1}}  \B)  \B] .
\end{align}
A straight forward calculation shows that $M$ at the maximum $\bar r_0$, denoted as $M_\varphi$, is generically nonzero.
\begin{align}
M_\varphi=\frac{2\sqrt{2 g_1 \alpha}}{\alpha-1}.
\end{align}
which means the first order nematic instability of $\varphi$ triggers a first order magnetic order transition.

\begin{figure}[htbp]
\begin{centering}
\includegraphics[width=\columnwidth]{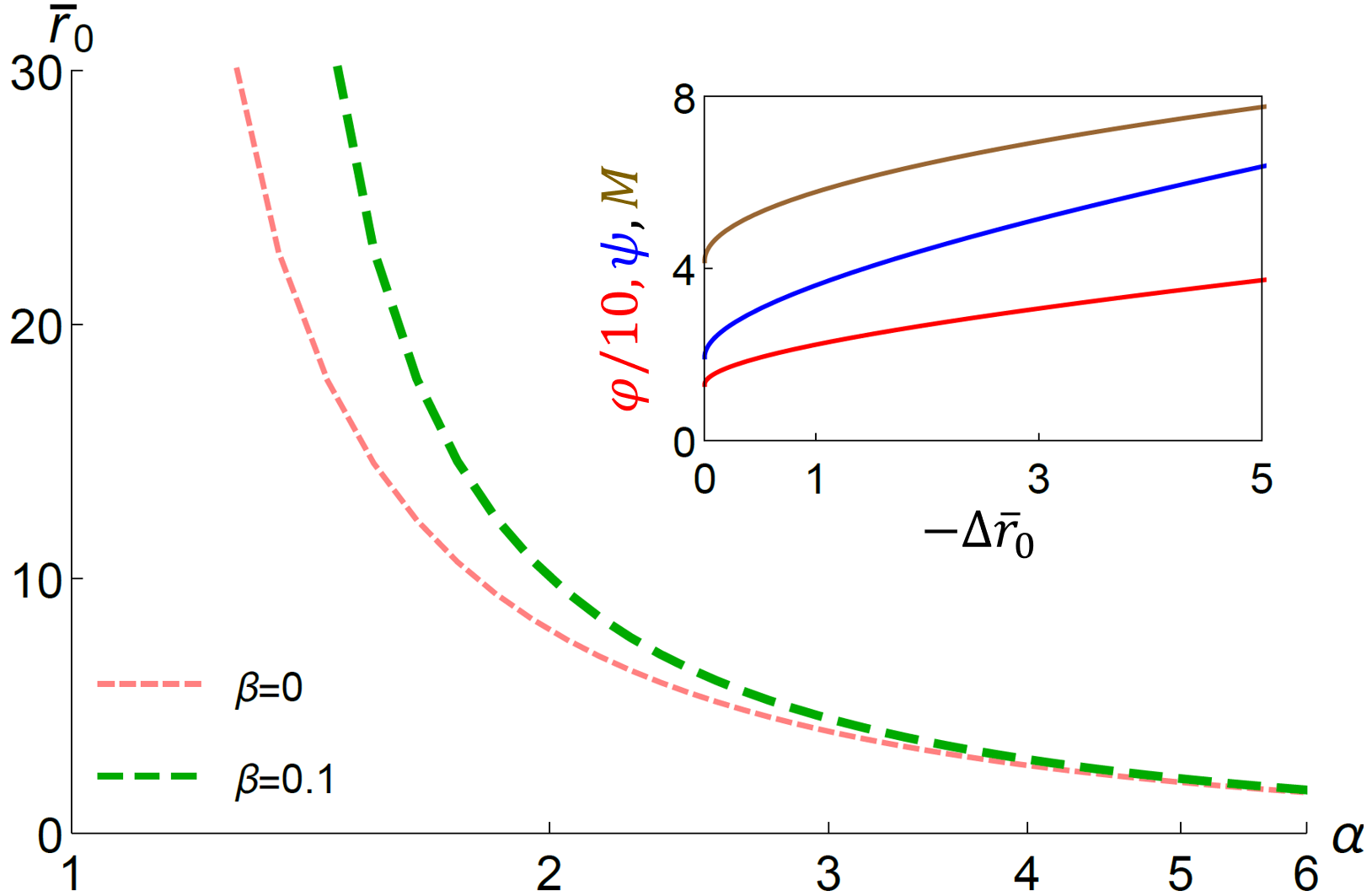} 
\par\end{centering}

\caption{(Color online) Three dimensional phase diagram for $\varphi, \psi $ and $M$, for two different values of $\beta = g_3/g_1$. At $\beta=0$ (dashed pink), $\psi$ of course does not turn on, and we have a simultaneous first order transition of $\varphi$ and $M$. At nonzero $\beta$, all three transitions are simultaneous and first order (thick double-dashed dark green), with increasing $\beta$ increasing the transition temperature($~\bar{r}_0$). Inset: all three order parameters $\varphi/10$(red), $\psi$(blue) and $M$(brown) as a function of $-\Delta \bar r_0$ for $\alpha = 2$ and $\beta=0.1$.   }
\label{fig:r0bar_alpha_3D}
\end{figure}

Next we turn to the finite $g_3$ problem, where we similarly find that the Ising-bond order transition for $\psi$ is accompanied by a simultaneous magnetic transition at $r=-\varphi+\psi$, which means that all three transitions are simultaneous. For conciseness, we will directly start with the saddle-point equations including $M$, \eqref{eq20}, and replace $r=-\varphi+\psi$:
\begin{eqnarray} \label{eq29}
\frac{\bar r_{0} - (\psi - \varphi)}{u} 	&  = & \sqrt{2 \psi}+ 2 \sqrt{\psi - 2 \varphi } - M^2 \cr
\frac{\varphi}{g_{1}} & = &   \sqrt{2 \psi}- 2 \sqrt{\psi - 2 \varphi} - M^2\cr
 \frac{\psi}{2 g_{3}} &=	&    \sqrt{2 \psi} + M^2 .
\end{eqnarray}
We can solve the third equation for $\psi(M)$,
\begin{align}
\psi= 2 g^2_3 \B( 1+\sqrt{1+\frac{M^2}{g_3}} \B)^2.
\end{align}
Substituting this expression into the second equation, we find $\varphi(M)$. At last, we substitute both $\varphi(M)$ and $\psi(M)$ into the first equation to get $\bar r_0 (M)$. 
\begin{align}
& \bar r_0 = g^2_1 \B[ 4(\alpha +1)  - (\alpha -2\beta -1) \frac{M^2}{g_1} \cr
& +  2 \beta (\alpha + 2 \beta -1)  \B( 1+\sqrt{1+\frac{M^2}{g_3}} \B) \cr
& \! + \! 4(\alpha \!+\! 1) \! \sqrt{1 \!+\! \frac{M^2}{g_1} \!+\!  \frac{1}{2} \beta (\beta \!-\! 1) \B( 1\! +\! \sqrt{1 \!+\! \frac{M^2}{g_3}} \B)^2} 
 \B].
\end{align}

$\bar r_0 (M)$ reaches its maximum value at a finite $M_\psi$, which turns on at a higher $\bar r_0$  than $M_\varphi$ for all $\beta \neq 0$, implying that $\psi$ and $\varphi$ transitions are always simultaneous, and coincident with the magnetic transition. All in all, for three dimensions, we will have only one single first order transition line in the phase diagram for any given $\beta$. Therefore, there are no preemptive Ising transitions any more, as in the SS case\cite{Fang2008, Xu2008, Mazin2009, Fernandes2012, Fernandes2014, Chubukov2015, Batista2011, Si2011, Capati2011, Brydon2011, Liang2013, Yamase2015}. Representative phase diagrams in 3D are shown in Fig. \ref{fig:r0bar_alpha_3D}. As $\beta$ decreases, the simultaneous first order transition approaches, but is always above the simultaneous transition line for $\beta=0$, indicating that the $\psi$ bond order enhances the transition temperature beyond that with only $\varphi$ and $M$, just as $\varphi$ enhances the transition temperature beyond that of only $M$, where $M$ orders at $r=-\varphi(+\psi)>0$, while the bare magnetic order emerges at $r=0$. This means that the emergence of the Ising-bond orders increase the ordering temperature of $M$. Therefore, even though all the transitions are simultaneous and first order, the Ising-bond order transitions are primary, and the magnetic transition is induced by their feedback. 

\subsection{Intermediate dimensions$(2<d<3)$} \label{subsec3_3}

\subsubsection{Generic solution}

For intermediate dimensions, we get a range of behavior that interpolates between the 2D and 3D results. As before, we begin with the simple case where $\psi = 0$, which we treat by setting $g_3$ and $\psi$ to zero.  Again, these results reproduce \citet{Fernandes2012}. These equations govern the region in the $(\alpha,\bar r_0)$ plane above the $\psi$ transition.  Eqs.\eqref{eq8} reduce to
\begin{eqnarray}
\frac{\bar r_{0} - r}{2 u} & = & \left(r +\varphi\right)^{\frac{d-2}{2}}+\left(r -\varphi\right)^{\frac{d-2}{2}} \nonumber \\
\frac{\varphi}{ 2 g_{1}} & = & \left(r +\varphi\right)^{\frac{d-2}{2}}-\left(r -\varphi\right)^{\frac{d-2}{2}} .
\end{eqnarray}
We again introduce $z \equiv \varphi/r$ and eliminate $r$ to obtain,
\begin{align}
\bar r_0 = (2 g_1)^{\frac{2}{4-d}} Q(\alpha, z ),
\end{align}
where
\begin{align} 
&Q(\alpha, z) = \B[  \frac{(1+z)^{\frac{d-2}{2}}-(1-z)^{\frac{d-2}{2}}}{z}  \B]^{\frac{d-2}{4-d}} \cr
&\qquad \times \B[ (\alpha + \frac{1}{z}) (1+z)^{\frac{d-2}{2}} + (\alpha - \frac{1}{z}) (1-z)^{\frac{d-2}{2}}  \B].
\end{align} 
As before, the transition occurs at the value of $z$ that maximizes $Q(\alpha,z)$.  There are three regions in $(\bar r_0, \alpha)$ separated by two critical values of $\alpha$.
\begin{align}
\alpha_{\varphi 1}=\frac{1}{3-d},\quad \alpha_{\varphi 2}=\frac{6-d}{6-2d}.
\end{align} 

In the region $1<\alpha<\alpha_{\varphi 1}$, $\bar r_0$ reaches its maximum when $|z|=1$. Here, $r=-\varphi$, and thus a simultaneous magnetic transition is triggered by $\varphi$. In this case, we use eqs.\eqref{eq20} to solve for both $\varphi$ and $M_\varphi$, where we use the subscript to indicate that this is the magnetization (and thus magnetic transition) that emerges when $\psi = 0$.
\begin{eqnarray} 
&	& \frac{\bar r_{0} - r}{u}  =  2 \left(r +\varphi\right)^{\frac{d-2}{2}}+ 2 \left(r -\varphi\right)^{\frac{d-2}{2}} - M_\varphi^2 \nonumber \\
&	& \frac{\varphi}{g_{1}}  =  2 \left(r +\varphi\right)^{\frac{d-2}{2}}-2 \left(r -\varphi\right)^{\frac{d-2}{2}} - M_\varphi^2 \nonumber \\
&	& (r + \varphi ) M_\varphi =0.
\end{eqnarray}
From the last equation, we find that $r=-\varphi$ or $M_\varphi=0$. We then substitute $r=-\varphi$ into the first two equations
and solve to find
\begin{align}
& \frac{\bar r_0 + 2 u M^2}{g_1 (1 + \alpha_1)} =  2\B[ \frac{2 (\bar r_0 + 2 u M^2)}{ (1 + \alpha_1)} \B]^{\frac{d-2}{2}} + M^2 \cr 
& \bar r_0 = 4 u (-2 \varphi)^{\frac{d-2}{2}} + (\alpha_1 -1)\varphi. 
\end{align}
Using the last equation, we can solve for the $\varphi_{cr}$ at which $\bar{r}_0$ is maximized.
\begin{align}
\varphi_{cr} =  - 2^{\frac{d}{4-d}} \Big( \frac{\alpha-1}{d-2} \Big)^{-\frac{2}{4-d}},
\end{align}
which is always finite, indicating that the simultaneous transition of $\varphi$ and $M_\varphi$ is always first order.

\begin{figure}[htbp]
\begin{centering}
\includegraphics[width=\columnwidth]{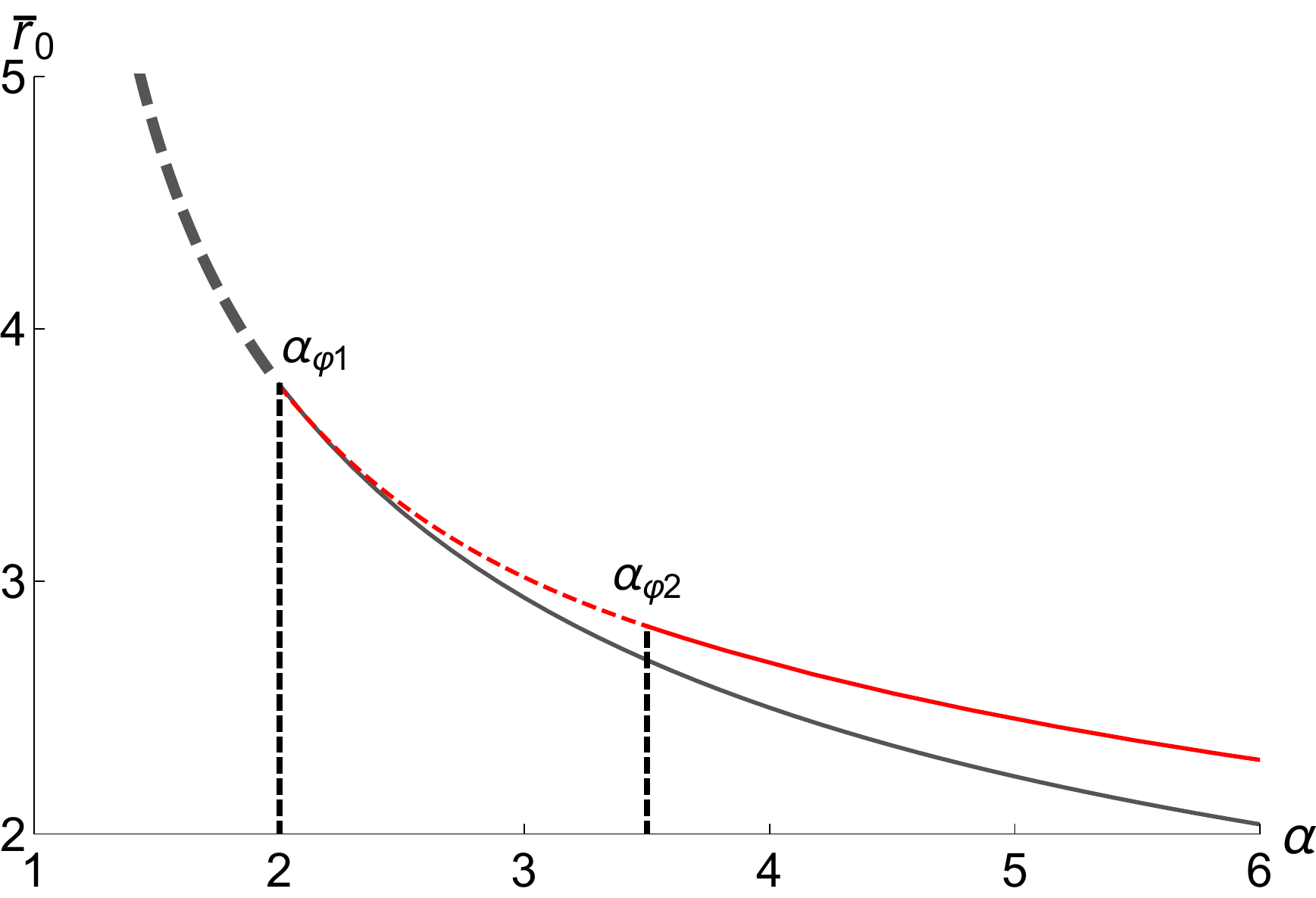} 
\par\end{centering}

\caption{ (Color online) The phase diagram for $g_3=0, d=2.5$ in the $(\alpha, \bar r_0)$ plane, showing $\varphi$(top, red) and $M_\varphi$(bottom, gray). First(second) order transitions are indicated by dashed(solid) lines. For $1<\alpha<\alpha_{\varphi 1}$, the Ising bond-order, $\varphi$ and magnetic order, $M_\varphi$ turn on simultaneously (thick dashed gray line). For $\alpha_{\varphi 1}<\alpha<\alpha_{\varphi 2}$, the transitions split. The transition of $\varphi$ remains first order while $M_\varphi$ is now second order. Finally, for $\alpha>\alpha_{\varphi 2}$, both transitions are second order. For $d = 2.5$, $\alpha_{\varphi 1}=2$ and $\alpha_{\varphi 2}=3.5$. }
\label{fig:r0bar_alpha1_phi_2dot5D}
\end{figure}

For $\alpha_{\varphi 1}<\alpha<\alpha_{\varphi 2}$, the first instability occurs for $0 < |z| < 1$. A second order magnetic transition then follows below the first order $\varphi$ transition. In the region $\alpha>\alpha_{\varphi 2}$, both transitions are second order. A representative phase diagram, for $d = 2.5$ is shown in Fig. \ref{fig:r0bar_alpha1_phi_2dot5D}.

Now we turn to the full problem, where we allow $\psi$ to be nonzero.  It can turn on simultaneously with or below the $\varphi$ and magnetic transitions.  In order to solve the saddle point equations here, we introduce $z \equiv \varphi/r$, as before and $z_1 \equiv \psi/r$. The saddle-point equations \eqref{eq8} become
\begin{eqnarray}
r^{\frac{2-d}{2}} \frac{\bar r_{0} \!- \! r}{u} & \!=\! & \left(1 \!+\!z\!-\!z_1 \right)^{\frac{d-2}{2}}\!+\!\left(1 \!+\!z \!+\!z_1\right)^{\frac{d-2}{2}}\!+\! 2 \left(1 \!-\! z \right)^{\frac{d-2}{2}} \nonumber \\
r^{\frac{4-d}{2}} \frac{z}{g_{1}} & \!=\! & \left(1 \!+\!z\!-\!z_1 \right)^{\frac{d-2}{2}}\!+\!\left(1 \!+\!z\!+\!z_1\right)^{\frac{d-2}{2}}\!-\! 2 \left(1 \!-\!z \right)^{\frac{d-2}{2}}\nonumber \\
r^{\frac{4-d}{2}} \frac{z_1}{2 g_{3}} &\! =\! & -\left(1+z- z_1 \right)^{\frac{d-2}{2}}\!+\!\left(1+z+z_1\right)^{\frac{d-2}{2}}.
\end{eqnarray}
We can again eliminate $r$ to find two equations: $\bar r_0$ as a function of $z$ and $z_1$,
\begin{align} \label{eq2}
\bar r_0 = &  g_1 ^{\frac{2}{4-d}} Q_1 (z, z_1),
\end{align}
and a constraint relating $z$ and $z_1$ via $\beta = g_3/g_1$.  
\begin{equation} \label{eq3}
\beta = Q_2 (z, z_1).
\end{equation}
Here, the two $Q$ functions are given by,
\begin{align} 
& Q_1 ( z, z_1)\! = \! \B[  \frac{\!(\! 1\!+\!z \!-\!z_1 \!)^{\frac{d-2}{2}}\!+\!(\!1\!+\!z \!+\! z_1 )^{\frac{d-2}{2}} \!-\!2 ( \!1\!-\!z \!)^{\frac{d-2}{2}}}{ z}  \B]^{\frac{d-2}{4-d}} \cr
& \quad \times \B[ (\alpha + \frac{1}{z}) (1+z -z_1)^{\frac{d-2}{2}} + (\alpha\! + \frac{1}{z}) (1+z +z_1)^{\frac{d-2}{2}} \cr
& \qquad + 2 (\alpha - \frac{1}{z}) (1-z)^{\frac{d-2}{2}}  \B],  \cr
& Q_2 (z, z_1) = \frac{z_1}{2 z} \frac{\!(\! 1\!+\!z \!-\!z_1 \!)^{\frac{d-2}{2}}\!+\!(\!1\!+\!z \!+\! z_1 \!)^{\frac{d-2}{2}} \!-\!2 (\! 1\!-\!z \!)^{\frac{d-2}{2}}}{-(1 +z-z_1 )^{\frac{d-2}{2}} +(1 +z+z_1 )^{\frac{d-2}{2}}}.
\end{align}

\begin{figure}[b]
\begin{centering}
\includegraphics[width=0.75\columnwidth]{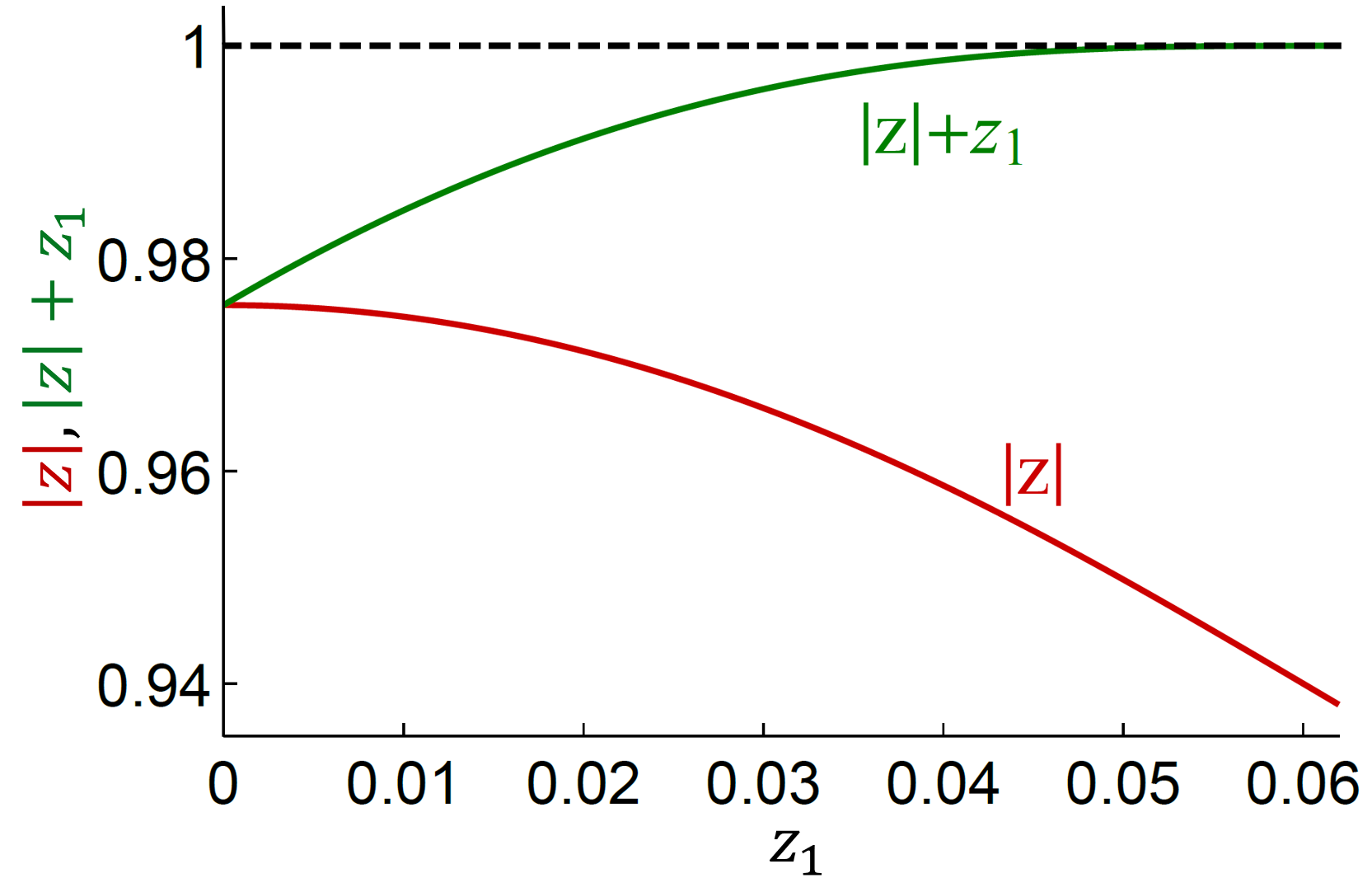} 
\par\end{centering}
\caption{(Color online) $|z|$(dark red) and $|z|+z_1$(dark green) as functions of $z_1$ for $d=2.5$ and $\beta=0.1$. Note that when $|z| + z_1 = 1$, magnetic order onsets.}
\label{fig:z_z1}
\end{figure}

\begin{figure}[h]\centering
   \begin{minipage}{0.48\columnwidth}
     \includegraphics[width=\linewidth]{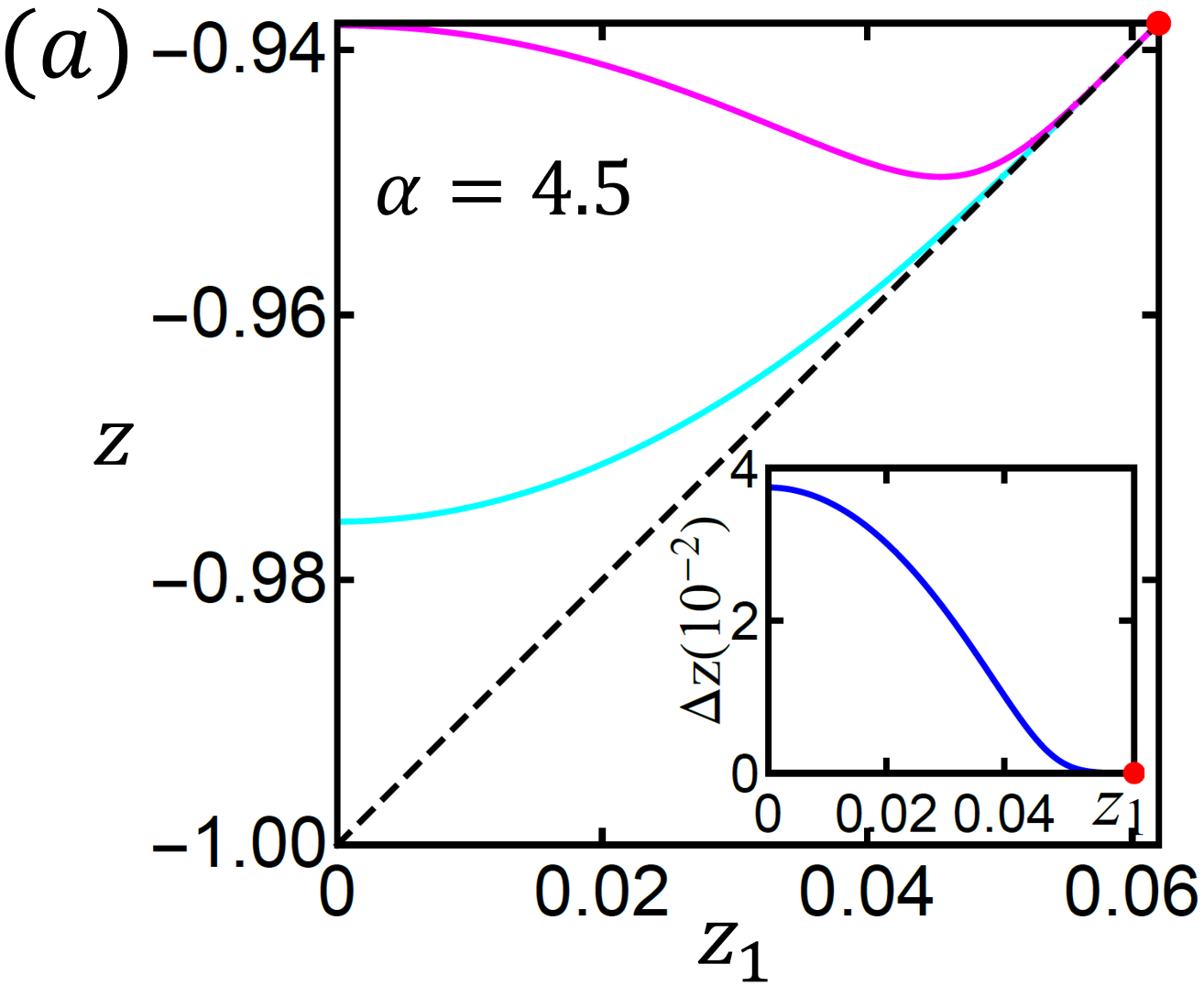}
   \end{minipage}
   \begin {minipage}{0.5\columnwidth}
     \includegraphics[width=\linewidth]{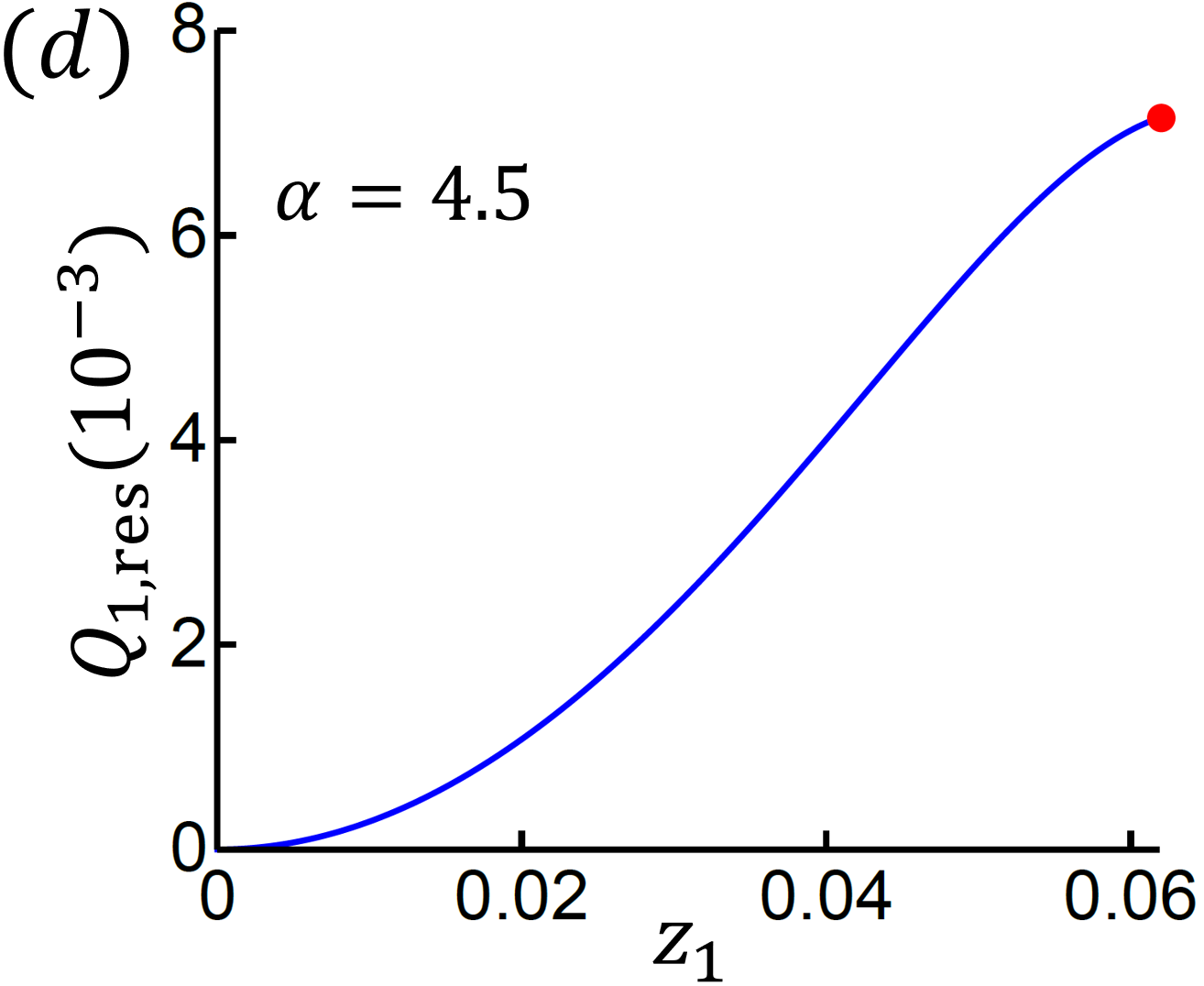}
   \end{minipage}
   \begin{minipage}{0.48\columnwidth}
     \includegraphics[width=\linewidth]{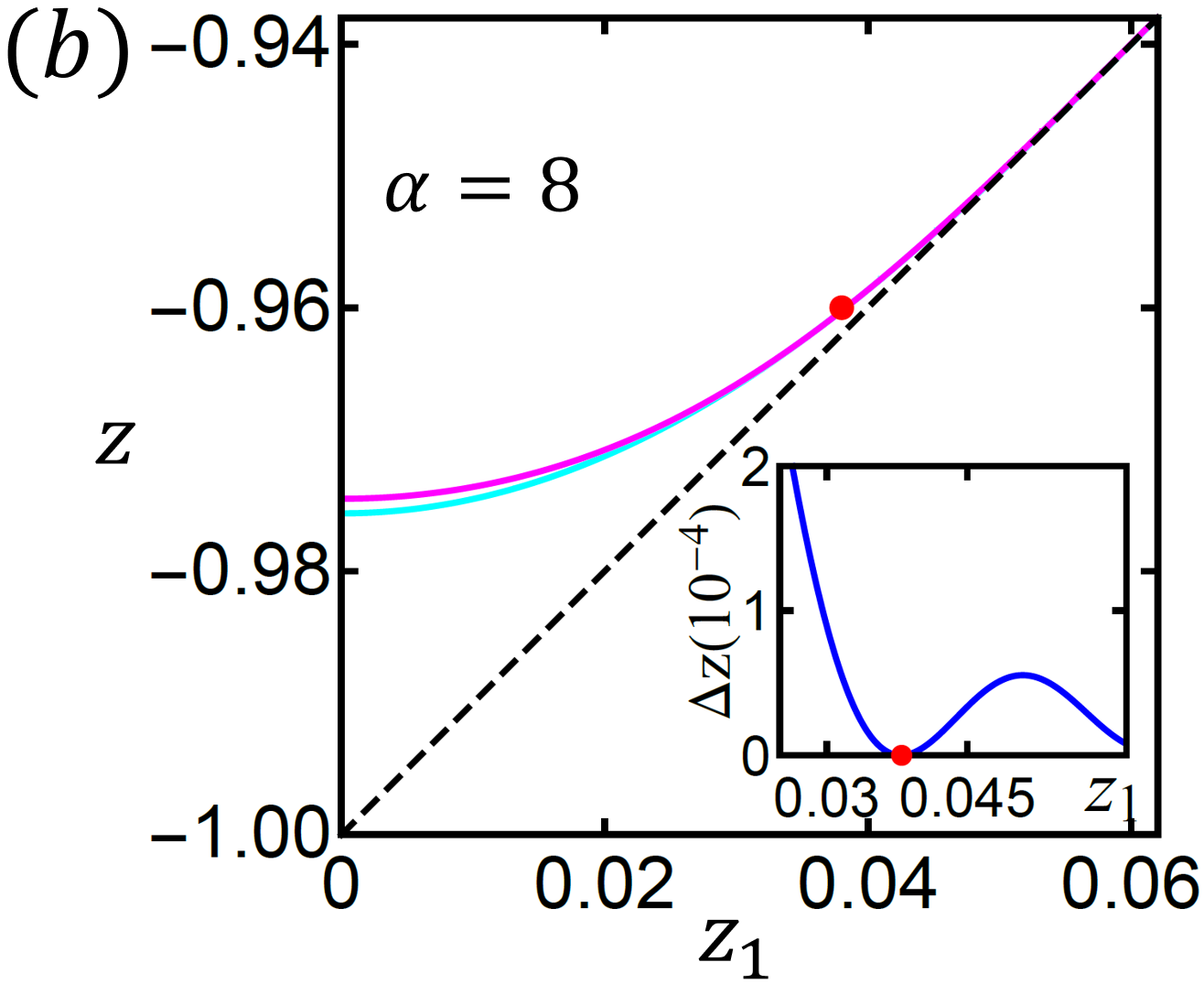}
   \end{minipage}
   \begin {minipage}{0.5\columnwidth}
     \includegraphics[width=\linewidth]{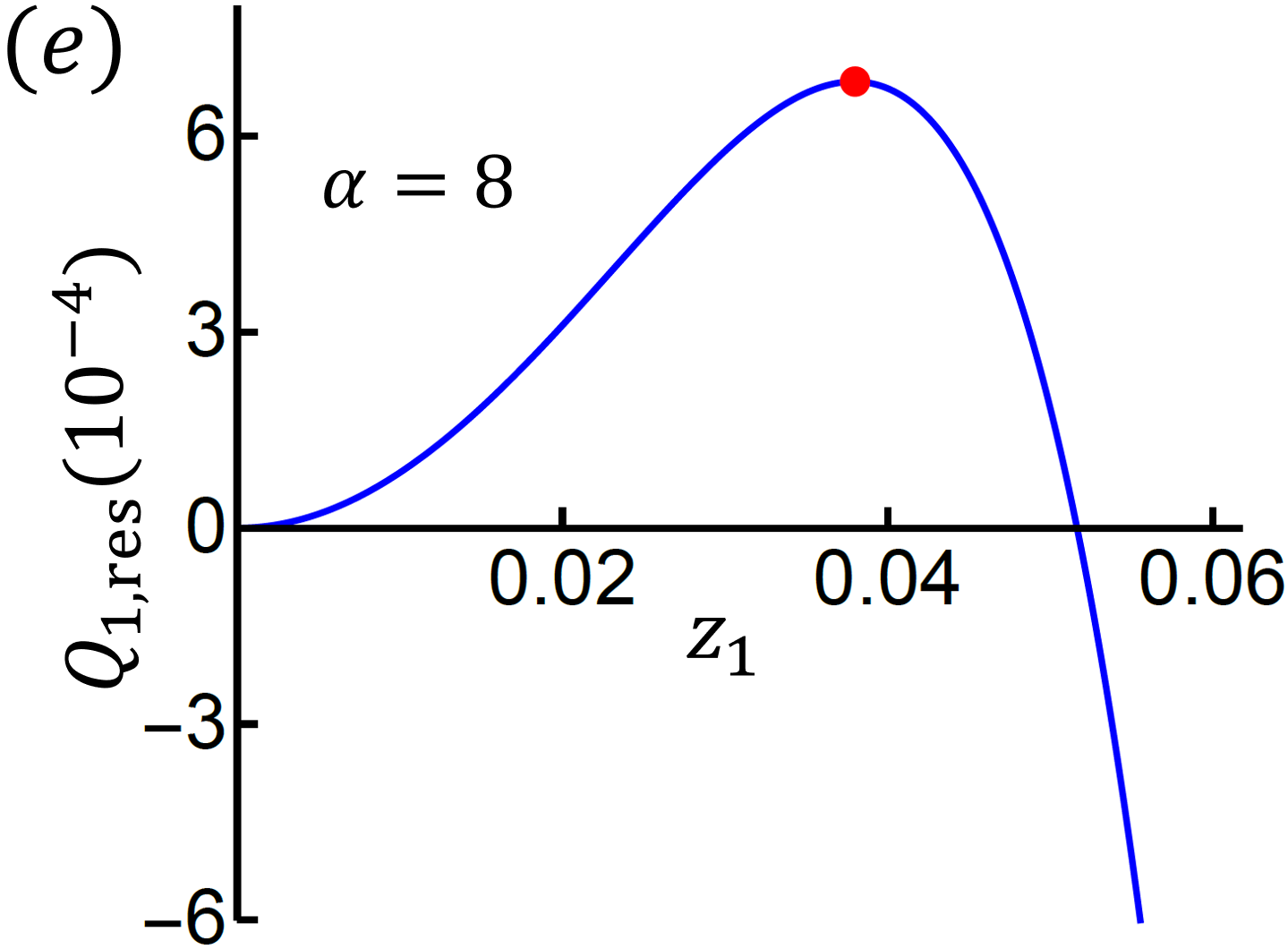}
   \end{minipage}
   \begin{minipage}{0.48\columnwidth}
     \includegraphics[width=\linewidth]{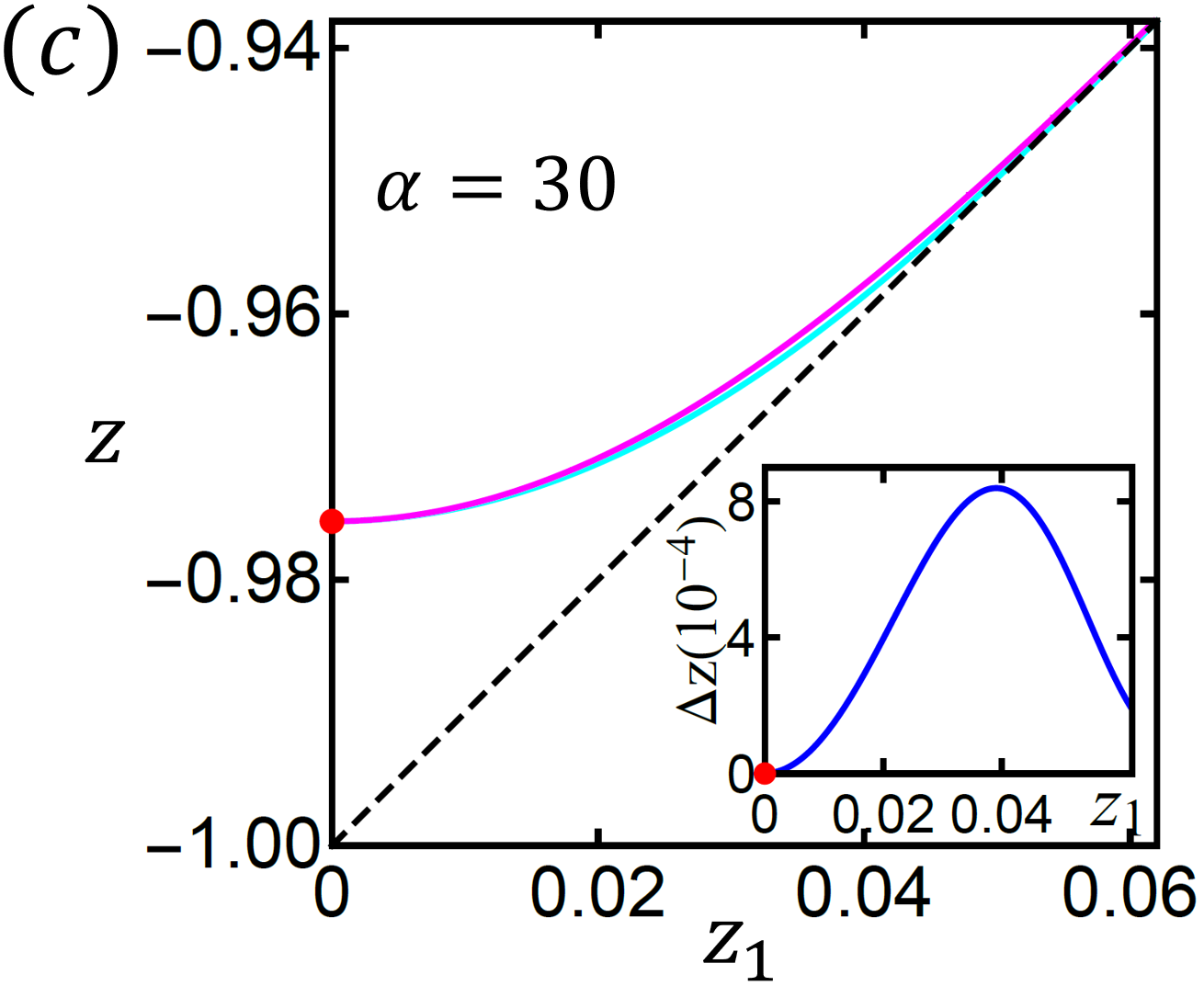}
   \end{minipage}
   \begin {minipage}{0.5\columnwidth}
     \includegraphics[width=\linewidth]{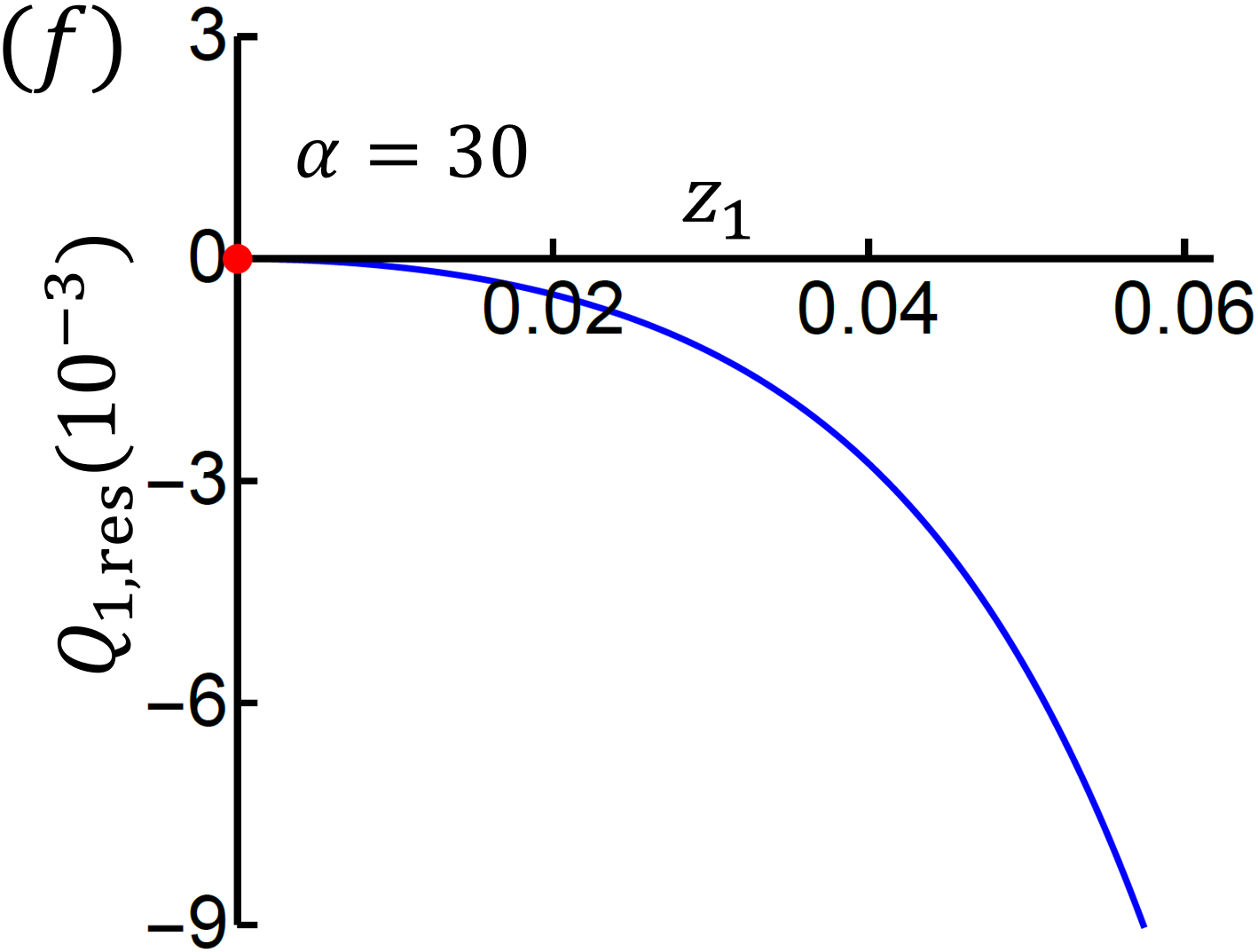}
   \end{minipage}
   \caption{(Color online) Examples of how the three regions may be resolved by considering several representative values of $\alpha = 4.5, 8$ and $30$ for $d = 2.5$ and $\beta = 0.1$. \emph{Left column}[(a)-(c)]: leading instabilities as shown in the $(z_1, z)$ plane. We show the solution of $z$ and $z_1$ at the maximum of $Q_1(z, z_1)$ for representative values of $\alpha$ in the three regions: $1<\alpha<\alpha_{\psi 1}$(top), $\alpha_{\psi 1}<\alpha<\alpha_{\psi 2}$(middle) and $\alpha>\alpha_{\psi 2}$(bottom). The blue line represents $Q_2(z, z_1)=\beta$, and the purple line represents the maxima of $Q_1(z, z_1)$. Their intersection is indicated with red dots. The dashed black line indicates the asymptotic line $|z|+z_1=1$, at which magnetic order develops. As the intersection point is difficult to resolve by eye, the inset shows the difference between the purple and blue lines $\Delta z$ as a function of $z_1$. \emph{Right column}[(d)-(f)]: leading instabilities as shown in the $(z_1, Q_1)$ plane. We plot $Q_{1, \mathrm{res}}$ as a function of $z_1$ to show the value of $z_1$ that maximize $Q_{1, \mathrm{res}}$. For this $d$ and $\beta$, $\alpha_{\psi 1}=4.8, \alpha_{\psi 2}=12.1$.  }
\label{fig:Q1_Q2_solution}
\end{figure}

The leading instability is determined by solving for $z_1(z)$ at a given $\beta$, and looking for the $z_1$ that maximizes the resulting $Q_1(z_1)$.   If this $z_1$ is zero, the transition is second order, while if it is finite, with $|z|+z_1 < 1$, the transition is first order.  Finally, if the maximum occurs where $|z|+z_1=1$, i.e. $r=|\varphi|+\psi$, the magnetic transition occurs simultaneously. Fig. \ref{fig:z_z1} displays $|z(z_1)|$ and $|z(z_1)|+z_1$ as determined from the constraint equation, \eqref{eq3}, which are used to determine the value of $z$ at the transition, and whether magnetic order is triggered. $|z|+z_1$ gradually increases and reaches one as $z_1$ increases from $0$ to its maximum value. For small $\beta$, $|z|$ decreases monotonically as $z_1$ increases, but for large $\beta$, $|z|$ undergoes an upturn before decreasing with increasing $z_1$. In Fig. \ref{fig:Q1_Q2_solution}, we present the leading instability in both the $(z_1, z)$ and $(z_1, Q_1)$ planes.
By investigating the slope of $Q_1(z_1)$ at the maximum $z_1$ and $z_1=0$, we find these three different regions of behavior.
For $1<\alpha<\alpha_{\psi 1}$(figs.(a) and (d)), $\psi$ and $M$ develop simultaneously at a first order transition. For $\alpha_{\psi 1}<\alpha<\alpha_{\psi 2}$(figs.(b) and (e)), $\psi$ remains first order, but $M$ develops at a second order transition. For $\alpha>\alpha_{\psi 2}$(figs.(c) and (f)), the two transitions are both second order.  Note that to obtain the full phase diagram, we must compare the $\psi = 0$ results with these. 

In the first region, where the $\psi$ transition is first order and simultaneous with magnetism, the magnetic transition will also be first order.  In order to see this, we once again go back to the effective action with the magnetic order parameters and solve eqs.\eqref{eq10} by substituting $r = \psi - \varphi$,
\begin{eqnarray}
\frac{\bar r_{0} - (\psi - \varphi)}{u} 	&  = & ( 2 \psi)^{\frac{d-2}{2}}+ 2 \left(\psi - 2 \varphi \right)^{\frac{d-2}{2}} - M^2  \label{eq14}\\
\frac{\varphi}{g_{1}} & = &   ( 2 \psi)^{\frac{d-2}{2}}- 2 \left(\psi - 2 \varphi \right)^{\frac{d-2}{2}} - M^2  \label{eq15}\\
 \frac{\psi}{2 g_{3}} &=	&    ( 2\psi)^{\frac{d-2}{2}} + M^2 \label{eq16}.
\end{eqnarray}

From \eqref{eq16}, we solve for $M$ as a function of $\psi$.
\begin{align} \label{eq17}
M = \B [ \frac{\psi}{2 g_3}  -( 2 \psi)^{\frac{d-2}{2}} \B]^{\frac{1}{2}},
\end{align}
which implies that $\frac{\psi}{2 g_3}  -( 2 \psi)^{\frac{d-2}{2}} \geq 0$, or that $\psi \geq \frac{1}{2} ( 4 g_3)^{\frac{2}{4-d}}$, which is consistent with a first order transition for $\psi$. From the first two equations, we get
\begin{align} \label{eq5}
\bar r_0 = & 4 u \left(\psi - 2 \varphi \right)^{\frac{d-2}{2}} + \psi + (\alpha -1 )\varphi\cr
\beta = & \frac{\psi / 2}{2 g_1 [( 2 \psi)^{\frac{d-2}{2}}-  \left(\psi - 2 \varphi \right)^{\frac{d-2}{2}}]-\varphi },
\end{align}
which we solve for $\psi(\varphi)$ and $\bar{r}_0(\psi)$. In the first region, where $\alpha<\alpha_{\psi 1}$, $M_\psi$ turns on simultaneously with $\psi$, meaning a first order $\psi$ transition triggers a first order magnetic transition. In the second region, $\alpha>\alpha_{\psi 1}$, $M_\psi$ becomes second order and appears below $\psi$. We find that for any $\beta$, $M_\psi$ always has a higher transition temperature ($\bar r_0$) than $M_\varphi$, meaning that the second Ising bond-ordering further boosts the magnetic transition temperature, and also that we need only consider the magnetic transition obtained with $\psi \neq 0$.

\begin{figure}[!htbp]
\begin{centering}
\includegraphics[width=0.9\columnwidth]{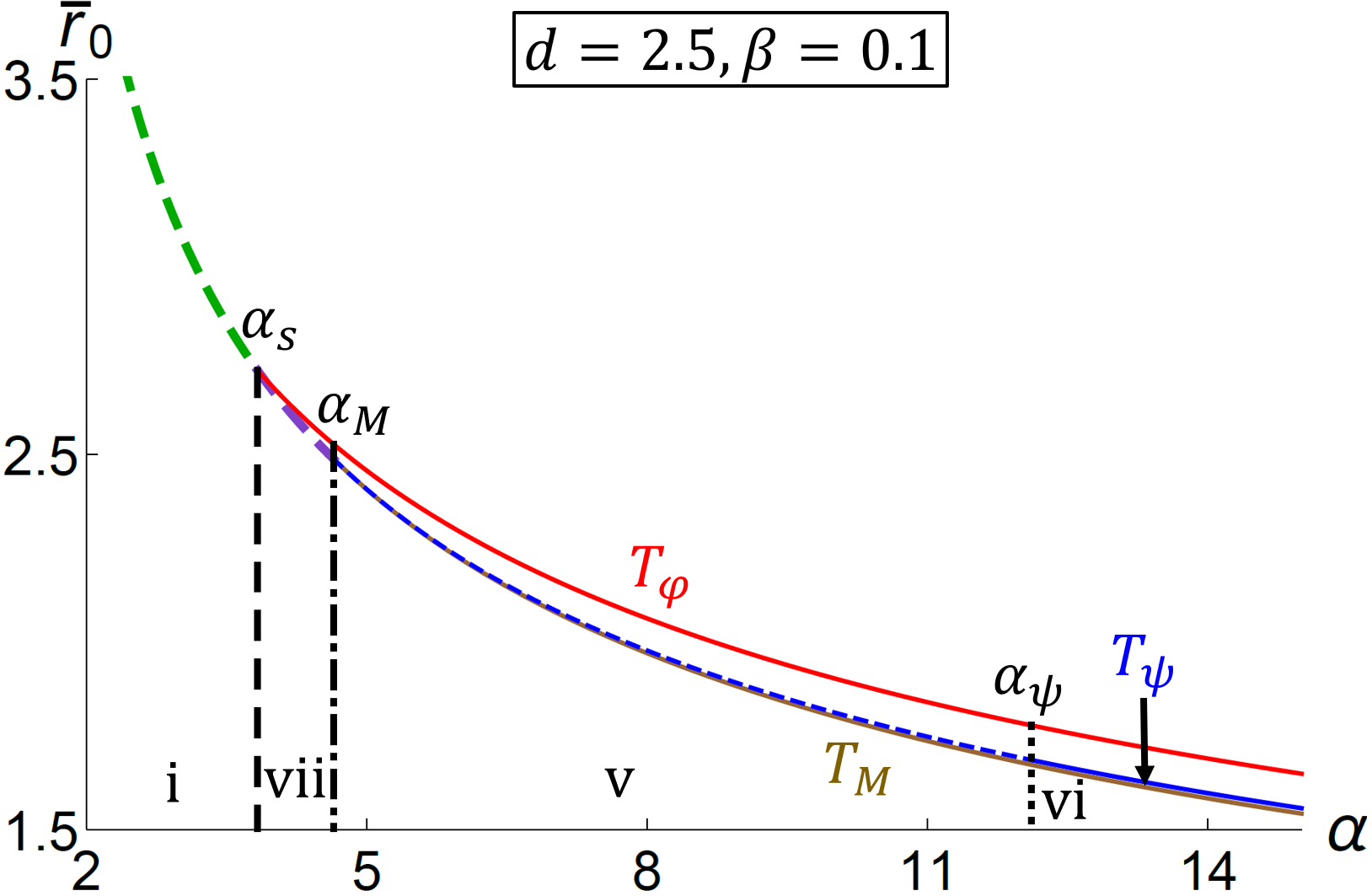} 
\par\end{centering}
\caption{(Color online) The phase diagram in the $(\bar{r}_0,\alpha)$ plane for $d=2.5$ and $\beta=0.1$, which shows first (dashed) and second (solid) order transitions of $\varphi$ (red), $\psi$ (blue) and $M$ (brown). The four regions of behavior are, \textbf{i}: $T_{\varphi 1} =T_{\psi 1} =T_{M 1}$; \textbf{vii}: $T_{\varphi 1} > T_{\psi 1} = M_1$; \textbf{v}: $T_{\varphi 2} >T_{\psi 1} > T_{M 2}$; \textbf{vi}: $T_{\varphi 2} > T_{\psi 2} > T_{M 2}$, where the notation is defined in Sec. \ref{subsec3_1}. The thick dashed green line represents simultaneous first order transitions of $\varphi$, $\psi$ and $M$ while the thick dashed purple line indicates simultaneous first order transitions of $\psi$ and $M$. In this figure, $\alpha_s=3.83$, $\alpha_M =4.64$ and $\alpha_\psi=12.11$. }
\label{fig:r0bar_vs_alpha_2dot5_beta=0dot1}
\end{figure}

To illustrate the general form of our results, we present an example phase diagram for $d=2.5$ and $\beta=0.1$ in Fig. \ref{fig:r0bar_vs_alpha_2dot5_beta=0dot1}. There are four regions in total. In region \textbf{i}, we have a simultaneous first order transition of $\varphi$, $\psi$ and $M$; region \textbf{vii} is a second order transition of $\varphi$, followed by simultaneous first order transitions of $\psi$ and $M$; region \textbf{v} is a second order transition of $\varphi$ followed by a first order transition of $\psi$ and later followed by a second order transition of $M$, where though the transitions of $\psi$ and $M$ are close, they are distinct; region \textbf{vi} contains three distinct second order phase transitions.  These phase diagrams are in general defined by a number of critical points.  For clarity, we now define: $\alpha_s$, where $T_\varphi = T_\psi$, and below which the two transitions are simultaneous and first order; $\alpha_\varphi$, where $T_\varphi$ becomes second order; $\alpha_\psi$, where $T_\psi$ becomes second order; and $\alpha_M$, where $T_M$ becomes second order, which always occurs when $T_M = T_\psi$.  In terms of the previous definitions, $\alpha_\varphi = \mathrm{Max}[ \alpha_{\varphi 2}, \alpha_s]$, $\alpha_\psi = \alpha_{\psi 2}$, and $\alpha_M = \alpha_{\psi 1}$, while $\alpha_s$ is new and requires comparing the $g_3 = 0$ and $g_3 \neq 0$ results.  Not all critical points will occur in all phase diagrams, or rather they will not always be distinct, as one can see in Fig. \ref{fig:r0bar_vs_alpha_2dot5_beta=0dot1}, where $\alpha_\varphi$ coincides with $\alpha_s$ and is thus not shown.

As the dimensionality and $\beta$ vary, the critical values of $\alpha$ evolve, leading to a number of different regions of behavior. In general, as the dimensionality increases, the phase space for magnetic order increases from zero in two dimensions to being everywhere (below any transition) in three dimensions. The phase space for second order transitions gradually vanishes as we approach three dimensions.  In the next section, we demonstrate this evolution, and the rich range of possible phase diagrams, by showing the results for several representative dimensionalities in detail.

\subsubsection{Evolution of the phase diagram for $2<d<3$}

\begin{figure}[h]\centering
   \begin{minipage}{\columnwidth}
     \includegraphics[width=\linewidth]{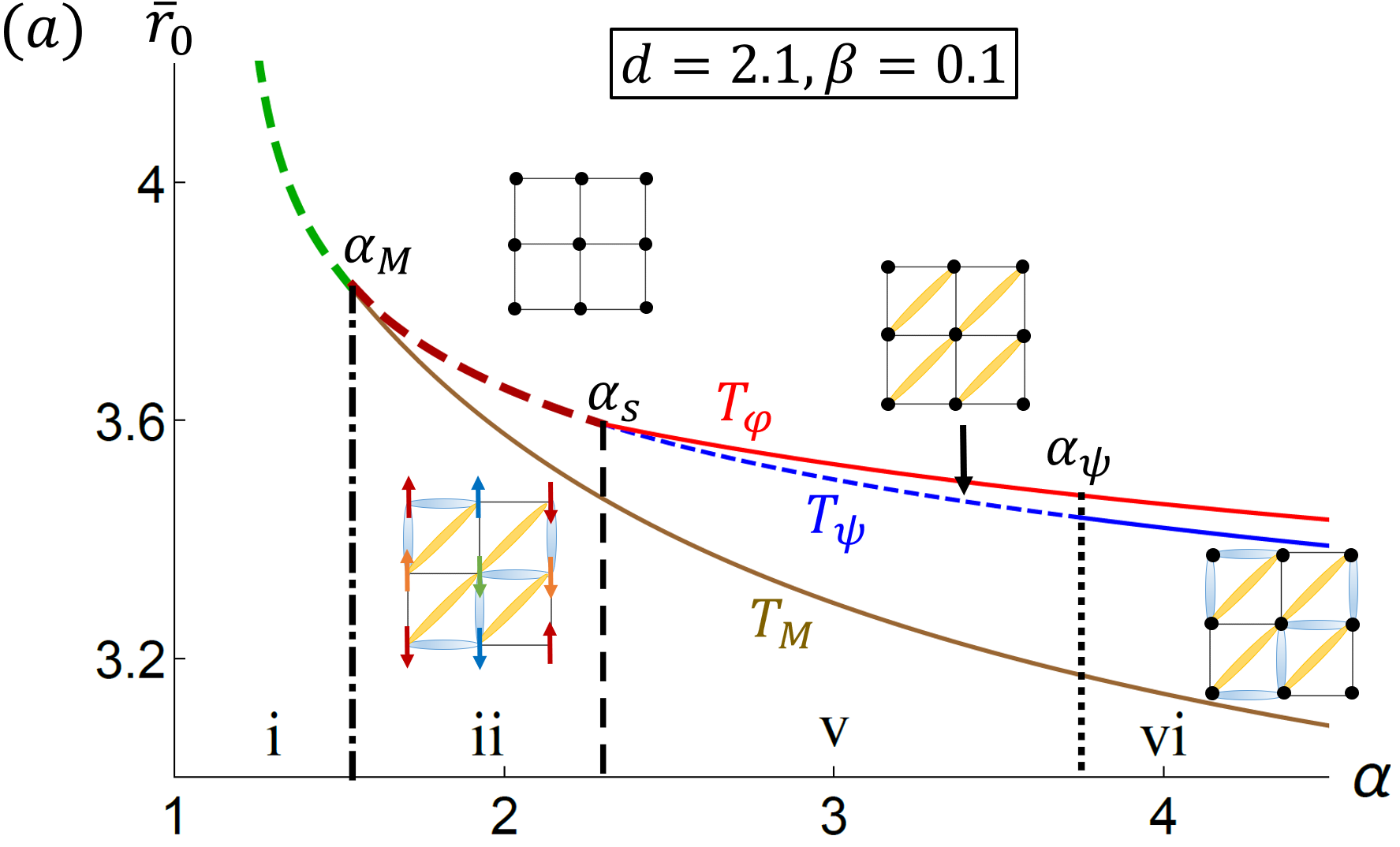}
   \end{minipage}
   \begin {minipage}{\columnwidth}
     \includegraphics[width=\linewidth]{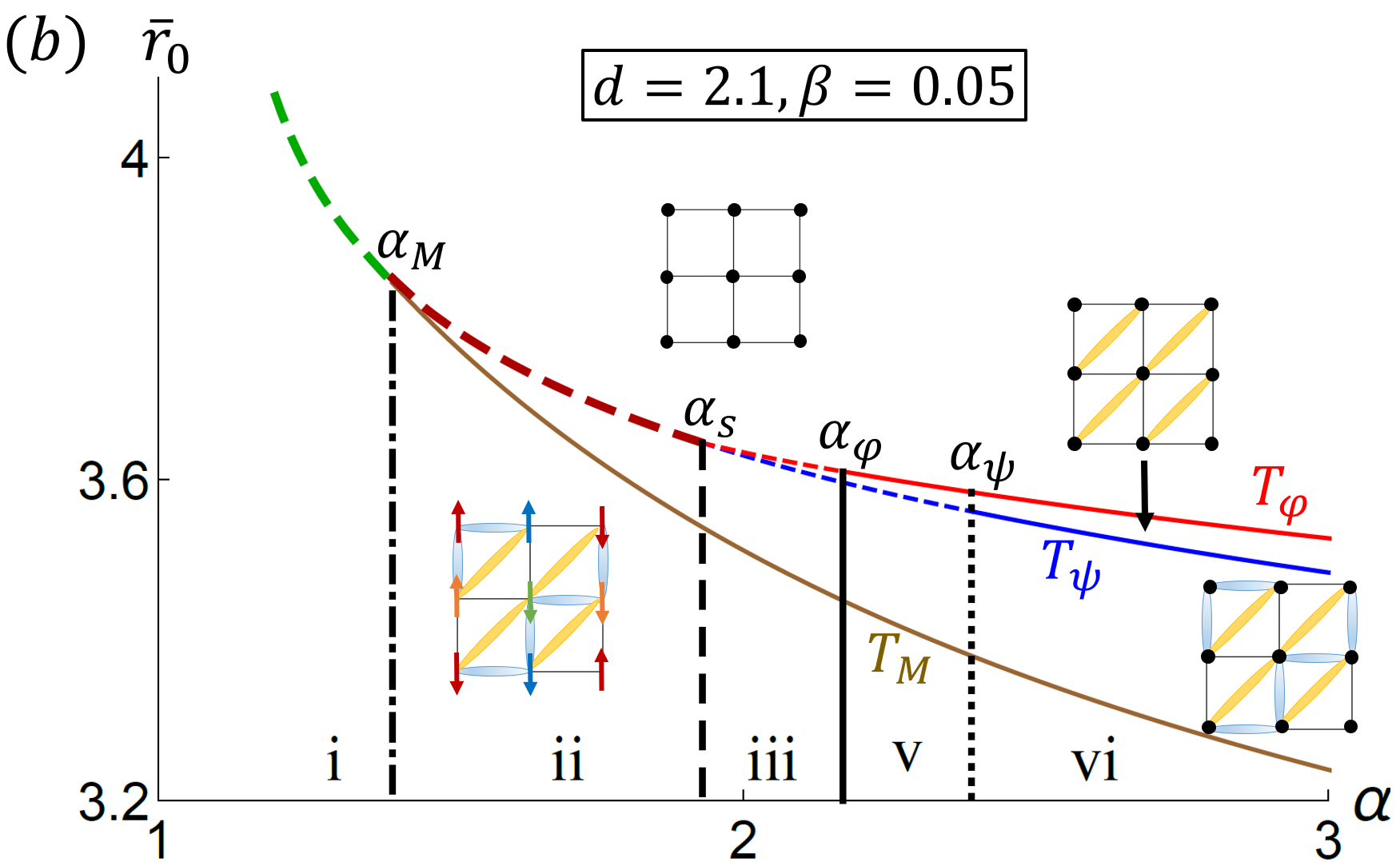}
   \end{minipage}
   \caption{(Color online) Two example phase diagrams showing how $\varphi$ (red), $\psi$ (blue) and $M$ (brown) develop as $\alpha$ varies, for $d=2.1$. $\bar{r}_0$ plays the role of temperature and the ratios of the biquadratic couplings are: (a) $\beta=0.1$; (b)$\beta=0.05$. Dashed(solid) lines indicate first(second) order transitions. The thick dashed green line indicates simultaneous first order transitions of $\varphi$, $\psi$ and $M$, while the thick dashed red line indicates simultaneous first order transitions of $\varphi$ and $\psi$ only.  The regions of different classes of behavior are indicated in Fig. \ref{fig:alpha1_vs_alpha2_2dot1}. The corresponding critical values of $\alpha$'s in the above figures are: (a)$\alpha_M =1.54 $, $\alpha_s = 2.3 $ and $\alpha_\psi = 3.75 $ ; (b)$\alpha_M =1.4 $, $\alpha_s = 1.93 $, $\alpha_\varphi = 2.17$ and $\alpha_\psi =2.39 $ .}\label{figtrans_2dot1}
\end{figure}

As the dimension increases above $d = 2$, magnetism is now allowed, but it is still relatively weak, and the magnetic transition temperature only reaches the bond-order transition temperatures for small $\alpha$, at which point the two bond-order transitions are already simultaneous and first order.  We show two example phase diagrams in Fig. \ref{figtrans_2dot1}, in the $(\alpha, \bar{r}_0)$ plane for two representative values of $\beta$.

\begin{figure}[h]
\begin{centering}
\includegraphics[width=\columnwidth]{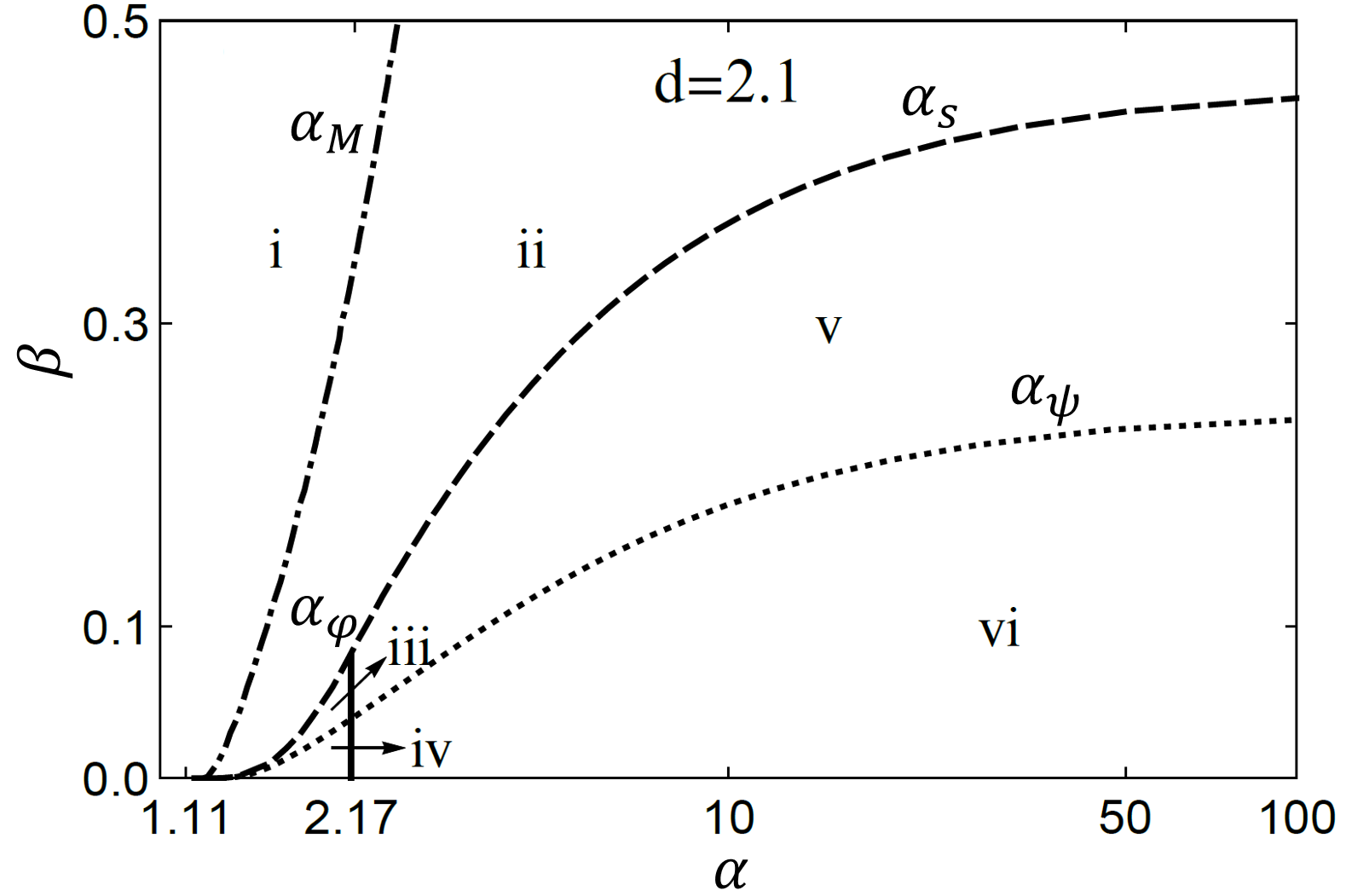} 
\par\end{centering}

\caption{The phase diagram in the $(\alpha, \beta)$ plane for $d=2.1$. The phase space is divided into six different classes of behavior by $\alpha_\varphi$ (vertical, solid), $\alpha_\psi$ (dotted), $\alpha_M$ (dot-dashed) and $\alpha_s$ (dashed): \textbf{i}: $T_{\varphi 1} =T_{\psi 1} =T_{M 1}$; \textbf{ii}: $T_{\varphi 1} =T_{\psi 1} > T_{M 2}$; \textbf{iii}: $T_{\varphi 1} >T_{\psi 1} >T_{M 2}$; \textbf{iv}: $T_{\varphi 1} >T_{\psi 2} > T_{M 2}$; \textbf{v}: $T_{\varphi 2} >T_{\psi 1} > T_{M 2}$; \textbf{vi}: $T_{\varphi 2} > T_{\psi 2} > T_{M 2}$. The notation in defined in Sec. \ref{subsec3_1}. As $\beta \rightarrow 0$, $\alpha_{\psi /M /s} $ approaches $\alpha_0 = 1.11$ for $d=2.1$. For $d=2.1$, $\alpha_{\varphi} = 2.17$ and intersects with $\alpha_\psi$ and $\alpha_s$ at $\beta_{\varphi \psi}=0.04$ and $\beta_{\varphi s}= 0.08$ respectively. }\label{fig:alpha1_vs_alpha2_2dot1}
\end{figure}

In Fig. \ref{fig:alpha1_vs_alpha2_2dot1}, we plot the four critical values of $\alpha$ versus $\beta$.  For $d = 2.1$, there are six possible classes of behavior, in contrast to the five classes for $d = 2$.  These are described in the caption and are separated by the critical $\alpha_{s/M/\varphi/\psi}(\beta)$'s discussed above.  Two of these critical lines asymptote to finite values of $\beta$ as $\alpha \rightarrow \infty$: the tricritical point where $\psi$ becomes first order, $\alpha_\psi$ asymptotes to $\beta_\psi = 0.245$; and the critical point where $T_\varphi = T_\psi$, $\alpha_s$ asymptotes to $\beta_s = 0.46$.  However, the intersection of magnetic and bond-order transitions, $\alpha_M$ does not asymptote to a finite value of $\beta$, at least not within the realm of validity of our approach, $\beta < 1$.  

As the dimensionality increases, the phase diagram in the $(\alpha, \beta)$ plane maintains the same topology up to $d = 2.4$, but with all lines pushed down and out to the right.  However, the $\alpha_M$ line decreases more rapidly and touches $\alpha_s$ for $d = 2.4$, as shown in Fig. \ref{fig:PD_all} (a).  Moreover, $\alpha_M$ begins to asymptote to a finite $\beta_M < 1$ for larger $d$'s.  As the dimensionality continues to decrease, $\alpha_M$ moves through $\alpha_s$, intersecting it at two points, and creating two new regions \textbf{vii} and \textbf{viii}, and a ``reentrant'' pocket of region \textbf{ii}, as is shown in Fig. \ref{fig:PD_all} (b), for $d= 2.45$.  
Region \textbf{vii}(\textbf{viii}) consists of a first(second) order transition of $\varphi$ followed by simultaneous first order transitions of $\psi$ and $M$.
Finally, at $d = 2.55$, the lower intersection point disappears, and $\alpha_M$ and $\alpha_s$ asymptote to the same $\beta_s = \beta_M$, causing region \textbf{ii} to vanish completely from the phase diagram.  As the dimensionality continues to increase, $\alpha_M$ is completely below $\alpha_s$, and while all lines continue to move out to larger $\alpha$ and shrink towards $\beta = 0$, the topology of the phase diagram remains the same out to three dimensions.  The phase space of region \textbf{i}, where all three transitions are simultaneous and first order continuously grows until it takes over the whole phase diagram in three dimensions. 

\begin{figure}[!ht]
\begin{centering}
\includegraphics[width=0.9\columnwidth]{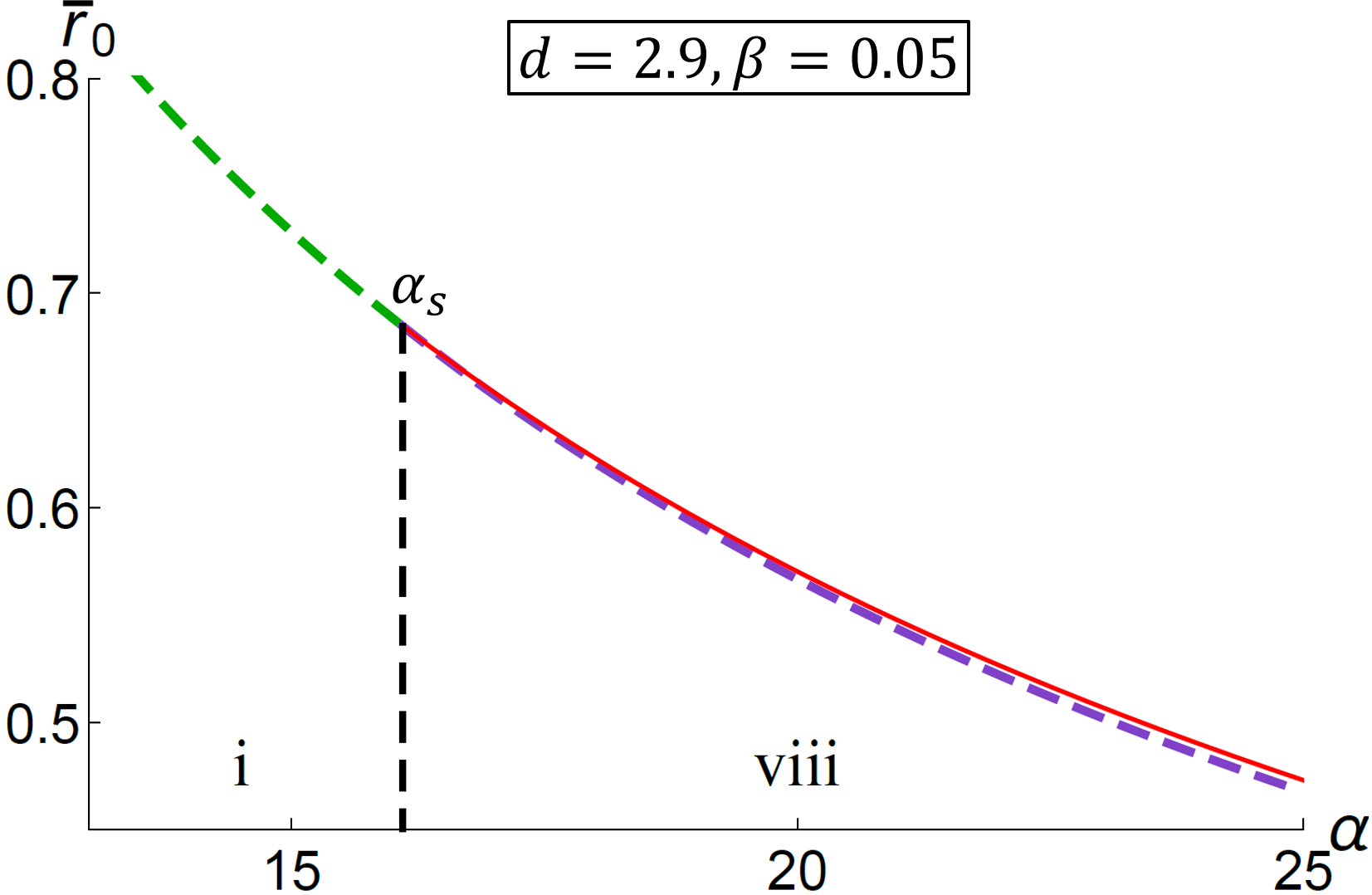} 
\par\end{centering}
\caption{(Color online) The phase diagram in the $(\alpha, \beta)$ plane at $d=2.9$ and $\beta=0.05$, which shows first(dashed lines) and second(solid lines) transitions of $\varphi$(red line), $\psi$ and the magnetic order. There are totally two regions of different classes of behavior separated by $\alpha_s=16.1$. \textbf{i}: $T_{\varphi 1} =T_{\psi 1} = T_{M 1}$; \textbf{viii}: $T_{\varphi 2} > T_{\psi 1} = T_{M 1}$. The notation is defined in Sec. \ref{subsec3_1}. The thick dark green dashed line represents simultaneous first order transitions of $\varphi$, $\psi$ and the magnetic order while the thick dark purple dashed line indicates simultaneous first order transitions of $\psi$ and the magnetic order $M$. }\label{fig:r0bar_vs_alpha_2dot9}
\end{figure}

\begin{figure*}[!htbp]\centering
   \begin{minipage}{\columnwidth}
     \includegraphics[width=\linewidth]{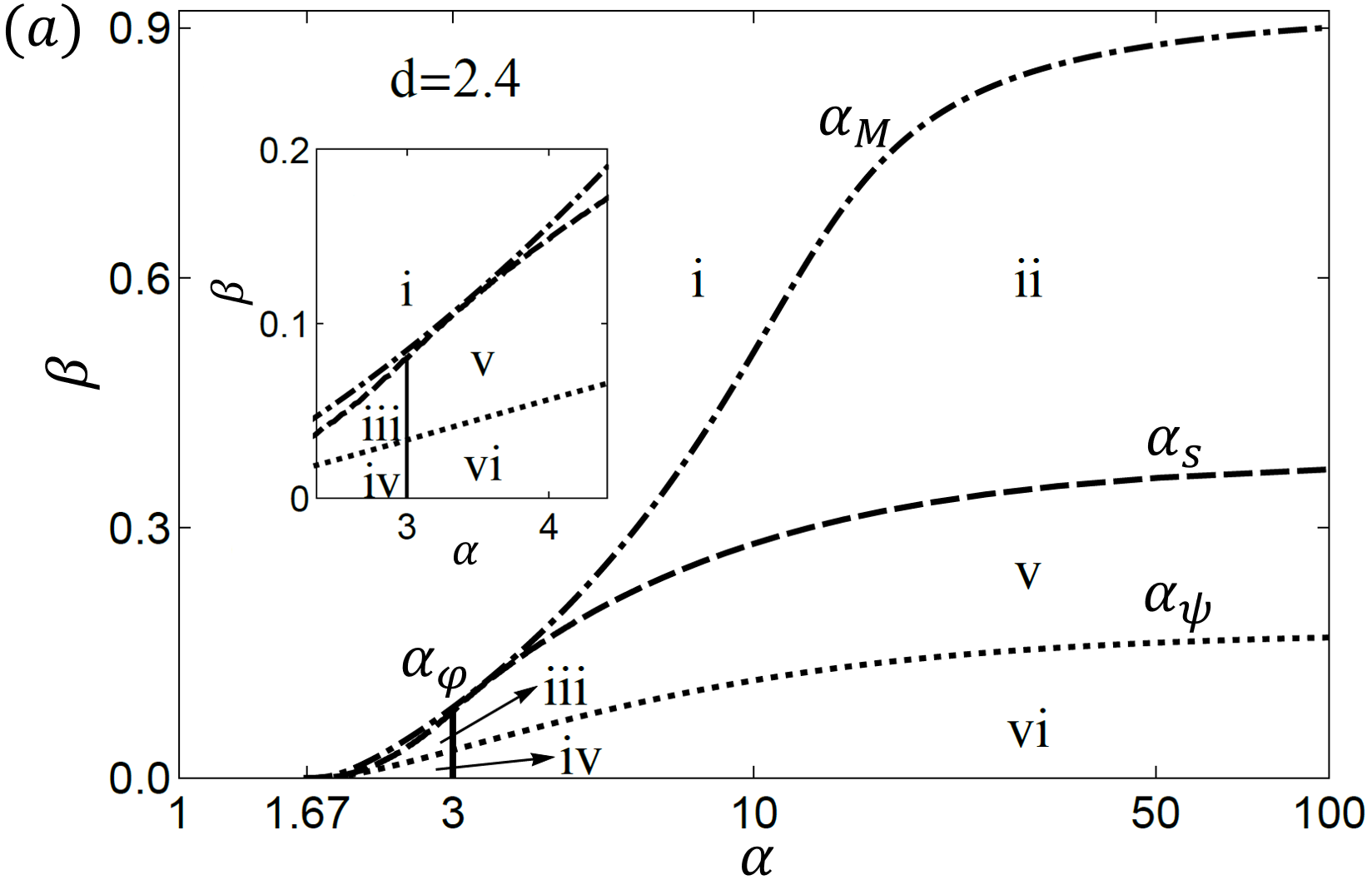}
   \end{minipage}
   \begin{minipage}{\columnwidth}
     \includegraphics[width=\linewidth]{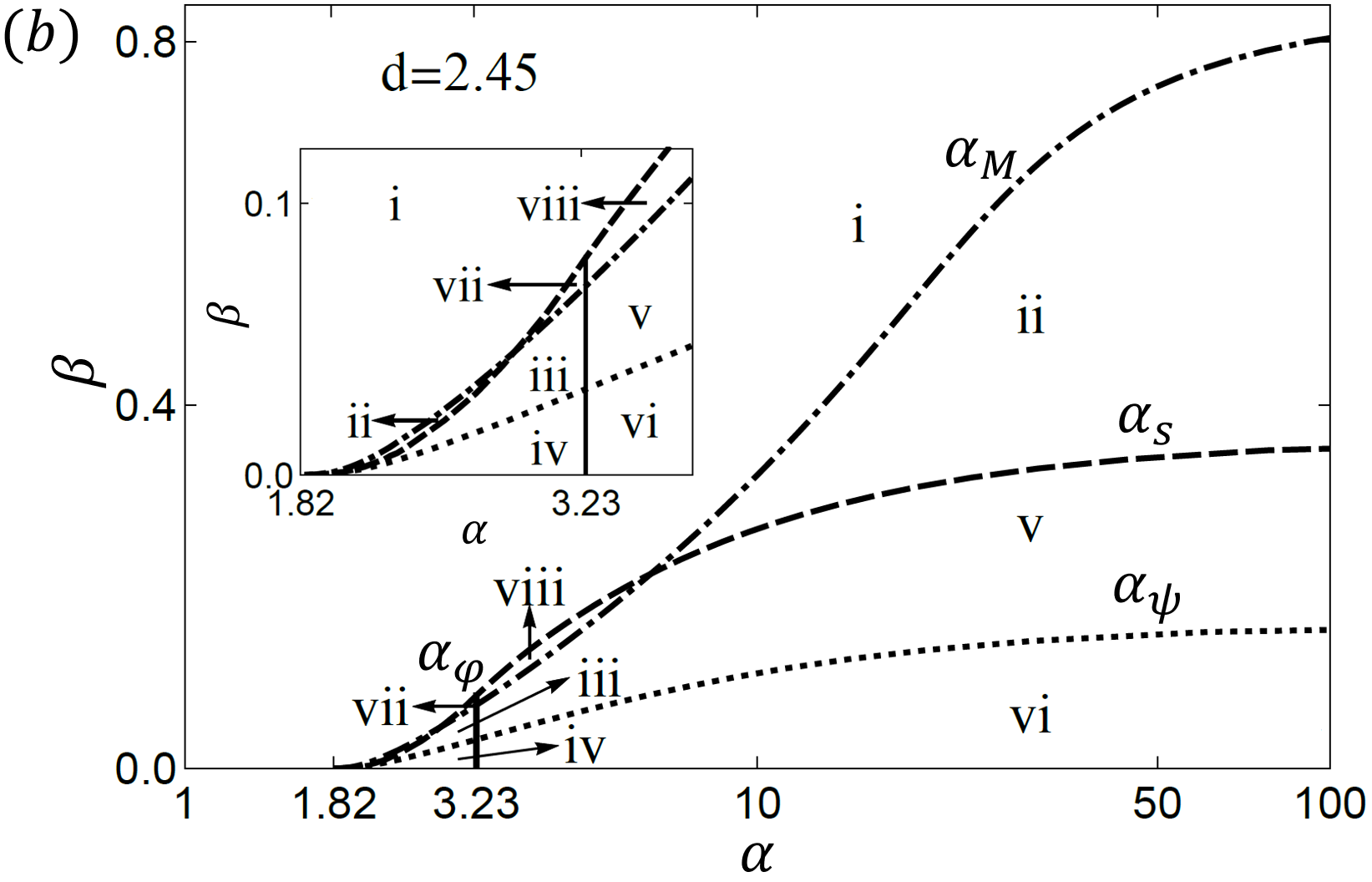}
   \end{minipage}
      \begin{minipage}{\columnwidth}
     \includegraphics[width=\linewidth]{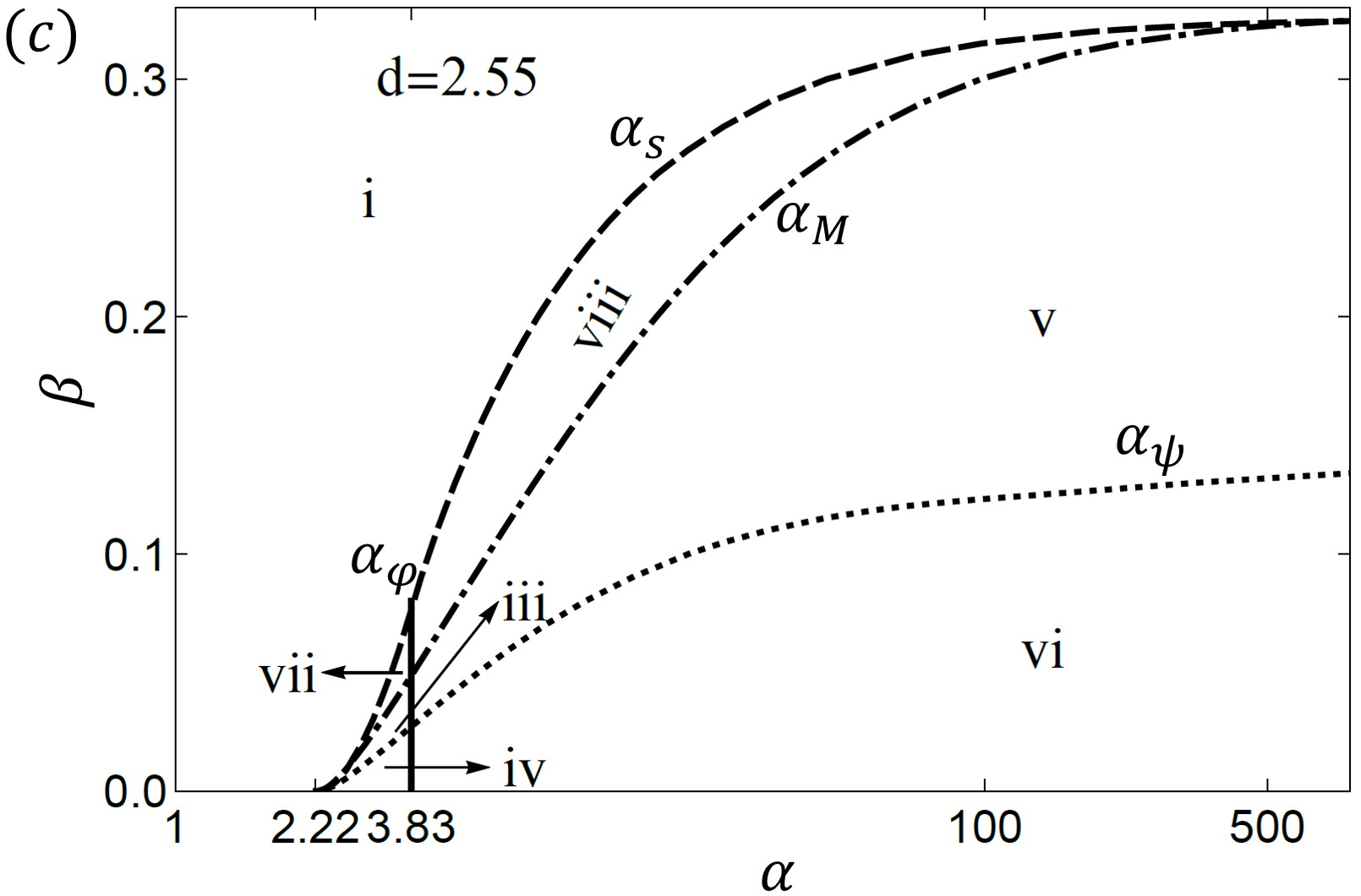},
   \end{minipage}
   \begin{minipage}{\columnwidth}
     \includegraphics[width=\linewidth]{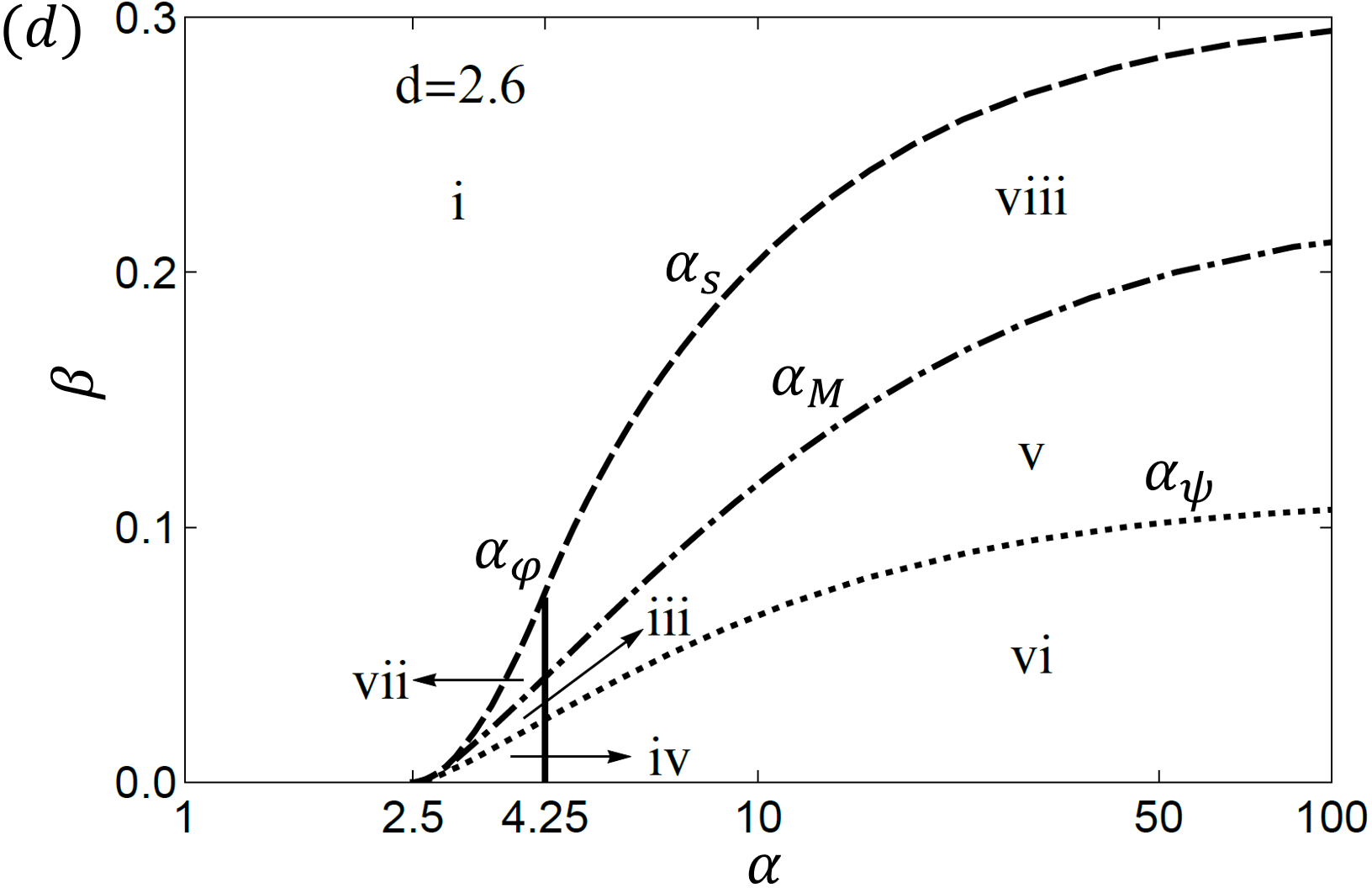}
   \end{minipage} 
\justify \caption{Evolution of the transition values separating regions of different phase transition behavior $\alpha_\varphi$(solid), $\alpha_\psi$(dotted), $\alpha_M$(dot-dashed) and $\alpha_s$(dashed) in the phase diagram in the $(\alpha, \beta)$ plane as the dimensionality increases for $2<d<3$. 
Figure (a) displays the critical situation where the dashed line touches the dot-dashed line at $d=2.4$, as seen more clearly in the inset. In general for $2<d<2.4$, there are totally six regions in the phase diagram labeled as \textbf{i} to \textbf{vii}. \textbf{i}: $T_{\varphi 1} =T_{\psi 1} =T_{M 1}$; \textbf{ii}: $T_{\varphi 1} =T_{\psi 1} > T_{M 2}$; \textbf{iii}: $T_{\varphi 1} >T_{\psi 1} >T_{M 2}$; \textbf{iv}: $T_{\varphi 1} > T_{\psi 2} > T_{M 2}$; \textbf{v}: $T_{\varphi 2} > T_{\psi 1} > T_{M 2}$; \textbf{vi}: $T_{\varphi 2} > T_{\psi 2} > T_{M 2}$. 
Figure (b) shows at $d=2.45$, two more regions emerge, thus giving rise to totally eight regions of different classes of behavior. \textbf{vii}: $T_{\varphi 1} > T_{\psi 1} = T_{M 1}$; \textbf{viii}: $T_{\varphi 2} > T_{\psi 1} = T_{M 1}$. The inset shows the dense regions at small $\alpha$ and $\beta$. The $\alpha_M$ transition line crosses the $\alpha_s$ transition line twice at $(\alpha, \beta) = (2.79, 0.045)$ and $(6.56, 0.215)$.
Figure (c) shows the seven phase regions at $d=2.55$ where the dashed line merges with the dot-dashed line at large $\alpha$. 
Figure (d) is for $d=2.6$, which has totally seven phase regions. The notation is defined in Sec. \ref{subsec3_1}. The corresponding asymptotic values of $\beta$ as $\alpha_{\psi /M /s}$ approaches infinity are: (a) $\beta_{\varphi }=0.175, \beta_M=0.92$ and $\beta_s=0.38$; (b) $\beta_{\varphi }=0.16, \beta_M=0.85$ and $\beta_s=0.36$; (c) $\beta_{\varphi }=0.135$ and $\beta_M=\beta_s=0.33$; (d) $\beta_{\varphi }=0.11, \beta_M=0.23$ and $\beta_s=0.3$. } \label{fig:PD_all}
\end{figure*}

\begin{figure}[H]
\begin{centering}
\includegraphics[width=\columnwidth]{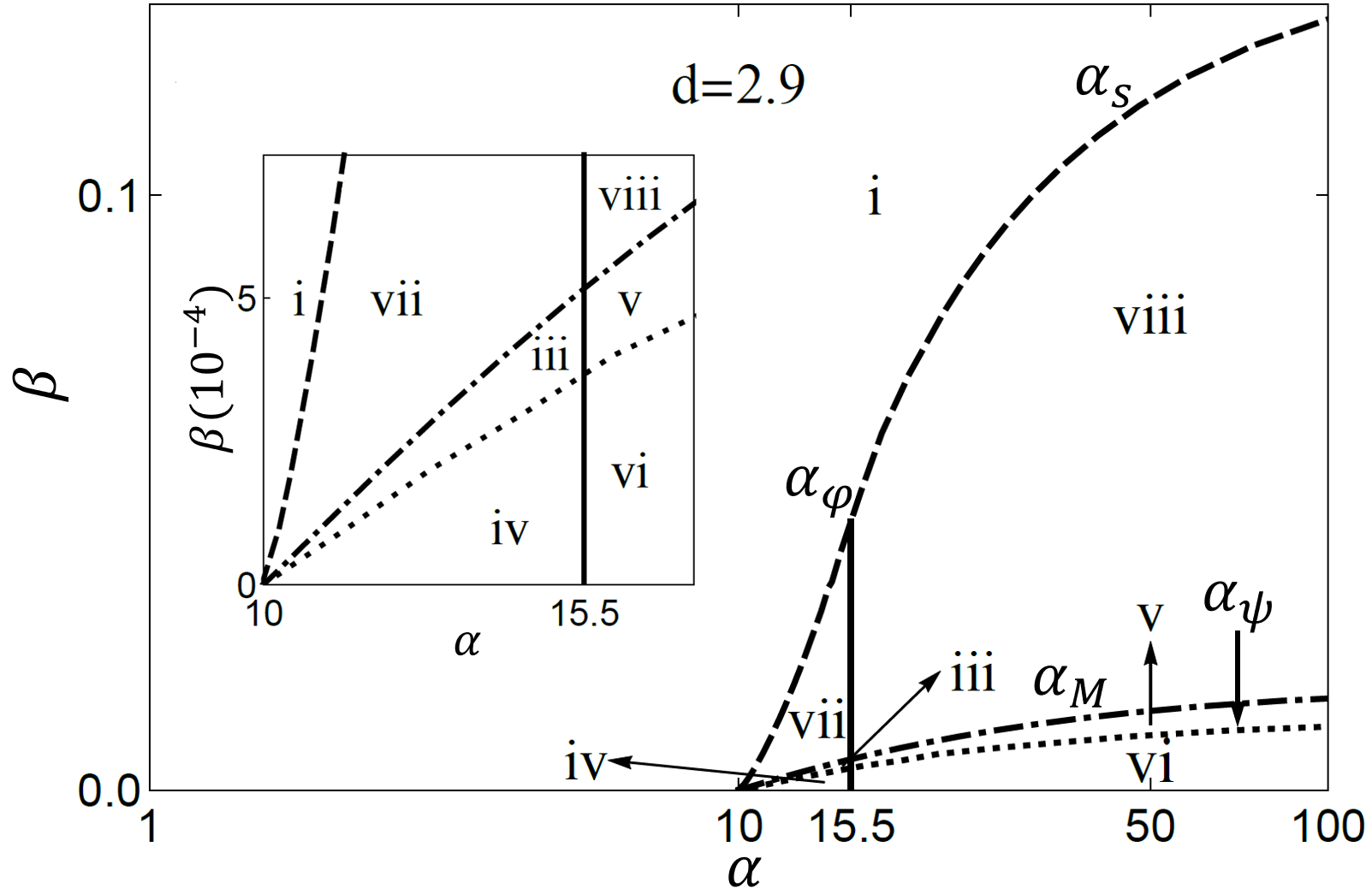} 
\par\end{centering}

\caption{The phase diagram in the $(\alpha, \beta)$ plane for $d=2.9$. The different regions are defined in Fig. \ref{fig:PD_all}. The inset shows the dense regions at small $\beta$. The corresponding asymptotic values of $\beta$ as $\alpha_{\psi /M /s}$ approaches infinity are: $\beta_{\varphi }=0.012, \beta_M=0.018$ and $\beta_s=0.14$. }\label{fig:PD_d=2dot9}
\end{figure}

We show the behavior for $d = 2.9$ in Figures \ref{fig:r0bar_vs_alpha_2dot9} and \ref{fig:PD_d=2dot9}, showing a representative phase diagram in the $(\alpha, \bar{r}_0)$ plane for $\beta = 0.05$ and the phase diagram in the $(\alpha,\beta)$ plane, respectively.

\section{Conclusions} \label{sec4}

In this paper, we explored how a double-stripe magnetic order that breaks two discrete lattice symmetries can be melted by fluctuations in up to three different stages, realizing two distinct spin-driven bond-order phases.  The first, nematic phase is captured by a next-nearest neighbor Ising bond order, $\varphi$ that breaks the $C_4$ rotational symmetry to $C_2$, while the second phase is captured by a dimerized nearest neighbor Ising bond order, $\psi$, which breaks both translation and diagonal mirror symmetries. As $\psi$ also breaks the $C_4$ rotational symmetry, it can only develop below or simultaneously with $\varphi$. We developed an effective field theory to study the interplay of these different transitions, as a function of changing dimensionality and relative biquadratic coupling strengths.  While in three dimensions, all three transitions are simultaneous and first order, in lower dimensions the phase diagram can become quite complex, with up to eight different regions of behavior classified by which transitions become simultaneous in addition to the first/second order nature of each transition. 

Double-stripe magnetism is realized in the ``11" iron-based superconductors $\mathrm{Fe}_{1+y} \mathrm{Te}_{1-x}\mathrm{Se}_x$, which has a simultaneous first order nematic and magnetic transition.  It has also been predicted by density functional theory as the ground state for $\mathrm{Ba Ti}_2 \mathrm{Sb}_2 \mathrm{O}$, which may show a weakly first order nematic ($\varphi$ and $\psi$) transition and no observed magnetic transition\cite{Zhang2017}.

\begin{acknowledgments}
This research was supported in part by Ames Laboratory Royalty Funds and Iowa State University startup funds. The Ames Laboratory is operated for the U.S. Department of Energy by Iowa State University under Contract No. DE-AC02-07CH11358. R.A.F. also acknowledge the hospitality of the Aspen Center for Physics, supported by National Science Foundation Grant No. PHYS-1066293 where this project was initiated. The authors also acknowledge valuable discussions with Rafael M. Fernandes, Igor I. Mazin, James K. Glassbrenner and John van Dyke.
\end{acknowledgments}

\bibliography{references}
\bibliographystyle{apsrev4-1}

\end{document}